\RequirePackage{fix-cm}
\documentclass[twocolumn,epjc3]{svjour3}  
\smartqed  % flush right qed marks, e.g. at end of proof

\RequirePackage{graphicx}
\RequirePackage{color}
\RequirePackage[colorlinks,citecolor=blue,urlcolor=blue,linkcolor=blue]{hyperref}
\usepackage{multirow}
\usepackage[T1]{fontenc}
\usepackage[latin1]{inputenc}
\usepackage{amsmath}
\usepackage{arydshln}
\usepackage[switch]{lineno}
\journalname{Eur. Phys. J. C}
\begin{document}

\title{Neutron-induced background in the CONUS experiment}

\author{J.~Hakenm\"uller\thanksref{e1,addr1,*} \and C.~Buck\thanksref{addr1} \and K.~F\"ulber\thanksref{addr2} \and G.~Heusser\thanksref{addr1} \and  T.~Klages\thanksref{addr3} \and M.~Lindner\thanksref{addr1} \and A.~L\"ucke\thanksref{addr3} \and  W.~Maneschg\thanksref{addr1} \and M.~Reginatto\thanksref{addr3} \and T.~Rink\thanksref{addr1} \and T.~Schierhuber\thanksref{addr1} \and D.~Solasse\thanksref{addr2} \and H.~Strecker\thanksref{addr1} \and R.~Wink\thanksref{addr2} \and M.~Zbo\v{r}il\thanksref{e2,addr3} \and A.~Zimbal\thanksref{addr3}
}

\thankstext{e1}{e-mail: janina.hakenmueller@mpi-hd.mpg.de}
\thankstext{e2}{e-mail: miroslav.zboril@ptb.de}
\thankstext{*}{corresponding author}
%\authorrunning{Short form of author list} % if too long for running head

\institute{Max-Planck-Institut f\"ur Kernphysik, Saupfercheckweg 1, 69117 Heidelberg, Germany \label{addr1}
           \and
          Preussen Elektra GmbH, Kernkraftwerk Brokdorf, Osterende, 25576 Brokdorf, Germany \label{addr2}
           \and
        Physikalisch-Technische Bundesanstalt, Bundesallee 100, 38116 Braunschweig, Germany \label{addr3}
}

\date{Received: date / Accepted: date}
% The correct dates will be entered by the editor
\maketitle

\begin{abstract}
CONUS is a novel experiment aiming at detecting elastic neutrino nucleus scattering in the almost fully coherent regime using high-purity germanium (Ge) detectors and a reactor as antineutrino source. The detector setup is installed at the commercial nuclear power plant in Brokdorf, Germany, at a close distance to the reactor core to guarantee a high antineutrino flux. A good understanding of neutron-induced backgrounds is required, as the neutron recoil signals can mimic the predicted neutrino interactions. Especially events correlated with the reactor thermal power are troublesome. 
On-site measurements revealed such a correlated, highly thermalized neutron field with a maximum fluence rate of (745$\pm$30)\linebreak[0]\,cm$^{-2}$d$^{-1}$. 
These neutrons, produced inside the reactor core, are reduced by a factor of $\sim$10$^{20}$ on their way to the CONUS shield.
With a high-purity Ge detector without shield the $\gamma$-ray background was examined including thermal power correlated $^{16}$N decay products and neutron capture $\gamma$-lines.
Using the measured neutron spectrum as input, Monte Carlo simulations demonstrated that the thermal power correlated field is successfully mitigated by the CONUS shield. The reactor-induced background contribution in the region of interest is exceeded by the expected signal by at least one order of magnitude assuming a realistic ionization quenching factor. 

\end{abstract}

%\vspace{1.0cm}
\section{Introduction}\label{sec:introduction}

Coherent elastic neutrino nucleus scattering (CE$\nu$NS) is a purely neutral weak interaction with a large variety of physics applications. These span from supernovae dynamics and nuclear form factors to the search for phenomena beyond the Standard Model: deviations from the Weinberg angle at MeV scale, electromagnetic properties of neutrinos as well as non-standard interactions in the neutrino-quark sector. Even though predicted in 1973 \cite{Freedman:1973yd}, CE$\nu$NS has eluded detection for more than four decades mainly due to technological difficulties in observing tiny nuclear recoils below few keV$_{ee}$ of ionization energy. It was observed for the first time in 2017 by the COHERENT experiment \cite{Coherent:2017}. Furthermore, the demand for very intense neutrino fluxes e.g. pion-decay-at-rest sources or commercial nuclear reactors, requires that the experiments are built close to these neutrino sources. At such shallow depth locations, several background components can aggravate the attempt of detecting CE$\nu$NS.\\
CONUS is a novel experiment which aims at detecting CE$\nu$NS signals using reactor antineutrinos. Since April 1, 2018, it is operational at the nuclear power plant in Brokdorf (Kernkraftwerk Brokdorf; KBR) \cite{PreussenElektra}, Germany, where it is located at an average depth of 24\,m of water equivalent (m w.e.) and 17.1\,m distance to the reactor core center. Four ultra-low threshold, high-purity germanium (HPGe) detectors are embedded in a multi-layer shield, profiting from decades-long developments for low-background Ge $\gamma$-ray spectroscopy \cite{Heusser:1995wd,Heusser:2015} at the Max-Planck-Institut f\"ur Kernphysik (MPIK) in Heidelberg, Germany. While for most applications (such as the selection of intrinsic radiopure materials) neutron-induced backgrounds were not critical so far, these become relevant for CONUS-like experiments. Thus, all potential neutron sources at the KBR reactor site had to be inquired first: cosmogenic neutrons induced by muons in the reactor building and in the CONUS shield; neutrons from the spent fuel storage pond above the experiment; ($\alpha$,n) reactions from natural radioactivity in the surrounding concrete walls and basements; neutrons from the reactor core; and $\gamma$-radiation from neutron-induced isotopes decaying along the primary coolant of the pressurized water reactor. Whereas the first three classes are steady-state sources, the latter two are potentially troublesome. Both can mimic CE$\nu$NS signals, since they are correlated with the thermal power and can contribute counts to the region of interest. Quantifying these backgrounds via independent measurements and determining their impact on the CONUS HPGe detector energy spectra are of fundamental importance. In order to achieve a high accuracy, the CONUS collaboration and the Neutron Radiation Department of the Physikalisch-Technische Bundesanstalt (PTB) in Braunschweig, Germany, developed an extensive measurement program and validation procedures. The multi-variate approach included neutron and $\gamma$-ray detection techniques, multiple measurement campaigns during reactor ON/OFF times at the experimental site, scans of different room positions, measurements inside and outside the CONUS shield, measurements at the reactor site and at the MPIK underground laboratory under similar overburden conditions, and the deployment of high and low activity $^{252}$Cf neutron sources. In addition, Geant4-based Monte Carlo (MC) simulations were performed for all these configurations. In a first step, the measurements helped to validate the MC code in terms of neutron generation and propagation. In a second step, the MC simulations were used to support and interpret the neutron measurement results in detail. Finally, they were used to predict the impact of the measured thermal power correlated neutrons and $\gamma$-ray flux on the CONUS HPGe detectors.\\
This article focuses on the neutron (direct) and neutron-induced $\gamma$-ray (indirect) measurements and related MC simulations for the CONUS experimental site. The article is structured as follows: Section~\ref{sec:neutron-detectors} describes the direct and indirect neutron detection techniques and the thermal power determination. Section~\ref{sec:neutron-measurements} presents the reactor environment and Section~\ref{sec:neutron-MC} the implementation of the reactor simulation. In Section~\ref{sec:Bonnersphereall}, the Bonner sphere neutron measurements and the results including the comparison to MC expectations are discussed in detail. In the same way, the measurements with the HPGe spectrometer CONRAD including the comparison to MC are summarized in Section~\ref{subsec:conradoutsideshield}. Finally, Section~\ref{sec:impact-on-cevns} describes the MC simulation of the measured neutron field and $\gamma$-ray background passing through the CONUS shield to investigate the impact of the neutron-induced signals on the energy spectra of the CONUS detectors.

\section{Description of neutron sensitive devices used in this work}\label{sec:neutron-detectors}

\subsection{Direct neutron detection}
\label{subsec:directndet}
\begin{figure}
	\centering
	\includegraphics[width=1.0\columnwidth]{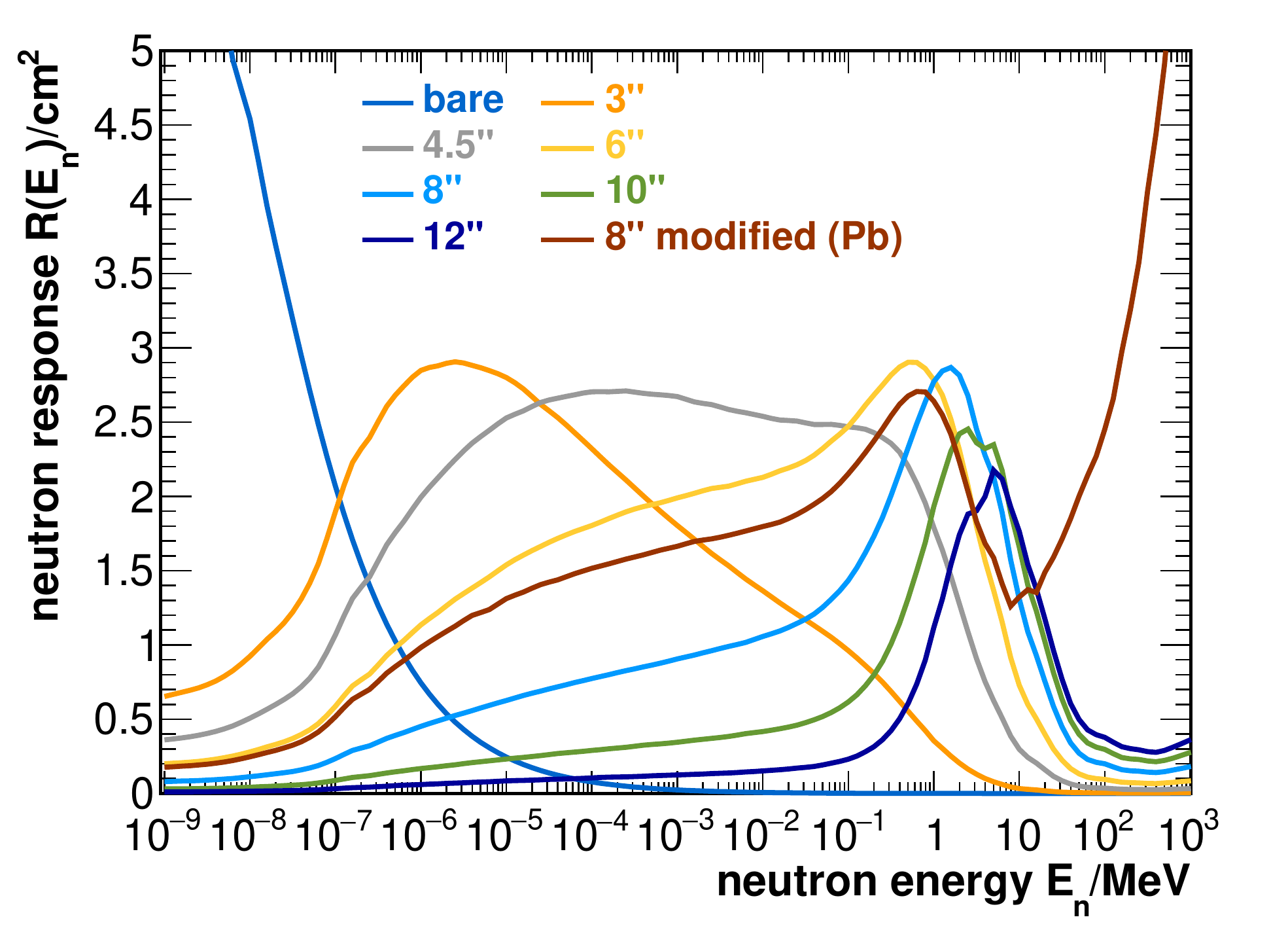}
	\caption{Neutron response functions of the Bonner spheres used for the measurement. At thermal neutron energies $E_\text{n}\sim10^{-9}$\,MeV the response of the bare detector rises to $R_\text{th}=6.2\,\text{cm}^2$}
	\label{fig:NEMUS_RF}
\end{figure}

A Bonner sphere Spectrometer (BSS)~\cite{Thomas2002,Alevra2003} consists of a set of moderating spheres with different diameters and a thermal neutron sensor that is placed at the centre of each sphere. Each sphere plus thermal sensor combination has a different energy-dependent response to neutrons. The peak of the neutron response function shifts to higher neutron energies as the size of the moderator increases (the responses of the Bonner spheres used for the measurements at KBR are shown in Figure~\ref{fig:NEMUS_RF}). It is usual practice to measure also with the thermal neutron sensor without a moderating sphere (i.e. the bare detector). %Any combination of sphere and thermal neutron sensor is usually called a ``sphere'' and this terminology has also been extended to the bare detector.

The measurements of the neutron background at KBR were carried out with the BSS NEMUS~\cite{Wiegel2002} of PTB. It consists of ten polyethylene (PE) spheres with diameters\footnote{In this paper we use the convention of labeling each Bonner sphere by its diameter in inches, $1\,\text{in}=2.54\,\text{cm}$.} 3, 3.5, 4, 4.5, 5, 6, 7, 8, 10 and 12\,''. The set also contains a bare detector (diameter 3.2\,cm), a Cd-covered detector, and four modified spheres with lead (Pb) and copper (Cu) shells. Thanks to the metal, embedded in the PE spheres, the response functions dramatically increase for neutron energies above $E_\text{n}\sim50$\,MeV. The central thermal neutron sensors are spherical $^3$He-filled proportional counters (type SP9, company Centronic Ltd. \cite{he3counter}), detecting the thermalized neutrons via the reaction:
\begin{equation}
\text{n} + {}^{3}\text{He} \rightarrow {}^{3}\rm{H} + p + Q\text{, with Q=764\,keV.}
\end{equation}
For the measurements carried out at KBR, we used SP9 counters with $^3$He pressure of $\sim200$\,kPa. In order to cover the complete energy range of the expected neutron field and yet minimize the required measurement time, we chose a minimal subset of spheres, namely a bare counter, 3, 4.5, 6, 8, 10 and 12\,'', plus the modified sphere of 8\,'' diameter containing a Pb shell of 1\,'' thickness. The inclusion of the modified sphere improves the spectrometric properties of the system at higher energies and allows to check for the presence of high-energy cosmic-ray induced neutrons ($E_\text{n}\sim100$\,MeV \cite{Wiegel2002b}), even though due to the massive concrete shield above the CONUS site their contribution is expected to be very small.

\begin{figure}
	\centering
	\includegraphics[width=1.0\columnwidth]{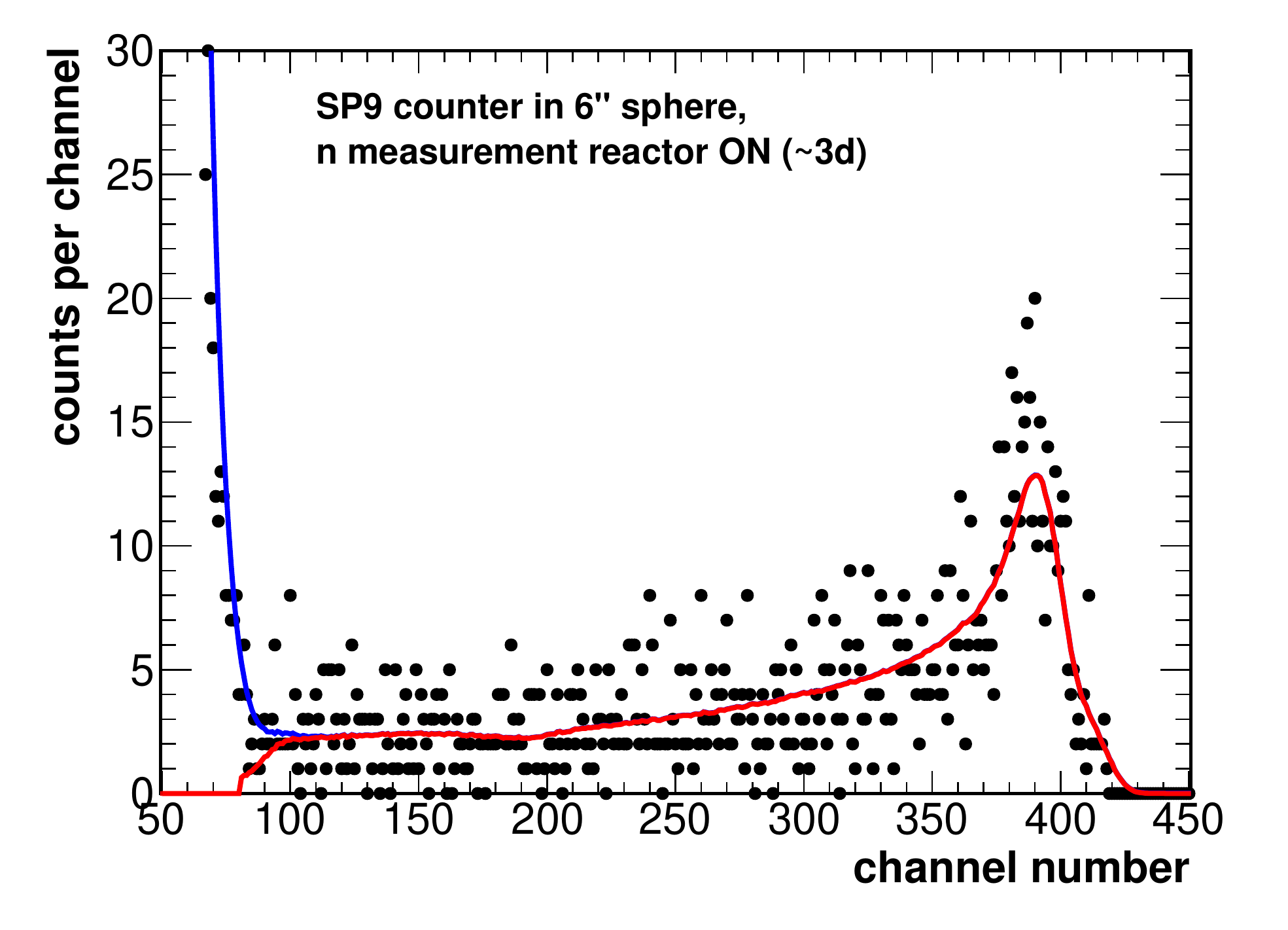}
	\caption{Pulse height spectrum of a $^3$He-filled proportional counter acquired after 3\,days of measurement during the reactor ON period. The solid blue line depicts the total fit function including the background. The solid red line shows the fit function describing the neutron-induced component of the PHS}
	\label{fig:PHS}
\end{figure}

In general, the neutron count rates measured at KBR were very low, of the order of $\sim10$\,counts per hour and detector, or less. To determine the number of neutron-induced events in the pulse height spectra (PHS) recorded with the SP9 counter, the procedure, previously developed for measurements in underground laboratories \cite{Zimbal2013,Reginatto2013,Reginatto2018}, was applied. Figure~\ref{fig:PHS} shows a typical PHS, together with the fit function to describe the PHS shape and to extract the neutron signal from background.

\subsection{Indirect neutron detection}
\label{subsec:indirectndet}
\begin{figure}[h]
    \includegraphics[width=0.52\textwidth]{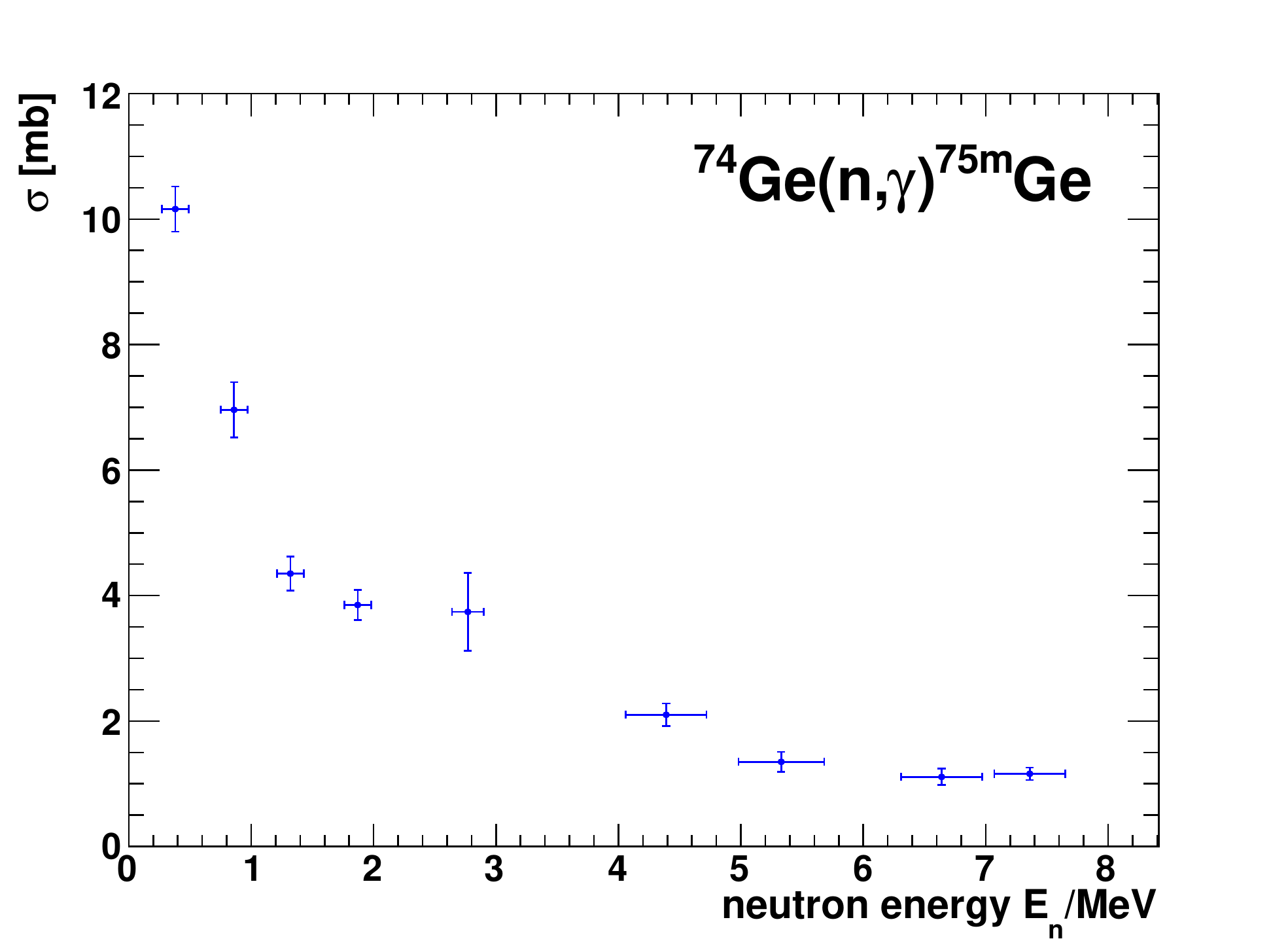}
    \caption{Cross section for neutron capture on $^{74}$Ge to the 139.7\,keV state of $^{75m}$Ge, data from \cite{bhike:2015}. The cross section for thermal neutrons with (130.5$\pm$5.6)\,mb \cite{Meierhofer2010} exceeds the values at higher energy shown in the plot by one order of magnitude}
    \label{fig:Ge_cap_crosSection}
    \end{figure}

HPGe spectrometers with and without shield can be used to indirectly gain information on neutron fluences.
This is possible by the detection of the $\gamma$-rays emitted after neutron capture either in the vicinity of the Ge detector or in the detector material itself. Mostly thermal neutrons are captured, but there is also a contribution from higher energetic neutrons (see Figure~\ref{fig:Ge_cap_crosSection}).  

Without any shield, $\gamma$-rays from neutron capture in the building structure (e.g.~reinforced concrete with steel) can be seen by the HPGe detector. The spectrum below 2700\,keV$_{ee}$ is dominated by natural radioactivity, but for neutron captures, higher energetic lines at up to $\sim$10\,MeV are emitted, where nearly no other background is expected. 
%For example for $^{56}$Fe(n,$\gamma$)$^{57}$Fe, the line with the maximum branching ratio lies at 7631\,keV \cite{capgamdatabase}. 
From the resulting spectrum, the isotopes found in the environment can be identified and from the $\gamma$-line count rate the neutron fluence rate can be estimated. In order to do this, MC simulations including the geometry of the detector and the location with the correct material compositions are required. In the MC simulation, the neutron captures are reproduced including the capture probability over the whole neutron energy range. The method is less precise than a direct measurement with Bonner spheres as described in Section \ref{subsec:directndet}, but it allows to estimate independently a fluence, to support results from direct measurements or to validate a MC simulation. For the CONUS experiment, this is done in a measurement with the CONRAD (CONus RADiation) HPGe spectrometer at the location of the nuclear power plant. The results can be found in Section \ref{subsec:conradoutsideshield}. 

Gaining information from neutron capture is mandatory, if a direct measurement is not possible e.g.\,to study the neutron fluence at a HPGe diode within a shield that cannot be opened anymore. 
While the $\gamma$-radiation from outside is highly suppressed, lines from neutron capture within the shield material as well as in the Ge of the detector become visible. At shallow depth, many of these lines are induced by the neutrons created via muon capture inside the shield. There can also be contributions from neutrons fluences from outside propagated through the shield. To be able to detect these $\gamma$-lines, a low background level within the shield is required. This is usually achieved by applying a muon anti-coincidence system (so-called "muon veto"). In this way, especially $\gamma$-lines from metastable Ge states 
%($^{71m}$Ge with T$_{1/2}$=20.4\,ms, $^{73m}$Ge with T$_{1/2}$=0.5\,s and $^{75m}$Ge with T$_{1/2}$=47.7\,ms) 
with half-lives longer than a veto window in the range of a few hundred ${\mu}$s become clearly visible \cite{Heusser:2015}. Once again, the neutron fluence can be estimated from the $\gamma$-line count rate or measurements of the line count rates can be used to validate the neutron production inside the shield in the MC simulation. This has been done for a detector at the MPIK laboratory in detail \cite{hakenmueller:masterthesis}. For the CONUS experiment, this will be featured in an upcoming publication. %If there is hydrogenous or borated materials in the vicinity of the detectors, the 2.2\,MeV $\gamma$-line from neutron capture on hydrogen or $\gamma$-rays emitted by neutron capture on boron (B) \cite{capgamdatabase} can be studied as well to support the results from the Ge lines \cite{hakenmueller:masterthesis}. 
 %During the commissioning of the CONUS setup at the MPIK laboratory even a direct measurement of the neutron fluence inside the shield using a Bonner sphere had been possible.
 
In addition to neutron capture, neutrons in Ge can also undergo inelastic and elastic scattering depending on their energy. In these interactions, there will be Ge recoils with an identical signature as the one for CE$\nu$NS. Thus, this highly relevant background for the CONUS experiment will be discussed in detail in Section \ref{sec:impact-on-cevns}. %The characteristics of the CONUS detectors are discussed below.

\subsubsection{Specifications of the HPGe spectrometer CONRAD}

For background studies without shield at the reactor site the ultra-low background p-type coaxial HPGe detector CONRAD ($m=$2.2\,kg), with the diode refurbished by Mirion Technologies, Canberra Olen \cite{canberraolen}, is employed within the CONUS experiment. It has been used furthermore for background studies inside the \linebreak[4]CONUS shield during the commissioning phase of the experiment at the MPIK underground laboratory. The large detector mass is beneficial to especially detect high energetic $\gamma$-lines up to $\sim$11\,MeV$_{ee}$ as the detector also has a high geometric detection efficiency at these energies. To be allowed to set it up at KBR, the detector was upgraded with an electrical cryocooling system.

The detector has an active mass of (1.90$\pm$0.15)\,kg, which was determined as described in \cite{Heusser:2015}, \cite{budjas2009}, \cite{TobiasMasterarbeit}. The thickness of the inactive layer at the diode surface has been evaluated from the ratio of the absorption of $^{241}$Am $\gamma$-lines at different energies compared to a MC simulation and amounts to (2.5$\pm$0.1)\,mm. 
Moreover, to adjust to the measured line count rates from $^{60}$Co measurements in different positions, the bore hole dimensions were adapted. All in all, 85\% of the Ge crystal are active.%, because previous to its use in the CONUS experiment the detector had been stored at room temperature for at least 10\,years causing the Lithium from the p-type contact layer to diffuse further inside the diode. This increases the dead layer by about 1\,mm per year \cite{Debertin1984}. 

With the help of pulser scans over the whole energy range up to 11\,MeV$_{ee}$ it could be confirmed that the detection efficiency due to electronics is constant over the whole spectral range. Furthermore, the peak position has been found stable within 1\,keV$_{ee}$ over a period of 2\,months. 
A small non-linearity within the energy scale has been discovered making it necessary to calibrate separately the two energy regions, where $\gamma$-lines have been observed (below 2700\,keV$_{ee}$ and above 4500\,keV$_{ee}$), with two linear functions.

\subsubsection{Specifications of the CONUS HPGe spectrometers}

The CONUS experiment employs four 1\,kg ultra-low-background p-type point contact HPGe spectrometers equipped with an electrical cryocooling system (manufactured by MPIK and Mirion Technologies, Canberra Lingolsheim\linebreak[0]\cite{canberralingolsheim}). This is beneficial for a low noise threshold required to detect CE$\nu$NS as well as for the reactor environment, as no cryogenic liquids for cooling the diodes are allowed there. 
With various source and background measurements, the characteristics of the detectors have been determined and will be described in an upcoming publication. In the course of this publication, the specifications of detector 1, referred to as C1, are used exemplary to determine the expected measured spectrum of reactor neutrons at the diode. 
As for the CONRAD detector, the active volume has been determined from the $^{214}$Am source measurements at different positions compared to MC simulations. An inactive layer on the side and on top of the diode at the opposite of the point contact is assumed. While the detector is completely inactive at the diode surface, i.~e. inside the so-called "dead layer", this is not true for the transition layer, i.~e. the volume in between the dead layer and the active volume of the diode. In this transition layer, the charge collection efficiency decreases continuously towards the diode's surface. Thus, energy depositions outside of the active volume can induce counts in the spectrum, but these so-called "slow pulses" will be reconstructed at energy values below the original energy. The effect can be observed clearly in the energy range below the 59.54\,keV peak of the $^{214}$Am source measurements. Assuming a sigmondial shape for the decreasing charge collection efficiency in the MC of the source measurements as suggested in \cite{Queching:Bscholz}, the thickness of the transition layer can be evaluated by varying it and comparing the resulting shape to the source measurements. For C1, this amounts to about 30\% of the total layer thickness. The information is employed to correctly describe the spectral shape of a background contribution. Subtracting transition and dead layer, an active mass of (0.94$\pm$0.03)\,kg is determined. %, slightly below the value of 0.96\,kg provided by the manufacturer has been determined.

Moreover, with the help of pulser measurements the detector response towards the noise threshold was studied. The detection efficiency decreases towards the noise edge at around 300\,eV$_{ee}$.

%%%%%%%%%%%%%%%%%%%%%%%%%%%%%%%%%%%%%%%%%%%%%%%%%%%%%%%%%%%%%%%%%%%%%%%%%%%%%%%%%%%%%%%%%%%%%%%%%%%%
\subsection{Neutron flux-correlated reactor instrumentation}
\subsubsection{Absolute thermal power}
\label{subsec:absthermalpowe}
%\begin{itemize}
%	\item thermal power from secondary circuit: discussion of uncertainties (Dr. Fuelber: syst 4 
%    percent from error propagation (wants it to be checked by someone) and two documents (see Wiki),
%    stat error neglible, Christian: 0.4 per cent, which value to use?)
For a good estimation of the neutrons emitted in the reactor core as well as a precise prediction of the neutrino flux, one of the crucial reactor quantities is the thermal power. The thermal power in a nuclear reactor is given by the number of fissions times the energy released per fission summed over all fission isotopes. The relevant contributions are coming from the two uranium (U) isotopes, $^{235}$U and $^{238}$U, as well as the two plutonium (Pu) isotopes, $^{239}$Pu and $^{241}$Pu.

%The absolute thermal power of a commercial pressurized water reactor (PWR) as in Brokdorf can be estimated  by a calculation of the energy balance around the steam generator. 
The absolute thermal power of a commercial pressurized water reactor (PWR) as in Brokdorf is determined by monitoring the heat flow in the secondary circuit.
The most relevant parameters in this calculations are the mass flow of the feed-water in the secondary circuit of the reactor and the specific enthalpy rise in the steam generator. Corrections have to be made for losses in the primary and secondary circuits e.g.~due to radiation and convection or for contributions of the circulation pumps. Those have only minor impact on the final uncertainty of the thermal power estimation, since they account for less than 1\% of the total power. The maximal thermal power of KBR at full operation is 3.9~GW corresponding to a gross electrical power output of 1.47~GW.%gross, da Brutto Leistung ohne Abzug des Eigenbedarfs

The systematic uncertainties on the thermal power estimation are summarized in Table~\ref{tab_therm}.
The enthalpy rise can be calculated from steam tables using measured values of pressure, temperature and moisture content around the steam generator. The feed-water is circulated at a rate of about 2000~kg~s$^{-1}$. The temperatures of the water and steam are determined before and after the steam generator. The systematic uncertainty on the thermal power associated to those measurements is 0.54\%. The moisture content of the steam contributes with 1.56\% to the total uncertainty. 

\begin{table}[h]
\caption{Contributions to the uncertainty of the thermal power P$_{th}$ estimation from the energy balance in the secondary circuit.\label{tab_therm}}
%	\centering
	\begin{tabular}{lc}
	\hline\noalign{\smallskip}
	syst.~uncertainty	& rel.~uncertainty\\
		& on P$_{th}$(\%)\\
	\noalign{\smallskip}\hline\noalign{\smallskip}
		temperature & 0.54\\
	    flow meter & 1.64\\
		moisture & 1.56\\
	\noalign{\smallskip}\hline\noalign{\smallskip}
	\end{tabular}
	
\end{table}

The dominant contribution on the uncertainty of the thermal power is given by the mass flow measurement. The flow meter used at KBR is operated on the principle of the Venturi effect and has an uncertainty of 1.64\%. From the combination of those uncorrelated contributions by quadratic summation, a total uncertainty on the absolute thermal power of 2.3\% ($1 \sigma$) is obtained. The statistical variations of the thermal power measurements during a cycle are on a negligible level of about 0.1\%. 

The thermal power determination in the secondary circuit is rather insensitive to fast changes and provides no spatial information about the situation inside the reactor core. The spacial distribution and power variations are therefore determined using ex-core and in-core neutron flux instrumentation (see Section \ref{subsec:incoreinstrumentation} and Section \ref{subsec:excoreinst}). Especially the fast neutron flux in the ex-core instrumentation is an indicator for the local thermal power generation. 

\subsubsection{Ex-core instrumentation}
\label{subsec:excoreinst}
\begin{figure}
    \includegraphics[width=0.5\textwidth]{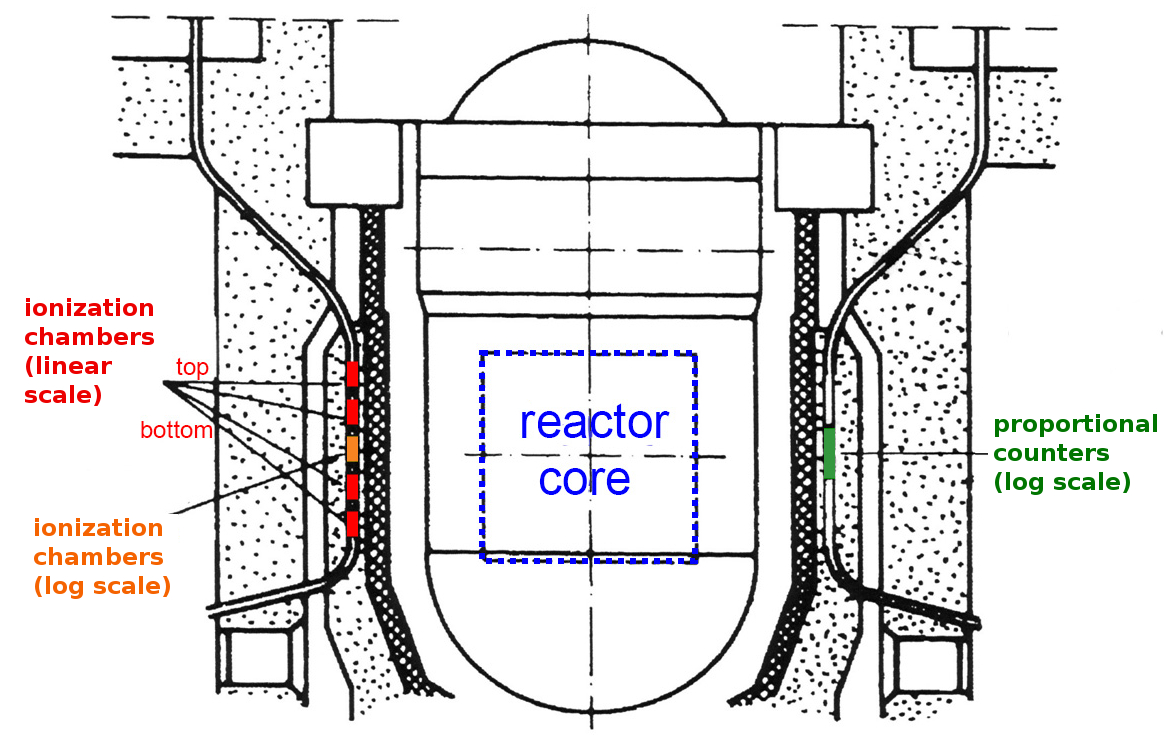}
    \caption{External neutron measurement system at nuclear power plant in Brokdorf (ex-core instrumentation, linear scale: standard reactor operation, log scale: when the reactor is turned on and off)}
    \label{fig:extneutron}
    \end{figure}

The ex-core instrumentation is situated in the concrete shield (so-called "biological shield") around the reactor core as shown in Figure~\ref{fig:extneutron}. To be able to cover more than 14 orders of magnitude of neutron flux the instrumentation consists of three different systems. Two of them use ionization chambers and one uses proportional counters. The counting gas in all systems is BF$_3$. The neutrons are detected via the reaction
\begin{equation}
{}^{10}\rm{B}\,+\,n \rightarrow {}^{7}\rm{Li}\,+\,\alpha\text{.}
\end{equation}

One of the systems based on ionization chambers is able to cover the full range of the thermal power from 0 to 100\% 
%a performance range from 0 to 100\% 
and provides a linear relation between the thermal power and the neutron flux. The chambers are placed in 4 radial positions around the core. The system consists of two chambers, connected in parallel, monitoring the upper half of the core and two chambers, connected in parallel, monitoring the lower half. There are basically three parameters having an impact on the detected signal: the relative power of neighboring fuel assemblies, the temperature of the coolant and the boron (B) concentration of the coolant. Therefore regular calibration of the system is needed, since the B concentration is decreasing over a cycle and the coolant temperature can vary e.g.\,in stretch-out operation of the reactor. In-between calibrations, these dependencies introduce a systematic uncertainty of up to 1.5\%. In combination with a 2.5\% statistical uncertainty, an overall uncertainty of the ex-core instrumentation of 3\% is estimated. The proportional counters and the logarithmic range are only used while inducing criticality at the beginning and end of each reactor cycle, when the reactor is turned on and off respectively.  

\subsubsection{In-core instrumentation}
\label{subsec:incoreinstrumentation}
%\item External neutron flux measurement in concrete shield (biological shield) next to pressure vessel
    %
 %   \begin{itemize}
 %   \item covers 14 orders of magnitude (the ones measuring performance range are important for us); four detectors 
  %  \item neutron detection via counting tubes and ionization chambers that are filled with 
  %  $BF_{3}$: $^{10}B+n\rightarrow ^{7}Li +\alpha$
  %  \item within performance range: neutron flux $\simeq$ local power extraction 
  %  \item regular calibration needed, e.g. due to decreasing boron concentration, varying coolant 
   % temperature, meaning that while the neutron flux decreases over one cycle (explicit value!) , it is ensured that the neutron flux stays proportional to the thermal power, \textcolor{red}{value for uncertainty} in our measuring period: no calibration
   % \end{itemize}
    %

%%%%%%%%%%%%%%%%%%%%%%%%%%%%%%%%%%%%%%%%%%%%%%%%%%%%%%%%%%%%%%%%%%%%%%%%%%%%%%%%%%%%%%%%%%%%%%%%%%%%
	
%%%%%%%%%%%%%%%%%%%%%%%%%%%%%%%%%%%%%%%%%%%%%%%%%%%%%%%%%%%%%%%%%%%%%%%%%%%%%%%%%%%%%%%%%%%%%%%%%%%%

\begin{figure}
    \begin{center}
    \includegraphics[width=0.44\textwidth]{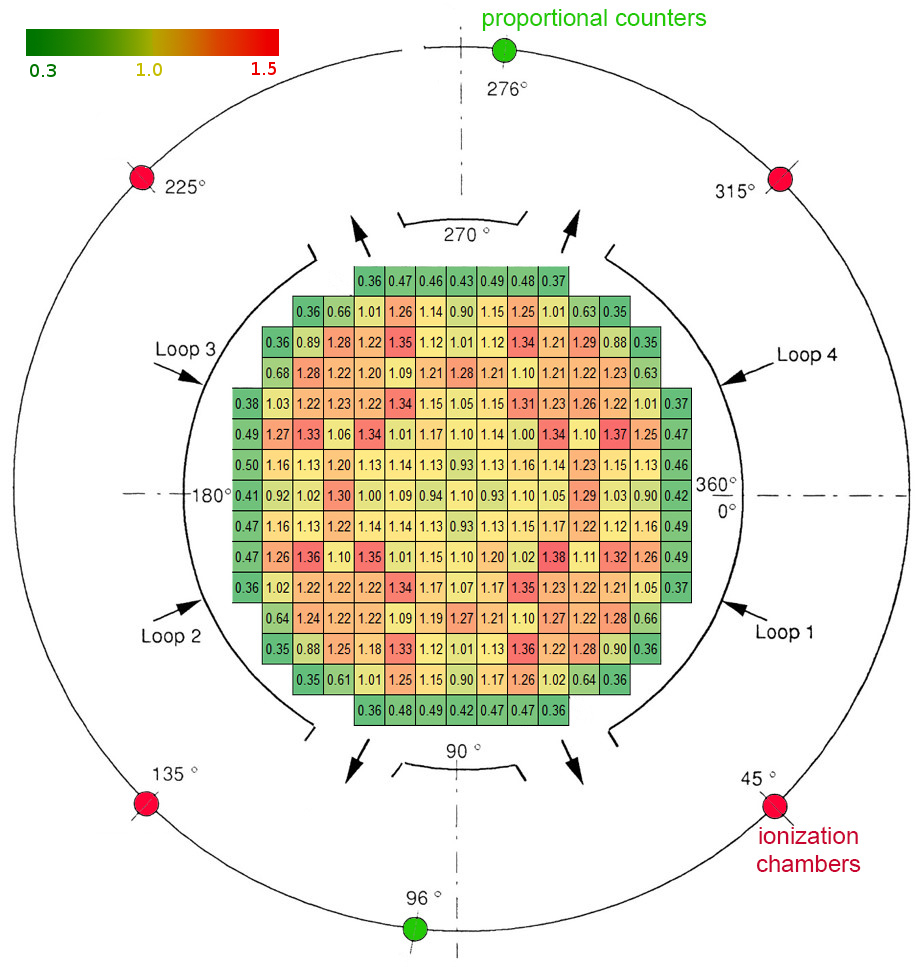}
    \end{center}
    \caption{Measured power density distribution using the in-core instrumentation during December 2016. 
    For each of the 193 fuel assemblies the relative contribution to the total thermal power is denoted. Green corresponds to a small contribution, red to an high contribution, yellow covers the regime in-between}
    \label{fig:coreMCboc}
    \end{figure}

The in-core instrumentation is positioned inside the guide tubes of certain fuel assemblies not occupied by control rods. Two systems exist. Eight fuel assemblies are each equipped with six so-called "self-powered neutron (SPN)" detectors, distributed axially over the length of the core. They rely on the reaction
\begin{equation}
 {}^{59}\rm{Co} + n \rightarrow {}^{60}\rm{Co} + \gamma\text{,}   
\end{equation}
where the $\gamma$-radiation generates an electrical potential due to the photoelectric effect. The subsequent current in the measuring chain is proportional to the neutron flux. The 48\,SPN detectors continuously monitor the radial and axial power distribution in the core. After appropriate calibration they show the maximum power per rod length unit (W/cm) in their respective surveillance region.
Furthermore, in the so-called "aeroball measuring system" (AMS) guide tubes of 28\,fuel assemblies (one tube in each assembly) are equipped with double pipes entering through the reactor vessel lid down to the lower end of the fuel assembly.
%, covering the whole active zone and part of the fuel free reflector zone of the core. 
During measurement a column of about 3000 steel balls (diameter 1.7\,mm) containing 1.5\% of Vanadium (V) is inserted into the core, where the $^{51}$V is partially activated to $^{52}$V (T$_{1/2}$=3.75\,min) for about three minutes. The amount of activation is proportional to the thermal neutron flux at the point of activation and hence to the local reactor power. % Afterwards the columns are transferred pneumatically to a measuring table using nitrogen at 10$^{6}$\,Pa. 
At 32\,axial measuring points (so-called "parcels") the 1.4\,MeV $\gamma$-ray emitted by $^{52}$Cr as $\beta$ decay product of $^{52}$V is measured. Semiconductor detectors are used to discriminate the signal delivered by interfering nuclides such as $^{56}$Mn and $^{51}$Cr. Thus, a 3D map of relative power distribution in the core can be created. The values for fuel assemblies not instrumented are extrapolated from the 28 instrumented ones. An example for such a (radial/polar) distribution is shown in Figure~\ref{fig:coreMCboc}. The relative contribution of each fuel assembly to the total power is given. Summing up all contributions amounts to 193, the total number of fuel elements. The AMS is used on demand, typically twice per week, to calibrate the SPN detectors and to calculate the position dependent reactor burn-up. %Furthermore, a comparison to core design data is performed to check the accuracy of calculations and simulations.
KBR operates a core simulator (POWERTRAX/S \cite{powertrax}, designed by FRAMATOME GmbH), relying on the same technology that is used for core design. Based on various plant data like temperatures, thermal reactor power and control rod insertion depth, the simulator is capable of providing an online 3D image of the power distribution in the core. %The simulation also considers the local burn-up situation, meaning how much energy is extracted from each fuel assembly over its live time split into the most important contributing isotopes. 
%The simulation also considers the local burn-up situation and the Xenon distribution. 
Calculations are done automatically, usually every two hours and more frequently in case of transients. The results are stored and can be displayed down to the level of a single fuel rod (236 being contained in each fuel assembly) and the 32 axial parcels. Thus, the local origin of neutrons and neutrinos escaping the reactor core and arriving at the CONUS experimental site can be calculated in detail.

%%%%%%%%%%%%%%%%%%%%%%%%%%%%%%%%%%%%%%%%%%%%%%%%%%%%%%%%%%%%%%%%%%%%%%%%%%%%%%%%%%%%%%%%%%%%%%%%%%%%
	
%%%%%%%%%%%%%%%%%%%%%%%%%%%%%%%%%%%%%%%%%%%%%%%%%%%%%%%%%%%%%%%%%%%%%%%%%%%%%%%%%%%%%%%%%%%%%%%%%%%%

%core simulation data: would be required if the comparison between the different campaigns and the plot of the thermal power along the z axis would be included, name of program, which kind of information: relative contribution to thermal power of each fuel element, each split into 32 pieces, uncertainty? \textcolor{red}{(plot of calculated power distribution?)}

%\end{itemize}
 
%%%%%%%%%%%%%%%%%%%%%%%%%%%%%%%%%%%%%%%%%%%%%%%%%%%%%%%%%%%%%%%%%%%%%%%%%%%%%%%%%%%%%%%%%%%%%%%%%%%%

\section{Description of the environment}
\label{sec:neutron-measurements}

\subsection{Experimental site at KBR}

\subsubsection{Overburden}
\label{subsubsec:overburden}
\begin{figure}[h]
    \includegraphics[width=0.47\textwidth]{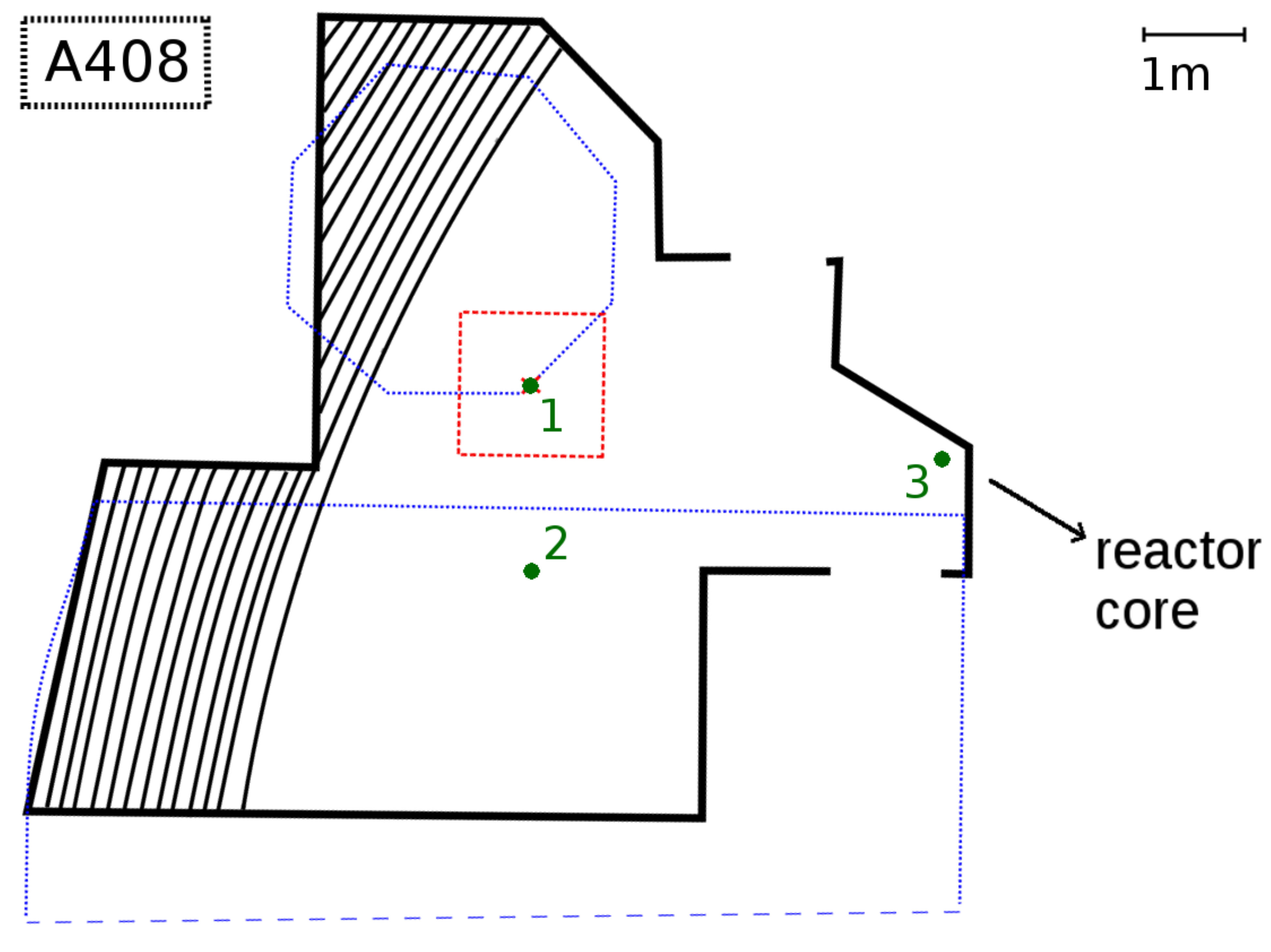}
    \caption{Room A408 at the nuclear power plant Brokdorf, where the CONUS experiment is located. Blue outlines mark the water ponds as overburden. (1) center of CONUS shield and location of BSS during DS-3, (2) location of monitor sphere during DS-3, (3) location of CONRAD detector}
    \label{fig:location_A408}
    \end{figure}

\begin{figure}[h]
    \includegraphics[width=0.5\textwidth]{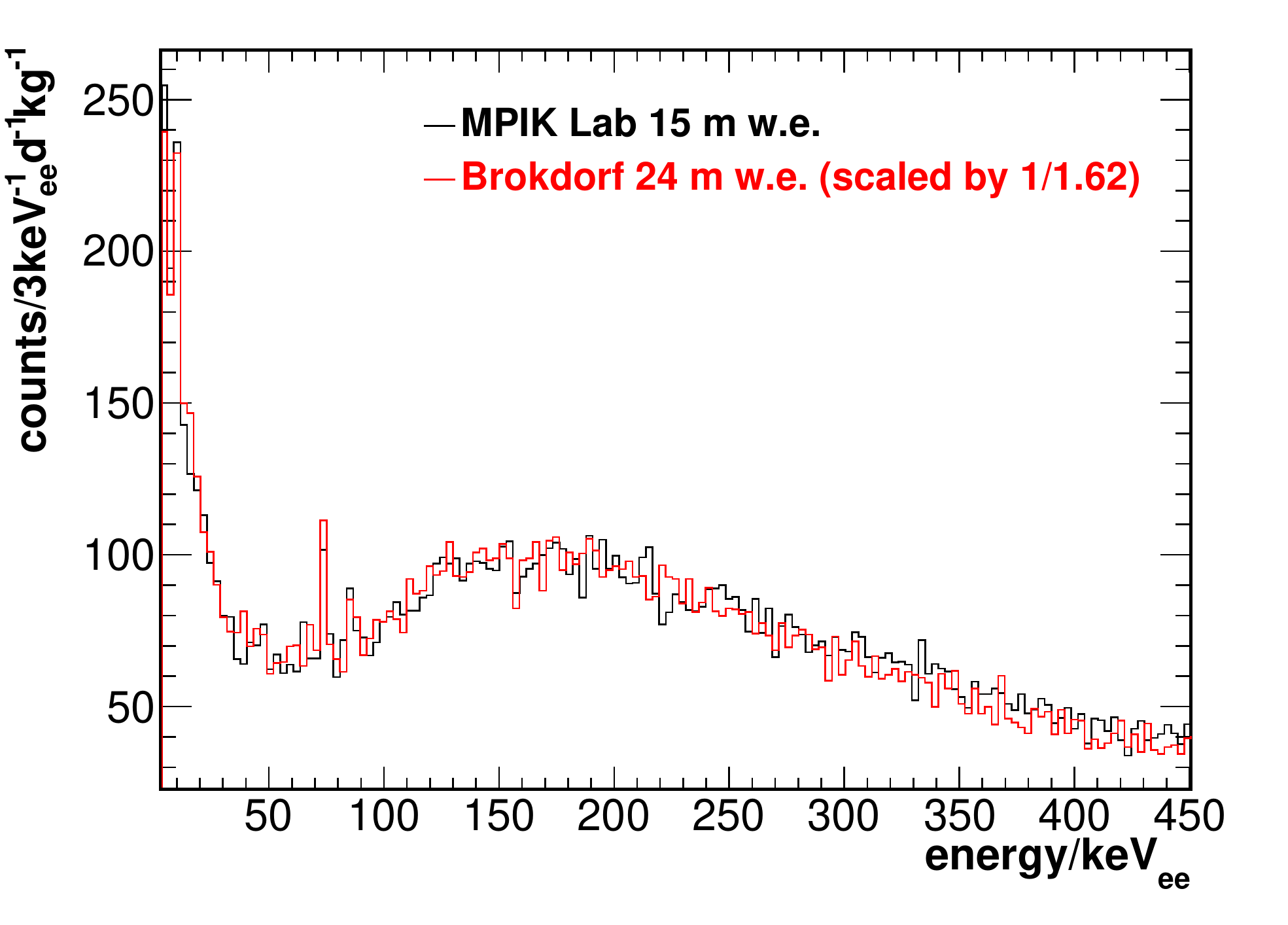}
    \caption{Comparison of measurements inside the CONUS shield without the active muon veto at the MPIK laboratory and at reactor site. The latter one was scaled with a factor of 1.62$^{-1}$}
    \label{fig:comparison_wovetospec_LLLtoKBR}
    \end{figure}

During the commissioning phase of the CONUS experiment, the shield with a varying combination of detectors has been set up at the underground laboratory at MPIK. In this way, it was possible to characterize the experimental site in Brokdorf relative to the well-known conditions at the MPIK laboratory.
At KBR, an overburden to shield against cosmic rays is provided by the concrete and steel structures of the reactor building. The whole building is enclosed in a concrete dome of 1.8\,m thickness and the safety containment consists of a steel sphere of 3\,cm wall thickness. The room A408, where the CONUS experiment is set up, is located in the lower hemisphere of the dome with more concrete around from the walls of the surrounding rooms and above.
The concrete density is 2.55\,g~cm$^{-3}$, with different steel contents ranging between 0.8\% for room divider walls and 3.2\% for the biological shield surrounding the reactor core. The hydrogen content, highly relevant for the moderation of neutrons, was determined via element analysis. The sample was taken from small concrete pieces, which were removed from the floor of room A408 during the installation of the CONUS experiment. %It is 0.8$\pm$0.1\% \cite{concrete_analysis}.
%The analysis reveals 1.55$\cdot$10$^{22}$\,H-atoms/cm$^{3}$ in the concrete \cite{concrete_analysis}.
The analysis reveals a hydrogen content of (0.8$\pm$0.1)\% \cite{concrete_analysis}.
Moreover, as it can be seen in the blue contours in Figure~\ref{fig:location_A408}, the room A408 is located partially below the water pond of spent fuel assemblies and fully below the smaller pond used for loading the spent fuel storing casks prior to shipment, with contributions to the overburden. They are permanently filled with borated water to a level of 13.4$\pm$0.1\,m (mean value over 7.5\,months), even if the filling with spent fuel assemblies is varying. This leads to a variability in the mean density between 1.0 and 1.55\,g~cm$^{-3}$ (maximal allowed filling of the pond). These changes in overburden are considered negligible for the CONUS experiment.
All contributions add up to an overburden of 10-45\,m~w.e. depending on the solid angle, meaning that the rather variable hadronic component of the cosmic rays at Earth`s surface is fully suppressed. The effective overburden for the suppression of the cosmic-ray muon component can be determined by comparing the measured spectra inside the passive shield of the CONUS experiment without the muon veto system measured at the MPIK laboratory and at the nuclear power plant as displayed in Figure~\ref{fig:comparison_wovetospec_LLLtoKBR}. The scaling factor over the whole energy range between both locations amounts to 1.62$\pm$0.02. The MPIK laboratory has a well-known overburden of 15\,m w.e \cite{hakenmueller:masterthesis} leading to an effective overburden of 24\,m w.e. at KBR. The spectral shape agrees over the whole energy range, meaning that the same physics processes for the muon and muon-induced neutron interactions are relevant at both places.

\subsubsection{Natural radioactivity}
\label{subsubnatrad}
Measurements with HPGe spectrometers without any shield are dominated by the environmental radioactivity and anthropogenic isotopes from the surroundings. Two comparable measurements at the MPIK underground laboratory as well as in room A408 at the nuclear power plant were carried out with a CONUS detector. Comparing the integral count rate in the range of [20,440]\,keV, it was found that the background level at KBR is higher by a factor of 4.2$\pm$0.1 with respect to the MPIK laboratory.
The difference between the locations is partially explained by the special attention that had been paid on employing concrete especially low in radioactivity at the MPIK laboratory \cite{location:Heusser1992}. 
%This means that at the MPIK laboratory, the environmental radioactivity lies below of what is expected from ordinary concrete.To test the concrete at reactor site, the activity of the small pieces of concrete also used for the determination of the hydrogen content was measured at the screening station in MPIK laboratory \cite{Heusser:1995wd}. In Table~\ref{tab:natradioactivity}, the results are compared to the known activities from the concrete of the MPIK laboratory, showing a factor of 3 higher Th and U concentrations in the concrete from the nuclear power plant, which is in range of standard concrete used for construction \cite{ordinaryconcrete}.
To test the concrete at reactor site, the activity of the small pieces of concrete was measured at the screening station in MPIK laboratory \cite{Heusser:1995wd}. In Table~\ref{tab:natradioactivity}, the results are compared to the measured activities from the concrete of the MPIK laboratory. For U and thorium (Th), comparable results were found, while the kalium (K) content is lower in the MPIK sample.
There is also a finite contamination of man-made $^{137}$Cs inside the concrete from Brokdorf. %, which has been avoided at the MPIK laboratory. 
Moreover, contrary to the MPIK laboratory, highly varying $^{60}$Co concentrations were observed in the samples and thus a range is given in Table~\ref{tab:natradioactivity}. This could be a surface contamination, as $^{60}$Co has also been observed in the dust in radio protection-related measurements similar to other nuclear reactors elsewhere \cite{texono2017}.

Furthermore, there is an additional background contribution at reactor site originating from reactor neutron interactions inside the water of the cooling cycle and neutron capture in concrete. 
%These resulting high-energetic $\gamma$-rays with several MeV energy increase the background at lower energies by pair production and Compton scattering. 
These $\gamma$-rays have been measured with the CONRAD detector and will be discussed in detail in Section \ref{subsec:conradoutsideshield}. 

\begin{table}
	\caption{Activity of concrete samples from room A408 and the MPIK laboratory, determined by HPGe spectrometer screening measurements at the underground laboratory in Heidelberg. The standard values are mean values from the activities of concrete from several countries \cite{ordinaryconcrete} (n.d.=no data).}
	\label{tab:natradioactivity}
	%\begin{center}
		\begin{tabular}{cllll}
			\noalign{\smallskip}\hline
			isotope	& KBR A408	& MPIK lab&standard\\
            		&  [Bq/kg]	&  [Bq/kg] & [Bq/kg]\\ \noalign{\smallskip}\hline\noalign{\smallskip}
		    $^{238}$U& $<$37   & 12$\pm$5&n.d.\\
            $^{226}$Ra& 13.2$\pm$0.2 & 12.8$\pm$0.4&44$\pm$21\\  
            $^{232}$Th& 15.3$\pm$0.3 &16.6$\pm$0.8 &30$\pm$14\\  
            $^{228}$Ra& 14.9$\pm$0.3 &17.2$\pm$0.8&n.d. \\  
            $^{137}$Cs& 1.37$\pm$0.07 &<0.03&n.d.\\  
            $^{60}$Co& 0.2--1.5&<0.03&n.d.\\  
            $^{40}$K& 433$\pm$12 &112$\pm$7&240$\pm$108 \\  
			\noalign{\smallskip}\hline
		\end{tabular}
	%\end{center}
\end{table}

\subsubsection{Distance to reactor core}
The distance of the CONUS experiment's shield center in room A408 of KBR to the middle of the reactor core amounts to (17.1$\pm$0.1)\,m, ensuring a high reactor antineutrino ($\overline{\nu}_{e}$) flux at the experimental site. The experiment is nearly at the same height as the reactor core with an offset along the z-axis of 0.25\,m. The HPGe diodes have a distance of $\sim$50\,cm from the floor of A408. The reactor core consists of 193\, fuel assemblies contained in a cylinder of 3.45\,m diameter (see Figure~\ref{fig:coreMCboc}) with an active length along the z-axis of 3.9\,m. Details about the materials and geometry between the reactor core and room A408 are given in Section \ref{sec:neutron-MC}.  

\section{Description of the MC simulation framework}\label{sec:neutron-MC}
\label{sec:neutron-simulation}
The MC simulation framework MaGe \cite{MC:mage}, based on Geant4 (version Geant4.9.6p04) \cite{MC:geant4_1,MC:geant4_2} is applied to support the understanding and to complement the neutron measurements via an ab initio calculation. In a first step, the neutron propagation from the reactor core to room A408 is simulated as well as the propagation of neutrons from the spent fuel assemblies inside the storage pool above room A408. In the second step, the neutrons arriving in A408 are propagated through the CONUS shield towards the HPGe diodes employing the measured neutron spectrum inside A408 as input. The relevant neutron interactions, models in Geant4 and the applied cross section data sets are listed in Table~\ref{tab:MCmodelsneutrons}. 

\begin{table*}[btp]
\caption{Geant4 MC models for neutron propagation and absorption. \label{tab:MCmodelsneutrons}}
\begin{tabular}{llcl}
\noalign{\smallskip}\hline
interaction & model & energy range & cross section \\ \noalign{\smallskip}\hline\noalign{\smallskip}
elastic & hElasticCHIPS    &19.5MeV$<$E$<$10\,TeV&GheishaElastic, ChipsNeutronElasticXS\\
        & NeutronHPElastic &$<$20MeV &GheishaElastic, ChipsNeutronElasticXS, NeutronHPElasticXS\\
inelastic &BertiniCascade&19.9MeV$<$E$<$9.9\,GeV&GheishaInelastic, Barashenkov-Glauber\\
        &NeutronHPInelastic&$<$20MeV&GheishaInelastic, Barashenkov-Glauber, NeutronHPInelasticXS\\
capture &G4LCapture&19.9MeV$<$E$<$2\,TeV&GheishaCaptureXS\\
        &NeutronHPCapture&$<$20MeV&GheishaCaptureXS, NeutronHPCaptureXS\\
fission &G4LFission&19.9MeV$<$E$<$2\,TeV&GheishaFissionXS\\
        &NeutronHPFission&$<$20MeV&GheishaFissionXS, NeutronHPFissionXS\\
\noalign{\smallskip}\hline
\end{tabular}
\end{table*}

\subsection{Implementation of geometry}
\subsubsection{Nuclear power plant and room A408}

\begin{figure}[h!]
    \includegraphics[width=0.45\textwidth]{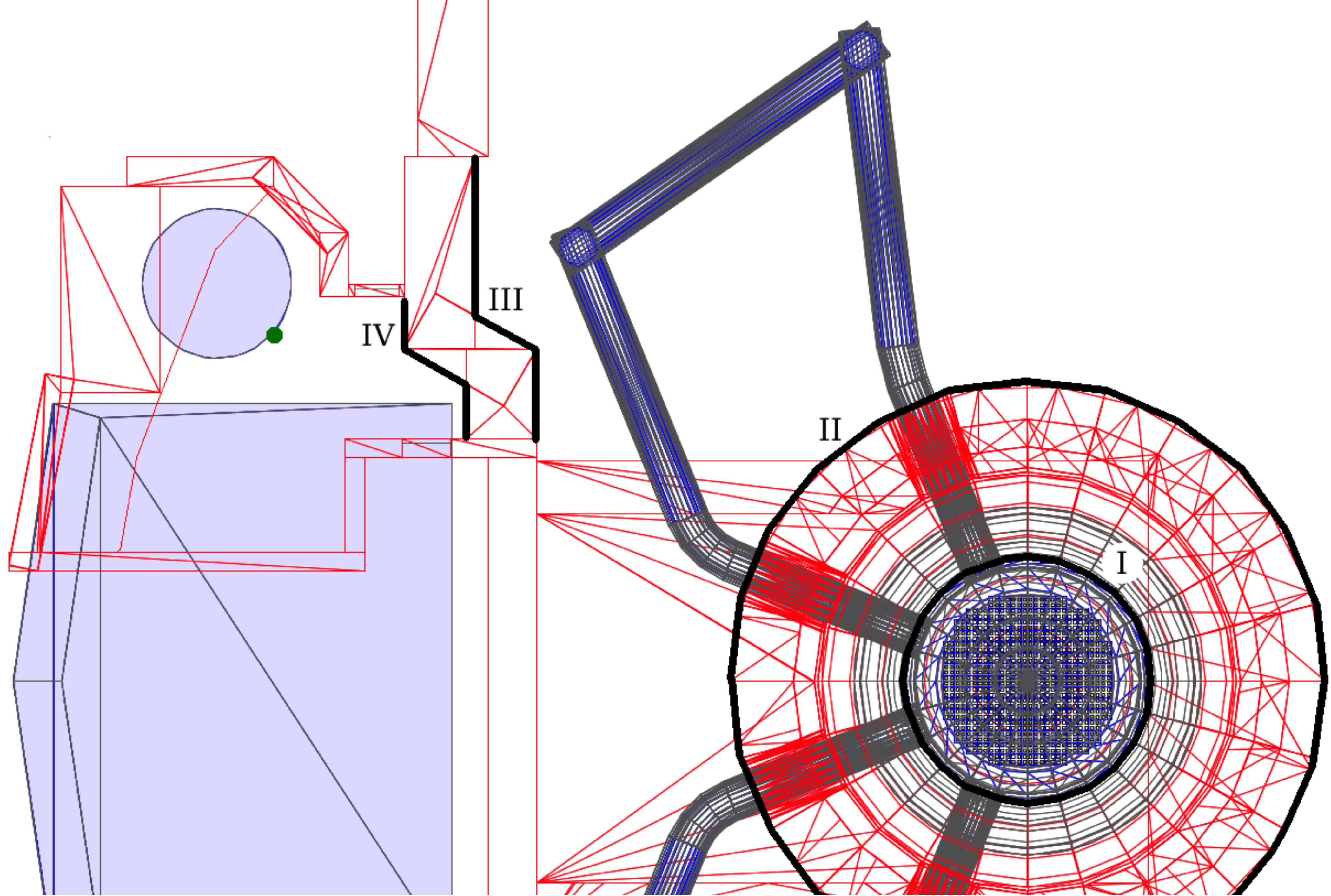}
    \caption{Top view of implemented MC geometry including room A408, the overburden and the reactor core with surrounding. Concrete structures are displayed in red, the blue areas correspond to the spent fuel storage pool and the cask for shipment of spent fuel assemblies. For simplification only two of the four loop pipes are implemented (only selected volumes are displayed for a better overview)}
    \label{fig:MCgeometry_display}
    \end{figure}

From construction plans, the overall structure and main concrete parts of the reactor building were implemented using the information on the concrete from Section \ref{subsubsec:overburden}. The reactor as starting point of the neutrons was modeled in detail, including all the 193 fuel assemblies. In the MC, these are approximated by four fuel rods instead of the 236 as in reality, each made up of Zirconium alloy cladding tubes filled with UO$_{2}$ pellets. The size of these fictive fuel rods was chosen so that the overall mass of a fuel assembly is reproduced correctly. The reactor core is filled with borated water with a 
B concentration
%boron acid content
of 500\,ppm of enriched $^{10}$B (1\% boric acid, 99\% water) as expected in the middle of a reactor cycle. A mean water temperature of 320$^{\circ}$C and a pressure of 15.7\,MPa have been assumed, leading to a water density of 0.687\,g~cm$^{-3}$ \cite{waterdensitytable}. The reactor core is contained inside the reactor pressure vessel (RPV) made from ferritic steel with a thickness of at least 25\,cm. Eight openings can be found at the top of the reactor core for the loop pipes leading the water from the core to the steam generators, where they heat up the water in the secondary cycle, and afterwards return it to the core. For simplicity, only the two loop pipes on the side of room A408 have been implemented into the MC geometry. The reactor core is enclosed by the biological shield and heat insulation amounting to more than 2\,m of concrete thickness in total. This is followed by an empty room around the biological shield. In the geometry, not all details of this space were implemented, but special attention was paid to ensure this area to be closed to all sides to allow for backscattering of neutrons.
Adjacent to this space behind a concrete wall of 1.3-1.45\,m thickness, the room A408 can be found. The interior was modeled as in Figure~\ref{fig:location_A408}. Also steel doors are included as well as the concrete walls of the neighboring room. 
Room A408 has a height of 2.8\,m and the concrete ceiling, which is also the floor of the spent fuel storage pool, has a thickness of 1.85\,m. The spent fuel storage pool and the cask loading pond are lined with several centimeters of steel and filled with 13.3\,m of borated water. Between the active part of the spent fuel assemblies and the floor of the pool, there is a distance of about 80\,cm. The amount of spent fuel assemblies within the storage pool is variable in the MC. The $^{10}$B content is constantly 2300\,ppm (5\% boric acid, 95\% water). The most important features of the implemented geometry including the location of the middle of the CONUS shield can be found in Figure~\ref{fig:MCgeometry_display}.

\subsubsection{Geometry of CONUS shield and HPGe detectors}

\begin{figure}[h]
    \begin{center}
    \includegraphics[width=0.4\textwidth]{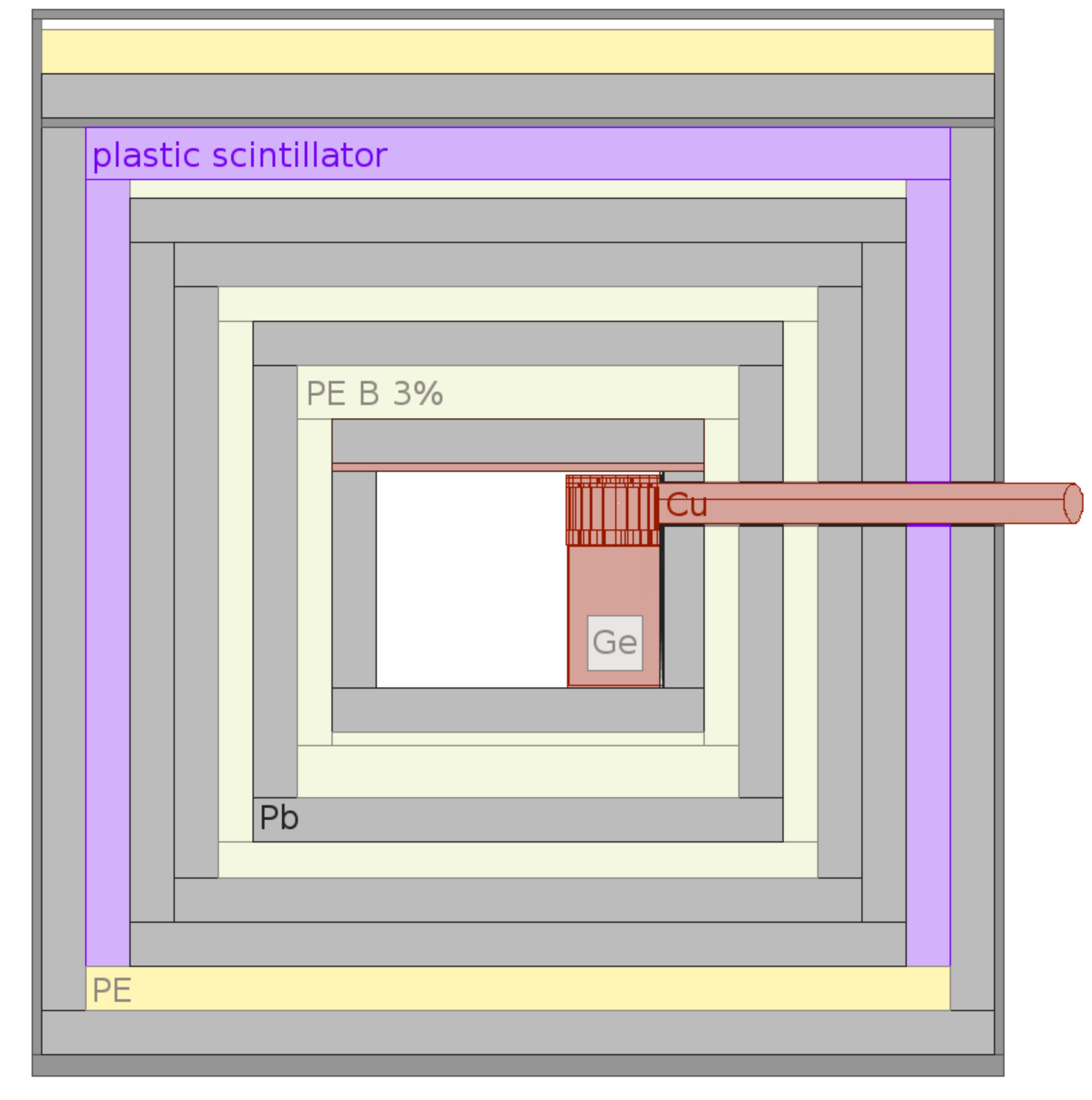}
    \end{center}
    \caption{Side view of the implementation of the CONUS shield geometry and exemplary one detector within. The HPGe diode is marked in gray inside the Cu cryostat. The shield is surrounded by a stainless steel cage assuring safety requirements (dark gray)}
    \label{fig:conushield_mc}
    \end{figure}

The CONUS shield was implemented in detail in the MC (see Figure~\ref{fig:conushield_mc}) inside the geometry of room A408 (see Figure~\ref{fig:MCgeometry_display}). To suppress exterior $\gamma$-radiation, 25\,cm of Pb in all directions are employed. Moreover, there are two layers of in total 10\,cm of borated PE (3\% equivalent of natural B) to moderate and capture neutrons from outside as well as neutrons created by muons in the Pb layers of the shield. The plates were produced from PE and boric acid H$_{3}$BO$_{2}$, enriched in $^{10}$B, which has an especially high neutron capture cross section compared to other isotopes. Neutrons are also moderated by two polyethylene plates (5\,cm each), one on top of the shield and one in the layers below the detector chamber. For the active muon veto system, organic plastic scintillator plates (thickness: 5.2\,cm) equipped with photo-multiplier tubes \linebreak[4](PMTs) are included in the shield. These plates also contributes to the moderation of neutrons. Inside this shield, the four CONUS detectors are placed within the detector chamber with a volume of 25\,l. The Cu cryostats and their interior, including the point-contact HPGe diodes and supporting structures are modeled in the MC geometry as in the technical drawings. 

Similarly, the coaxial HPGe CONRAD detector without any shielding is modelled by setting up the detector's Cu cryostat with cooling finger and its full interior inside the MC geometry in front of the wall adjoined to the space around the reactor core as it was positioned for the measurement (see Figure~\ref{fig:location_A408}).

\subsubsection{Input spectra and output for reactor neutron MC}

At a nuclear power plant, neutrons are created predominately inside the reactor core via fission and immediate evaporation from fission products. Over the whole cycle, more than 50\% of the fissile material is made up of $^{235}$U, while $^{238}$U, $^{239}$Pu and $^{241}$Pu contribute as well \cite{MCnorm:Kopeikin2004}. The $^{235}$U neutron fission spectrum according to a Watt distribution function is displayed as black line in Figure~\ref{fig:MCreactorcore_spectra_propagation}, with a mean neutron energy of 1.95\,MeV \cite{book:reaktortechnik}. As the neutron spectra of the main other isotopes undergoing fission such as $^{239}$Pu are very similar, in MC simulation the $^{235}$U fission spectrum is employed as initial spectrum for the neutrons.
Most of these neutrons are moderated within the reactor core and induce fission again, fueling the chain reaction employed to create the power output of the reactor. However, about 10$^{-4}$ of the neutrons will leave the reactor core before they either hit fissile material, are absorbed in the fuel assemblies' structures or are moderated enough to induce another fission (see Table~\ref{tab:MCsuppressionfactorpropagation}). This is most likely to happen at the border of the reactor core, thus justifying to start neutrons only within the volumes of the UO$_{2}$ pellets of the first and second outer-most ring of fuel assemblies in the MC simulation (in total 104 of 193 fuel assemblies). For the purpose of the propagation outwards from the reactor core, the ongoing fission reactions are not required and consume computation time, thus all fission products are killed immediately by the MC. Due to the huge decrease of the neutron flux along the way towards room A408, the MC simulation has been split into four steps where the spectrum of neutrons passing a certain geometric boundary was registered and used as a new input spectrum for the next part of the MC simulation (see Figure~\ref{fig:MCgeometry_display} for the single steps denoted with I-IV). In the end, the spectrum of the neutrons leaving the walls adjoined to the space around the reactor core is recorded. Moreover, the neutrons hitting a 6'' diameter air sphere at about the location of the middle of the shield of the CONUS experiment are tracked to be compared to and to be used in the analysis of the BSS data. The device itself is not simulated, since the conversion to the measured PHS is carried out by the response functions as described in Section \ref{subsec:directndet}. Alternatively, to represent several spheres set up at the same time, the neutrons passing through a fictional horizontal air plate (size: 2\,m$\times$3.5\,m) are accounted for. 
Furthermore, for the CONRAD detector measurements, the last step is repeated with this detector present inside room A408. All hits inside the HPGe detector are registered. The decreased charge collection efficiency outside the active volume is added in the post-processing.

Besides the reactor core, neutrons are also emitted by the spent fuel assemblies in the storage pool above the CONUS experiment. The majority of the neutrons are emitted by actinides, especially by $^{244}$Cm, while other isotopes only contribute to a few percent \cite{reviewspentfuel1986}. Thus the Watt distribution function for $^{244}$Cm is used as initial spectrum. Assuming the storage pool is filled with the maximum number of fuel assemblies, neutrons were started from the volume of those 192 fuel assemblies located above room A408 and registered in the horizontal air plate described above.

\subsection{Initial spectrum and output for the particle propagation through the CONUS shield}
Assuming a homogeneous neutron flux inside A408, neutrons are started isotropically from a hemisphere (diameter=1.4\,m) spanning around the CONUS shield towards the floor. To take into account backscattering effects, the walls, ceiling and floor of room A408 are included in the simulated geometry (see Figure~\ref{fig:MCgeometry_display}). The measured neutron spectrum in the exact location of the CONUS experiment is used as input for the neutron energy (see Figure~\ref{fig:ON-OFF_diff}). The neutrons are propagated through the CONUS shield. All neutrons arriving at the HPGe diodes are registered. Moreover, all energy depositions inside the HPGe volume are saved as well as the identity of the particles responsible for the energy deposition. No dead layer is assumed and the charge collection efficiency in the different sub-volumes of the HPGe diode is added in the post-processing. 
For the CONUS experiment the region of interest lies in the very low energy range of the spectrum (below 1\,keV$_{ee}$) and thus in this simulation the secondary production cuts are lowered to 1.2\,keV$_{ee}$ for $\gamma$-rays and 850\,eV$_{ee}$ for electrons and positrons, increasing the computation time. This means that below these thresholds, the particles will not produce further secondary particles, but the whole remaining energy will be deposited directly in one location. For hadronic processes there is no such threshold.

Additionally to the neutron propagation simulation, also the measured $\gamma$-ray background inside A408 (see section \ref{subsec:conradoutsideshield}) has been used as MC input. Mono-energetic $\gamma$-rays were started from the wall closest to the reactor core and the resulting spectrum inside the CONUS diodes was evaluated.
All in all, more than 10$^{4}$\,d in CPU time have been spent on the propagation of the neutrons through the reactor building geometry and the CONUS shield.

\subsubsection{Validation of MC}
\label{subsubsec:MCvallit}
For a reliable MC result, it is important to validate the physics processes involved. In MaGe, for electromagnetic interactions this has been done among others for source measurements in \cite{Heusser:2015} and \cite{budjas2009} as well as \cite{Poon2005}, \cite{Hurtado2004} and \cite{Amako2015}, for muon-induced interactions in \cite{hakenmueller:masterthesis}. 

For neutrons, however, there are in general much less validation campaigns available. The propagation of neutrons through shield materials have been examined at the MPIK by carrying out $^{252}$Cf source measurements within and in front of the CONUS shield and a similar shield of another HPGe spectrometer, GIOVE \cite{Heusser:2015}. The correct propagation of the neutrons through the shield was confirmed.   

An overall good agreement for the isotopes relevant here for the probability of the number of emitted $\gamma$-rays in neutron capture has been found (see table \ref{tab:MCbrcomparisontolit}). Especially the relative branching ratio between the different $\gamma$-lines is in excellent agreement with the literature values. However, additional $\gamma$-lines occur in the de-excitation spectrum in the MC, that are not supposed to be created. This has to be corrected for by removing all MC events containing such $\gamma$-lines.

Moreover, if the isotopes produced in neutron capture are metastable, they are not created in the MC using Geant4.9.6. It is especially of interest to be able to simulate these $\gamma$-lines for the metastable Ge states as described in Section \ref{subsec:indirectndet}. To do this, the cross section has to be implemented manually into the code. A separate simulation has to be run to study the $\gamma$-line count rate to avoid to add the energy depositions from this metastable decays to the prompt contribution.

\begin{table*}[h!]
	\caption{The branching ratio (br) in Geant4 of the main $\gamma$-lines from neutron capture on the isotopes relevant at KBR are compared with literature values. For the lines marked with (*) the MC generates two neighbouring $\gamma$-lines close in energy. Out of those one was recognized to be not physical and was neglected. The absolute branching ratio refers to the ratio of emitted $\gamma$-rays per neutron capture (given for strongest line), while the relative branching ratio is given in respect to the strongest emitted line.}
	\label{tab:MCbrcomparisontolit}
		\begin{tabular}{clll}
			\hline\noalign{\smallskip}
	        neutron capture&energy [keV] \cite{capgamdatabase}&br MC&br lit\\  \noalign{\smallskip}\hline\noalign{\smallskip}
            $^{54}$Fe(n,$\gamma$)$^{55}$Fe   &9297.80$\pm$1.00&abs. 49.9(*) &abs. 56.8$\pm$4.9 \cite{EGAF}\\      \noalign{\smallskip}\hline\noalign{\smallskip}
		    $^{56}$Fe(n,$\gamma$)$^{57}$Fe   &7645.58$\pm$0.10	&abs. 23.6&abs. 29.00$\pm$4.94 \cite{capgamdatabase}\\ \noalign{\smallskip}\hdashline\noalign{\smallskip} 
		    &7631.18$\pm$0.10&rel. 86.22(*)&rel. 86.21$\pm$19.94  \cite{capgamdatabase}\\ 
                    		                 & 7278.82$\pm$0.90&rel. 20.70&rel. 20.69$\pm$4.58 \cite{capgamdatabase}\\ \noalign{\smallskip}\hline\noalign{\smallskip}
            $^{63}$Cu(n,$\gamma$)$^{64}$Cu   &7916.26$\pm$0.08&abs. 28.74(*)&abs. 33.10$\pm$0.60 \cite{capgamdatabase}\\ \noalign{\smallskip}\hdashline\noalign{\smallskip}   
                                             &7638.00$\pm$0.09&rel. 48.94&rel. 48.99$\pm$1.50 \cite{capgamdatabase}\\
                    		                 &7307.31$\pm$0.06&rel. 27.07&rel. 27.18$\pm$0.61 \cite{capgamdatabase}\\
                                             &7253.05$\pm$0.06&rel. 12.54&rel. 12.48$\pm$0.27 \cite{capgamdatabase}\\ 
           % $^{53}$Cr(n,$\gamma$)$^{54}$Cr   &834.87$\pm$0.02&abs. 79.8&abs. 79.8$\pm$1.9 \cite{EGAF}\\ \noalign{\smallskip}\hdashline\noalign{\smallskip}   
        	%	                             &8884.81$\pm$0.18&rel. &rel. 55.70$\pm$3.96 \cite{EGAF}\\                     		
        \noalign{\smallskip}\hline 
		\end{tabular}
\end{table*}

\section{Bonner sphere measurements at KBR}
\label{sec:Bonnersphereall}
\label{subsec:campaigns}
\subsection{Measurement campaigns with Bonner spheres}
\begin{figure}[h]
	\centering
	\includegraphics[width=0.83\columnwidth]{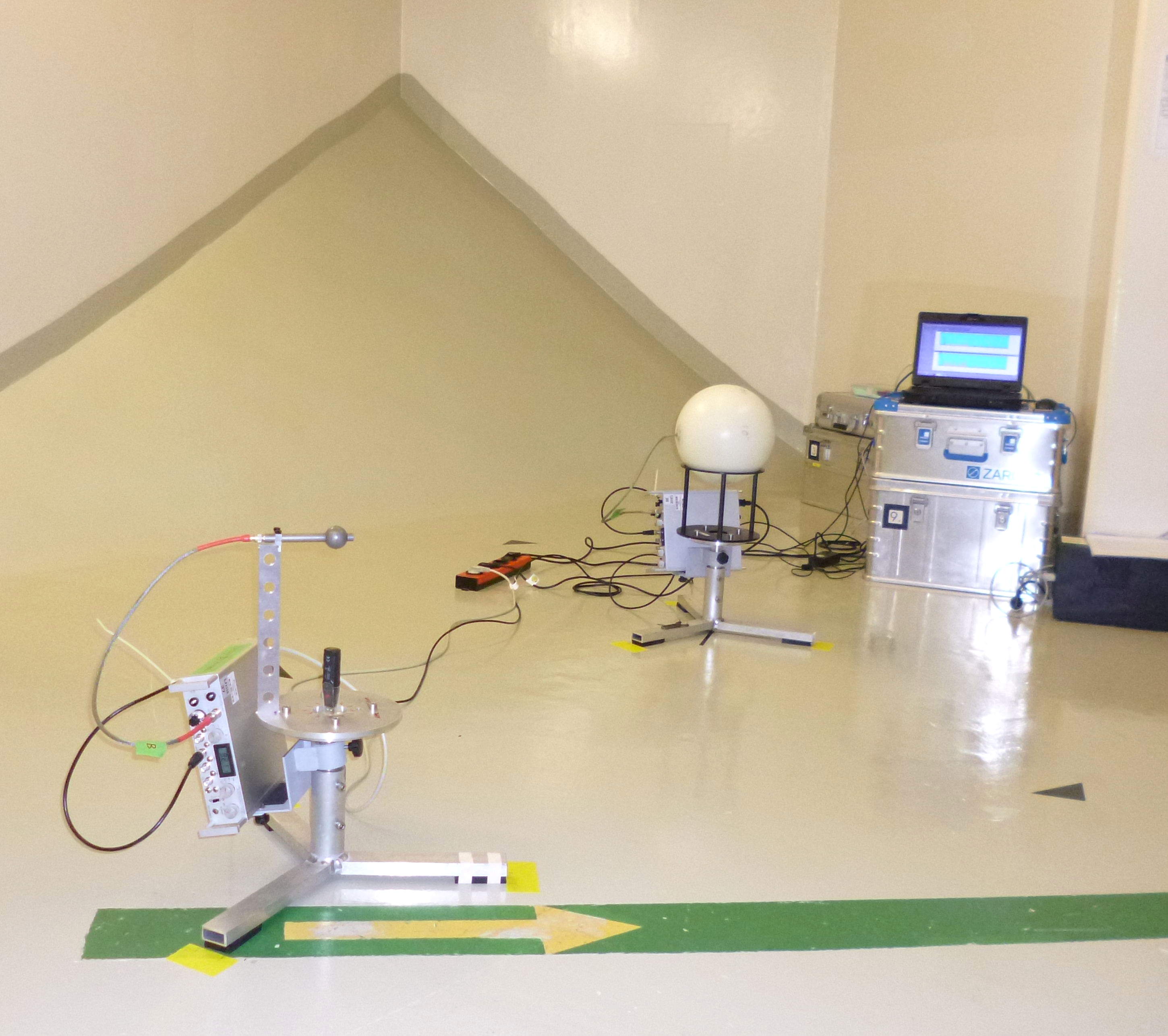}
	\caption{Experimental arrangement used for the DS-3 measurement (reactor ON). The bare detector in the foreground was used as a monitor of the thermal neutron fluence rate. The Bonner sphere in the back was placed in the central CONUS position. Other Bonner spheres were placed on this position one after another}
	\label{fig:photo_ON}
\end{figure}

Three data sets were collected during the measurement campaigns at KBR, summarized in Table ~\ref{tab:campaigns}. The data set DS-3 (reactor ON) was acquired with the arrangement shown in Figure ~\ref{fig:photo_ON}, where the Bonner spheres were placed one after another in the central position of the CONUS site (position 1 in Figure~\ref{fig:location_A408}). A bare detector, placed 1.8\,m away (position 2 in Figure~\ref{fig:location_A408}) from the measurement position, was used as a monitor of the thermal neutron fluence rate during the entire campaign. 
%The arrangement was chosen in order to characterise the neutron field exactly at the CONUS central position. 
The thermal neutron counters within the Bonner spheres were at a height of 51\,cm above ground, identical to the vertical centre of the future CONUS setup.

Due to the low neutron count rates, the measurement times amounted to 3--4\,d per Bonner sphere, with the exception of the 12\,'' sphere. The measurement time with this sphere was set to 9\,d, because an extremely low count rate was expected.

\begin{table}
	\caption{Overview of data sets collected during the measurement campaigns with the NEMUS spectrometer in room A408 at KBR.}
	\label{tab:campaigns}
		\begin{tabular}{llll}
		\hline\noalign{\smallskip}
			data	& reactor	& time					& Bonner spheres				\\
			set		& state	& period				& arrangement			\\
		\noalign{\smallskip}\hline\noalign{\smallskip}
			DS-1    & ON		& 08.12.16--04.01.17	& all spheres used		\\ 
					&			&						& simultaneously		\\  		
			DS-2	& OFF		& 09.02.--26.02.17	& all spheres used		\\ 
					&			&						& simultaneously		\\
			DS-3	& ON		& 31.08.--04.10.17	& spheres swapped		\\
					&			&						& at central position	\\
			\noalign{\smallskip}\hline
		\end{tabular}
\end{table}

\begin{figure}[h]
	\centering
	\includegraphics[width=0.83\columnwidth]{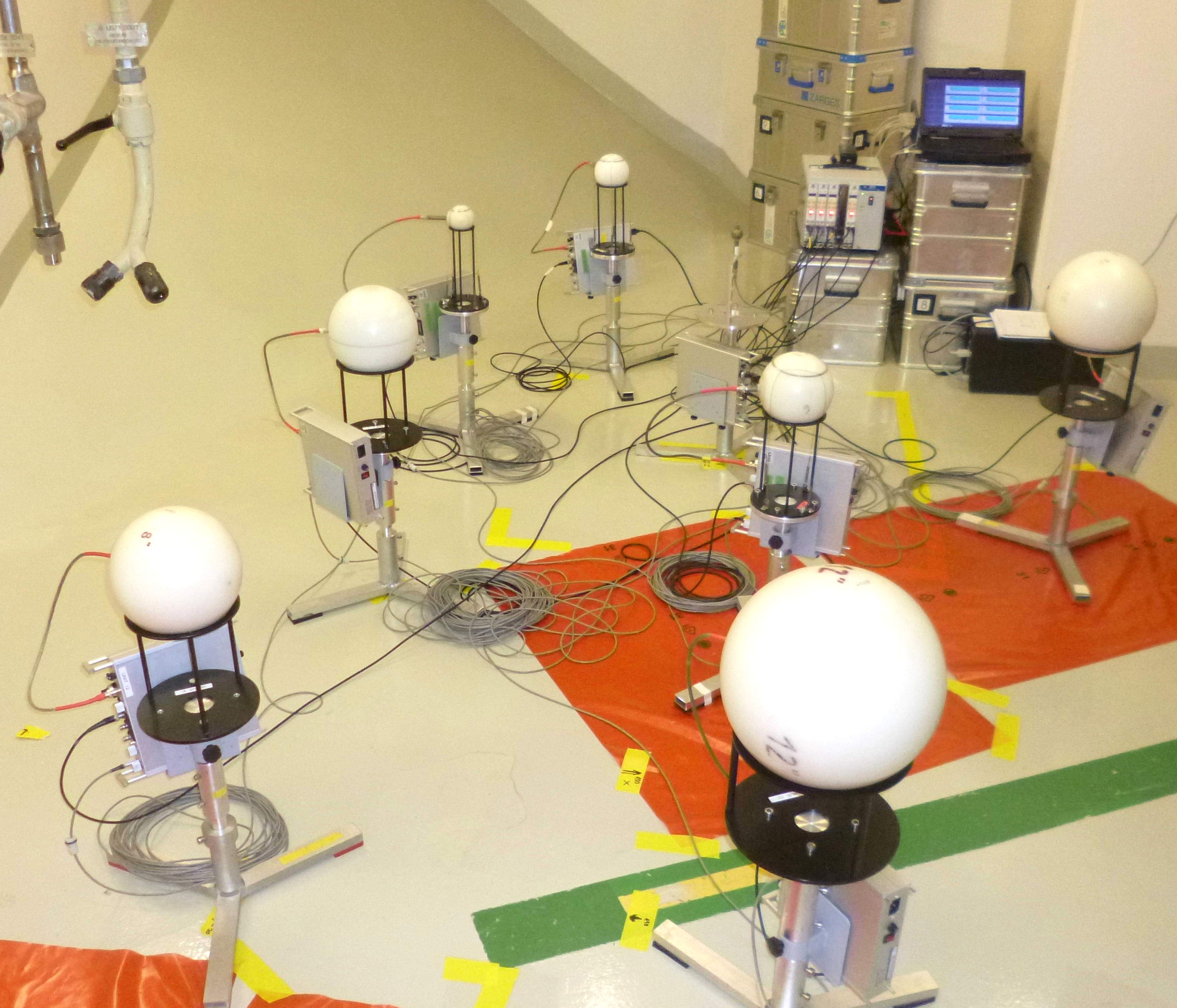}
	\caption{Experimental arrangement used for the DS-1 (reactor ON) and DS-2 measurement (reactor OFF). The Bonner spheres were distributed around the central position of the future CONUS setup. The bare detector is located at the central position}
	\label{fig:photo_OFF}
\end{figure}

The data sets DS-2 (reactor OFF) and DS-1 (reactor ON) were acquired with the Bonner spheres distributed around the central position of the future CONUS setup, as shown in Figure~\ref{fig:photo_OFF}. The distance between the individual spheres was chosen to be 1\,m or more, in order to reduce the effect of neutron scattering from one sphere to another. With the reactor OFF, the neutron count rates were of the order of $\sim1$\,count per hour. Therefore, given the time slot available during the KBR reactor outage, it was necessary to measure with all the spheres simultaneously. DS-1 was the first test run inside the room A408 before the actual measurement campaign and will only be used as a reference to illustrate the spatial inhomogeneity of the neutron field inside A408 and the correlation to the thermal power.

%The individual pulses of the counters were stored in list mode format, i.e., for each pulse the timestamp and the PHS channel number were stored. For the sake of convenience and data loss prevention, the complete PHS were stored automatically in a time loop every 4\,hours.

As the reactor power varies with time due to load follow operation, it was necessary to normalize the neutron counts from the individual BSS measurements of DS-3 to a quantity related to the reactor power. For this, the thermal power $P_\text{KBR}$ in units of [GW], with the time resolution of 1\,h is used. The evaluation of the thermal power with its uncertainty is described in Section \ref{subsec:absthermalpowe}. Based on the start and stop times of the BSS measurements, we were able to calculate the thermal energy output $E_\text{KBR}$ in units of [GW\,h] corresponding to a given neutron measurement.

\subsection{Measurement results with BSS}
\label{sec:res-all}
\subsubsection{Neutron energy distribution inside A408 during reactor ON time}
\label{sec:res-ptb}
\label{subsec:reactor_ON}

\begin{figure}[h]
	\centering
	\includegraphics[width=1.0\columnwidth]{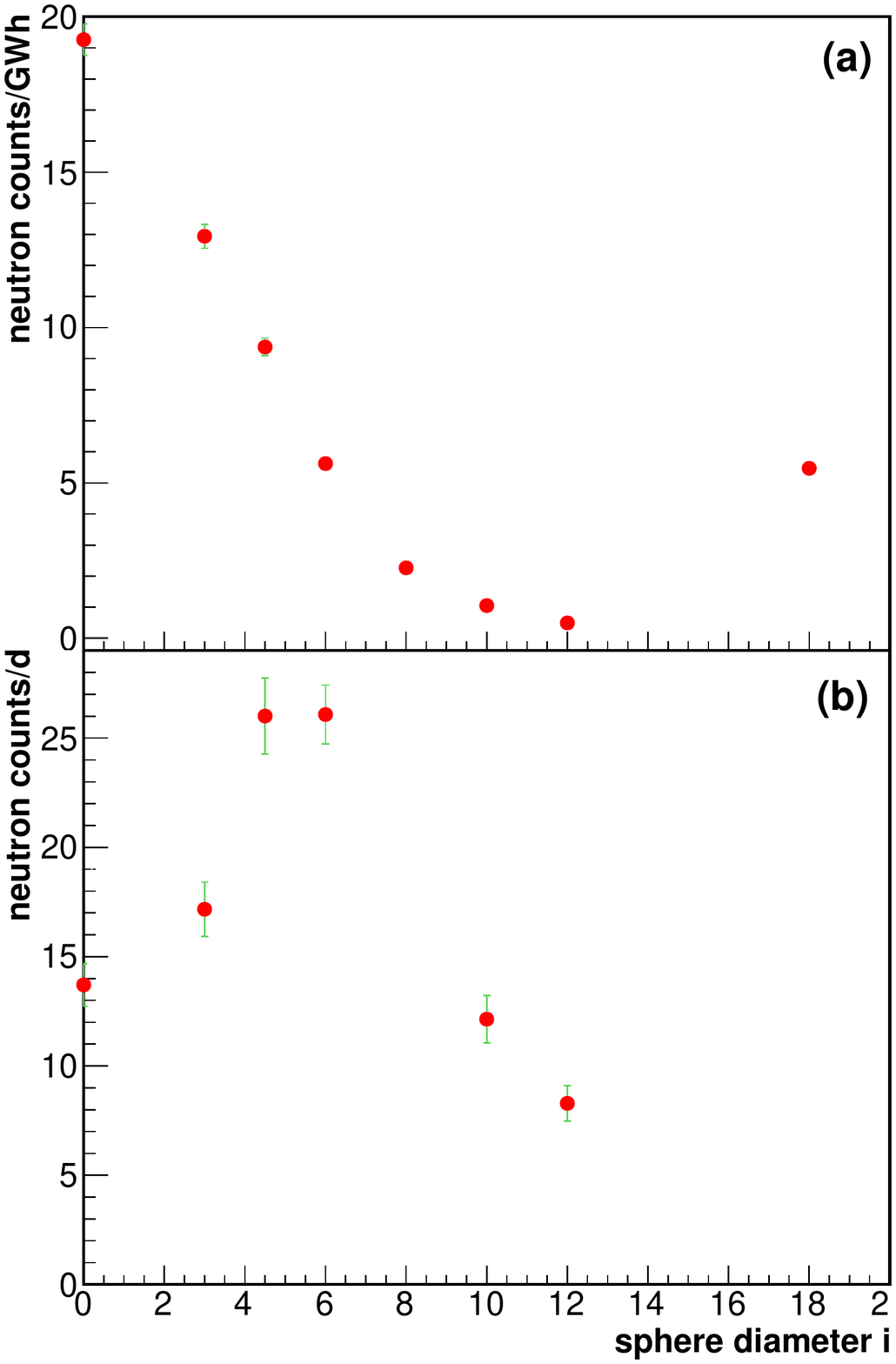}
	\caption{Neutron counts in the individual Bonner spheres, as a function of the sphere diameter $i$. The data point at $i=0$ corresponds to the bare detector. The data point at $i=18$ corresponds to the 8\,'' modified sphere. \textbf{(a)} Data from DS-3 (reactor ON), normalized to 1\,GW\,h of the reactor energy output $E_\text{KBR}$ according to Eq.~\ref{eq:ON_norm}. \textbf{(b)} Data from DS-2 (reactor OFF), normalized as counts per 24\,h. The error bars are in some cases smaller than the size of the symbols}
	\label{fig:count_rates}
\end{figure}

Figure~\ref{fig:count_rates} shows the data from the reactor ON measurement (DS-3). The neutron counts in the individual Bonner spheres were normalised as follows:
\begin{equation}
\label{eq:ON_norm}
N_{p,i}^{(1)}=\frac{C_i^{(1)}}{E_{\text{KBR,}\,i}}\,, \quad E_{\text{KBR,}\,i} = \sum_j P_{\text{KBR,}\,j} \cdot t_j\,,
\end{equation}
where $i$ denotes a given Bonner sphere from the subset used at KBR, $C_i^{(1)}$ is the number of neutron-induced counts determined via fits of the PHS and $N_{p,i}^{(1)}$ is the number of counts normalized to the thermal energy output $E_{\text{KBR,}\,i}$ of the KBR reactor during the measurement with the Bonner sphere $i$.

%The measurement uncertainties varied depending on the total number of counts in the spheres. 
Averaging the neutron counts $C_i^{(1)}$ per measurement times $t_i$, the neutron count rate in the bare counter amounted to $\sim$66\,counts per hour. In the 12\,'' sphere, the average neutron count rate was as low as $\sim$2\,counts per hour. Correspondingly, the relative uncertainty of the normalized neutron counts $N_i^{(1)}$ ranged from 2.6\% for the bare counters (3-4\,d measurement time) up to 5.8\% for the 12\,'' sphere ($\sim$9\,d measurement time). The values include the systematic uncertainty of the corresponding thermal energy output $E_{\text{KBR,}\,i}$ of 2.3\%.

The additional bare SP9 detector used as a neutron monitor for DS-3 allowed us to check the consistency of the data. The monitor readings were also analyzed using the fitting procedure of low-level neutron data mentioned in Section~\ref{subsec:directndet}. The number of neutron-induced counts in the monitor per GW\,h of thermal energy output was determined (weighted mean) as $f_{\text{mon}/E_\text{KBR}}=(16.00\pm0.15$)\,GW\,h$^{-1}$, averaged over the whole DS-3 data set. During all DS-3 measurements, this ratio was constant within $\pm3$\%. The same holds for the readings of the Bonner spheres: the two normalizations, either using the thermal energy output according to Eq.~\ref{eq:ON_norm}, or the monitor readings as a scaling factor, agreed to within $\pm3$\%.

To analyze the data, we used Bayesian parameter estimation~\cite{Sivia2006}. The shape of the result of the propagation of neutrons from the reactor core (see plot of $\mathit{\Phi}_{MC}(E_n)$ in Figure~\ref{fig:ON-OFF_diff}) indicates that the neutron energy spectrum extends up to a few hundred keV. This is consistent with previous Bonner sphere measurements at reactors behind shielding \cite{Fernandez2004,Lacoste2007}.
For the reactor ON measurement, there is an additional component in the form of a peak at around 1\,MeV, caused by neutrons induced by muons in the reactor building. This component is also present during reactor OFF time. Therefore, we introduced a parameterized model (similar to the one in \cite{regi2006}), which consists of a thermal peak, an intermediate region, which is flat in the lethargy representation $(d\Phi/d\log(E_n))$ up to a few hundred keV followed by a smooth drop, and a peak above. The fluence and mean energy of this peak corresponds to the peak at around 1\,MeV in the neutron energy spectrum measured under reactor OFF conditions (see Figure~\ref{fig:unfolded_OFF}). We considered two models, one allowing for a slope in the intermediate region and one setting this slope to zero (corresponding to a $1/E_\text{n}$ behavior). The complete model contained three free parameters: the magnitude of the thermal peak and the magnitude and slope of the intermediate region. The analysis was done using the software package WinBUGS~\cite{Lunn2000}.

\begin{figure}
	\centering
	\includegraphics[width=1.05\columnwidth]{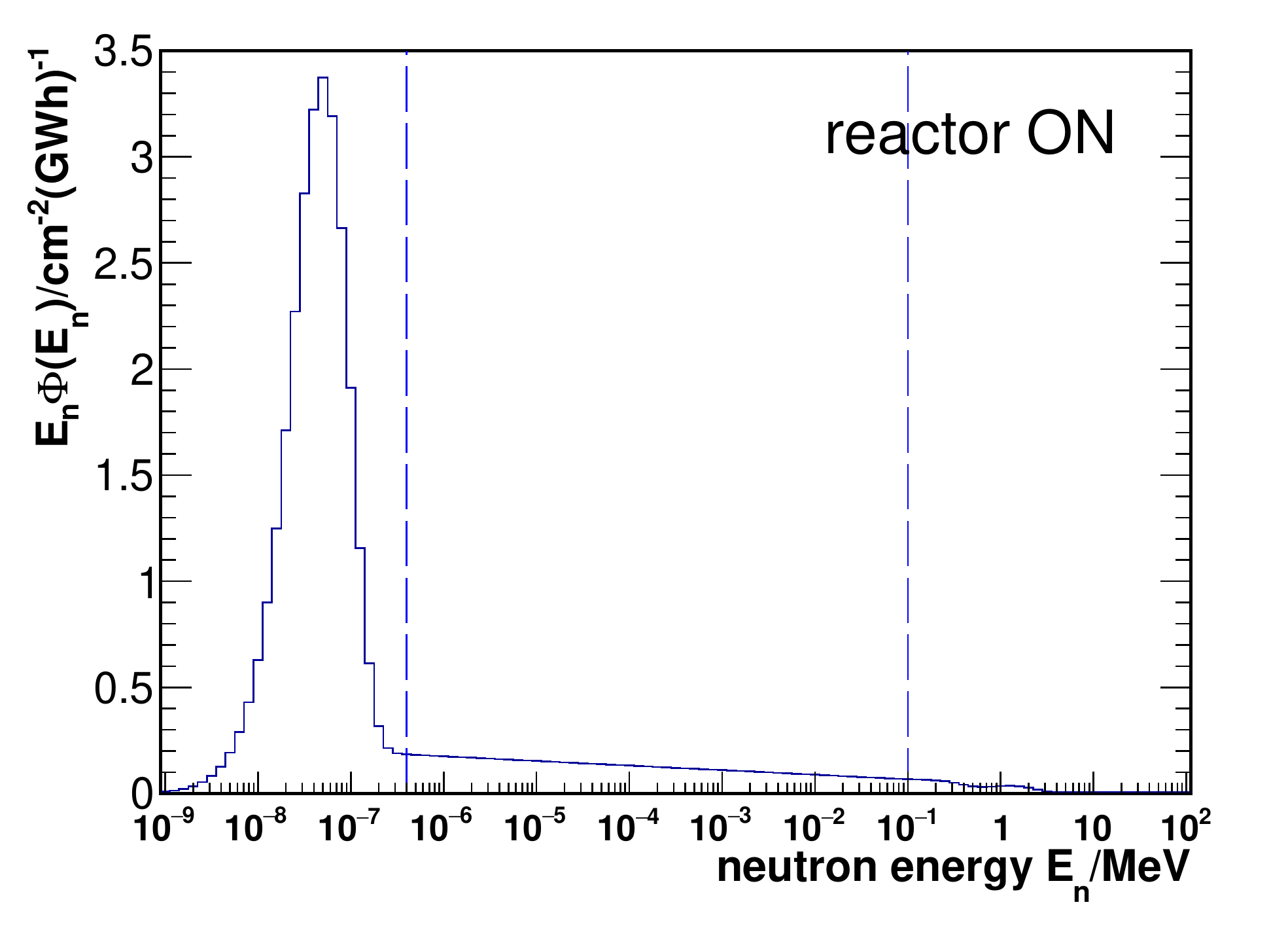}
	\caption{Solution of the neutron energy distribution $\mathit{\Phi}^{(1)}(E_\text{n})$ resulting from the analysis of the DS-3 data normalized to the energy emitted by the reactor}
	\label{fig:unfolded_ON}
\end{figure}

It was checked, whether the BSS data indicate the presence of high-energy cosmic-ray induced neutrons ($E_\text{n}\sim100$\,MeV). This was done by considering, in addition to the model described above, a second, preliminary model, which included a high-energy peak at this energy. The analysis showed that the BSS data favors the models not including such neutrons. This is expected due to the massive concrete shielding of the reactor building, surrounding the CONUS experimental site. From the BSS data alone it could not be determined whether there is a slope of the intermediate region significantly different from zero. The detailed MC calculations clearly indicate a non-zero slope (cf. Section \ref{subsec:resultreactorcoreMC}). Therefore we decided to use here the results of the non-zero slope model only. The resulting neutron energy distribution $\mathit{\Phi}^{(1)}(E_\text{n})$ is plotted in Figure~\ref{fig:unfolded_ON} in terms of the lethargy representation. The integral quantity of the neutron fluence normalized to the KBR thermal energy output $\mathit{\Phi}^{(1)}$, derived from the analysis, is stated in Table~\ref{tab:integral_quantities_ON} for the individual $E_\text{n}$ regions of the neutron energy distribution. The overall shape of the solution  $\mathit{\Phi}^{(1)}(E_\text{n})$ agrees reasonably well with the MC predictions, as discussed in Section \ref{subsec:resultreactorcoreMC}.

\begin{table}
	\caption{Neutron fluence $\mathit{\Phi}^{(1)}$ derived from the analysis of the DS-3 measurement, and normalized to the KBR thermal energy output of 1\,GW\,h. The uncertainties are stated as standard uncertainties (1~standard deviation). The $E_\text{n}$ regions are defined as follows: thermal $[1.0\times10^{-9},\,4.0\times10^{-7}]$\,MeV; intermediate $[4.0\times10^{-7},\,0.1]$\,MeV; fast $[0.1,\,19.6]$\,MeV.}
	\label{tab:integral_quantities_ON}
		\begin{tabular}{lc}
		\hline\noalign{\smallskip}
			$E_\text{n}$ region & $\mathit{\Phi}^{(1)}$ \\
			 					&   [$\text{cm}^{-2}\,(\text{GW\,h})^{-1}$]		\\
		\noalign{\smallskip}\hline\noalign{\smallskip}
			thermal				& $6.42 \pm 0.41$			\\
			intermediate		& $1.56 \pm 0.21$			\\
			fast				& $0.15 \pm 0.05$			\\ 	\noalign{\smallskip}\hdashline\noalign{\smallskip}
			total				& $8.13 \pm 0.32$			\\
		\noalign{\smallskip}\hline
		\end{tabular}
\end{table}

The solution indicates a highly thermalized neutron field, as about 80\% of the total fluence $\mathit{\Phi}^{(1)}$ is due to thermal neutrons of energies $E_\text{n}\le0.4$\,eV. Therefore, the count rate observed in a given Bonner sphere has a non-negligible contribution from thermal neutrons, even for large Bonner spheres (for example, 30--50\% of the measured counts in the 8--12\,'' spheres are due to thermal neutrons\footnote{These portions were determined via folding the solution of $\mathit{\Phi}^{(1)}(E_\text{n})$ with the response functions of the individual spheres, and then comparing the expected counts in the region of thermal neutron energies to the total expected counts.}). This is one reason why the fast component has a relatively large uncertainty of the order of 30\%.

The bare counter used for monitoring was running permanently during the data collection of DS-3. For a few days, at the central position of the CONUS experiment also a bare counter was placed. Comparing the neutron count rates measured by both counters and correcting them for slightly different sensitivities, a relative difference of (18.8$\pm$2.3)\% was found. This demonstrates inhomogeneities in the thermal field inside room A408 and underlines the importance to characterize the neutron spectrum at the exact location, where an experiment is planned.

%---------------------------------------------------------------
\subsubsection{Neutron energy distribution inside A408 during reactor OFF time}
\label{subsec:reactor_OFF}

The data from the reactor OFF measurement (DS-2) are depicted in Figure~\ref{fig:count_rates}. In this case the normalization was done simply as
\begin{equation}
\label{eq:OFF_norm}
N_i^{(2)}=\frac{C_i^{(2)}}{t_i}\,,
\end{equation}
where $C_i^{(2)}$ is the number of neutron-induced counts determined via fits of the PHS and $t_i\equiv t$ is the measurement time which was common for all Bonner spheres and amounted to 18\,d. The measurement uncertainties of the neutron count rates $N_i^{(2)}$ were in the range of 5--10\%. The maximal count rates were not observed in the bare detector, but instead in the 4.5 and 6\,'' spheres, indicating that the neutron field was no longer dominated by thermal neutrons during reactor OFF time.

\begin{figure}[h]
	\centering
	\includegraphics[width=1.05\columnwidth]{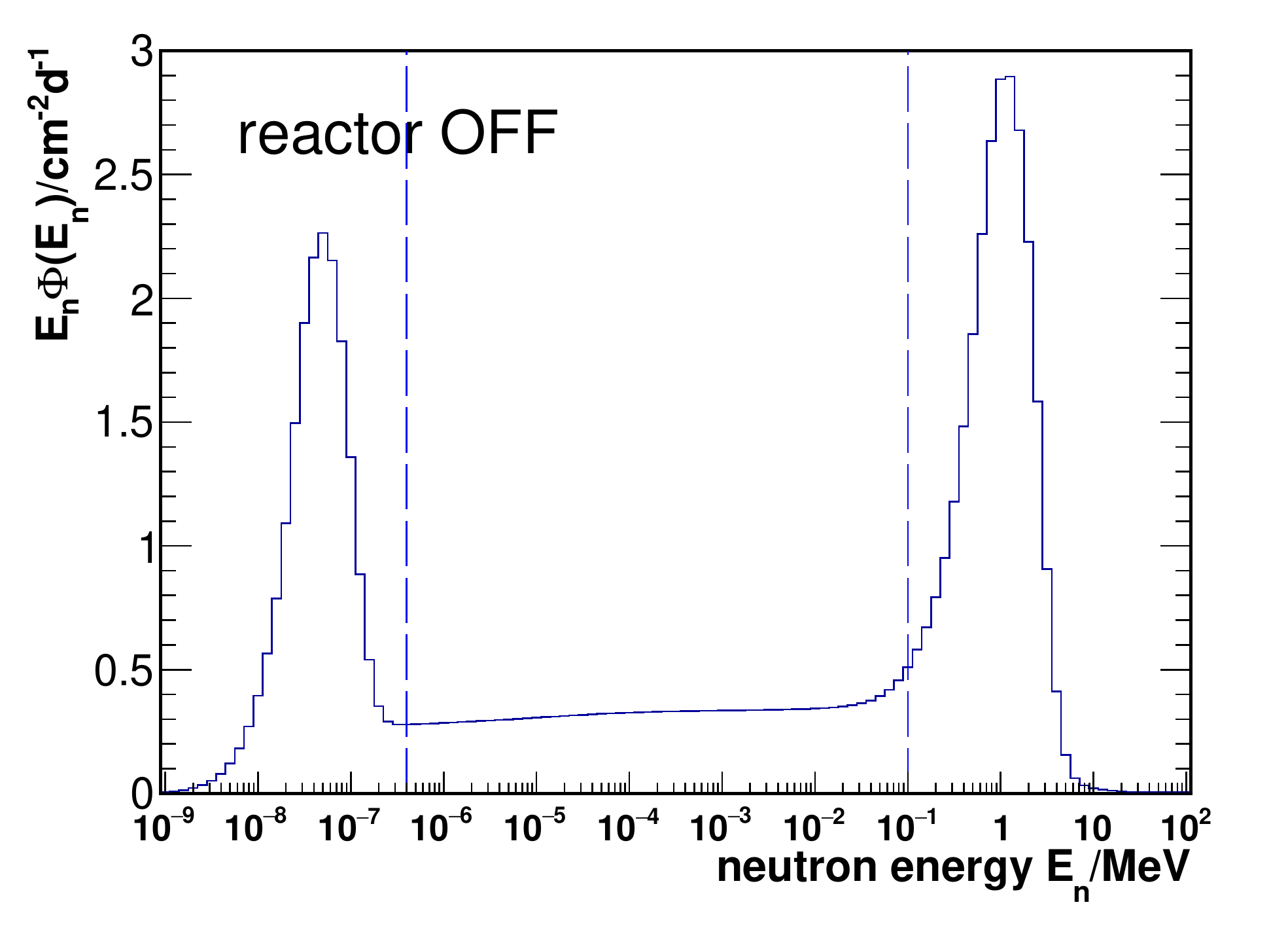}
	\caption{Solution of the neutron energy distribution $\mathit{\Phi}^{(2)}(E_\text{n})$ resulting from the analysis of the DS-2 data normalized to the measurement time}
	\label{fig:unfolded_OFF}
\end{figure}

For the analysis, we used a parameterized model consisting of a thermal peak, a peak at $E_\text{n}\sim$1\,MeV, and an intermediate region, which is flat in the lethargy representation. Thus, the complete model contained five free parameters: the magnitude of the thermal peak, the magnitude and the slope of the intermediate region, and the magnitude and mean energy of the $E_\text{n}\sim$1\,MeV peak. The solution $\mathit{\Phi}^{(2)}(E_\text{n})$ is plotted in Figure~\ref{fig:unfolded_OFF}. The integral quantity of the neutron fluence rate $\mathit{\Phi}^{(2)}$ derived from the analysis is listed in Table 7. The largest contribution ($\sim$40\%) to the total neutron fluence is now in the fast neutron region $E_\text{n}=[0.1,\,19.6]$\,MeV.

\begin{table}
	\caption{Neutron fluence rate $\mathit{\Phi}^{(2)}$ derived from the analysis of the DS-2 measurement (reactor OFF), normalized per 24\,h. The uncertainties are stated as standard uncertainties (1~standard deviation). The definition of the $E_\text{n}$ regions is identical to Table~\ref{tab:integral_quantities_ON}.}
	\label{tab:integral_quantities_OFF}
		\begin{tabular}{lr}
		\hline\noalign{\smallskip}
	$E_\text{n}$ region & \multicolumn{1}{c}{$\mathit{\Phi}^{(2)}$} \\
	&   \multicolumn{1}{c}{[$\text{cm}^{-2}\,\text{d}^{-1}$]}		\\
	\noalign{\smallskip}\hline\noalign{\smallskip}
	thermal				& $4.47 \pm 0.67$			\\
	intermediate		& $4.19 \pm 1.15$			\\
	fast				& $6.35 \pm 0.96$			\\ \noalign{\smallskip}\hdashline\noalign{\smallskip}
	total				& $15.03\pm 0.99$			\\
	\noalign{\smallskip}\hline 
		\end{tabular}
\end{table}

\subsubsection{Correlation to thermal reactor power}
\label{subsec:meas-kbr}
\begin{figure*}[h]
    \includegraphics[width=0.9\textwidth]{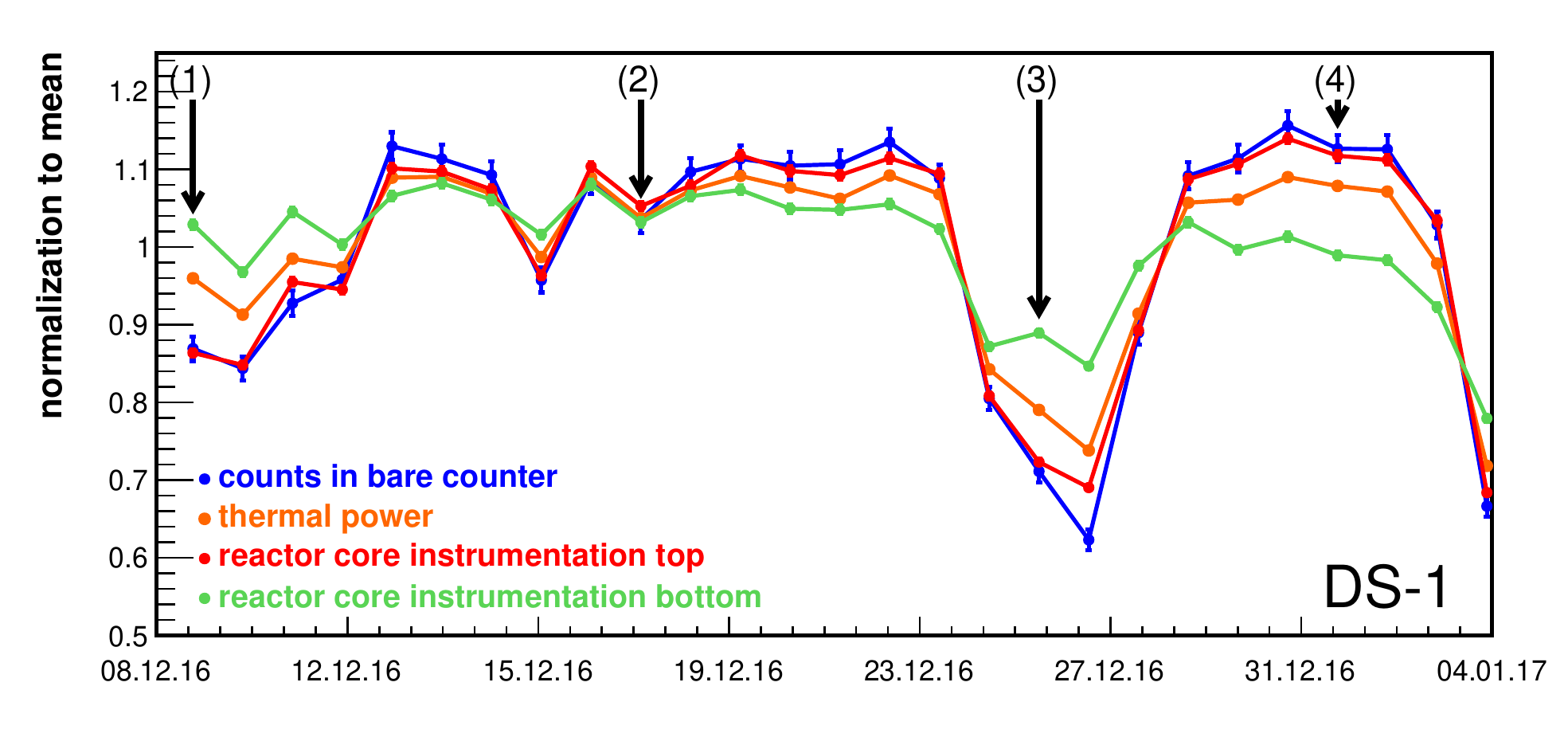}
    \caption{DS-1 at the end of a reactor cycle; counts in bare BSS compared to the thermal power and the core instrumentation. The arrows indicate the days, where the thermal power distribution along the z axis is shown in Figure~\ref{fig:thermpoweralongzaxis1_simple} }
    \label{fig:ncountsvsthermpower1}
    \end{figure*}

\begin{figure*}[h]
    \includegraphics[width=0.9\textwidth]{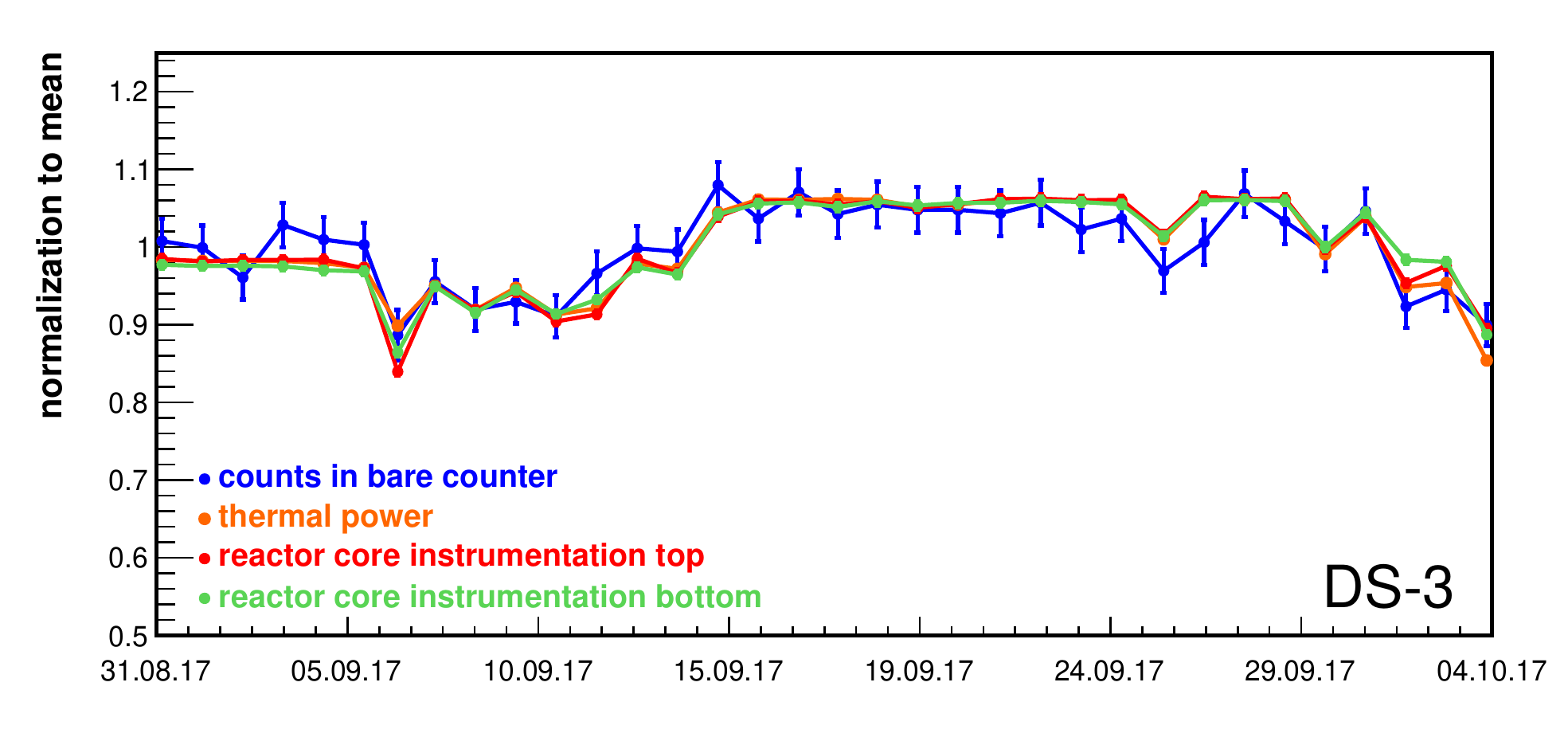}
    \caption{DS-3 at the beginning of a reactor cycle; counts in bare BSS compared to the thermal power and the core instrumentation}
    \label{fig:ncountsvsthermpower2}
    \end{figure*}

The mean value of the neutron count rate per day is compared to the mean thermal power per day as well as the neutron flux measured with the ex-core instrumentation in Figure~\ref{fig:ncountsvsthermpower1} and Figure~\ref{fig:ncountsvsthermpower2} for both measurement campaigns. Both curves are normalized to their mean value during the campaign. Overall, for both cases a correlation between the neutron fluence rate in A408 and the data from the reactor was observed.
DS-3 took place immediately after the refueling of the reactor in 2017 and the reactor power was nearly constant except for the increase in the middle of the month. The thermal power data as estimated from the energy balance in the secondary circuit and the ex-core instrumentation are in very good agreement. Both match the course of the thermal neutron fluence in room A408 as well. 
During DS-1 at the end of the previous reactor cycle were more variations in the thermal power. This is due to less demand for electricity during holiday seasons and competing renewable energy production.

\begin{figure}[h]
    \includegraphics[width=0.5\textwidth]{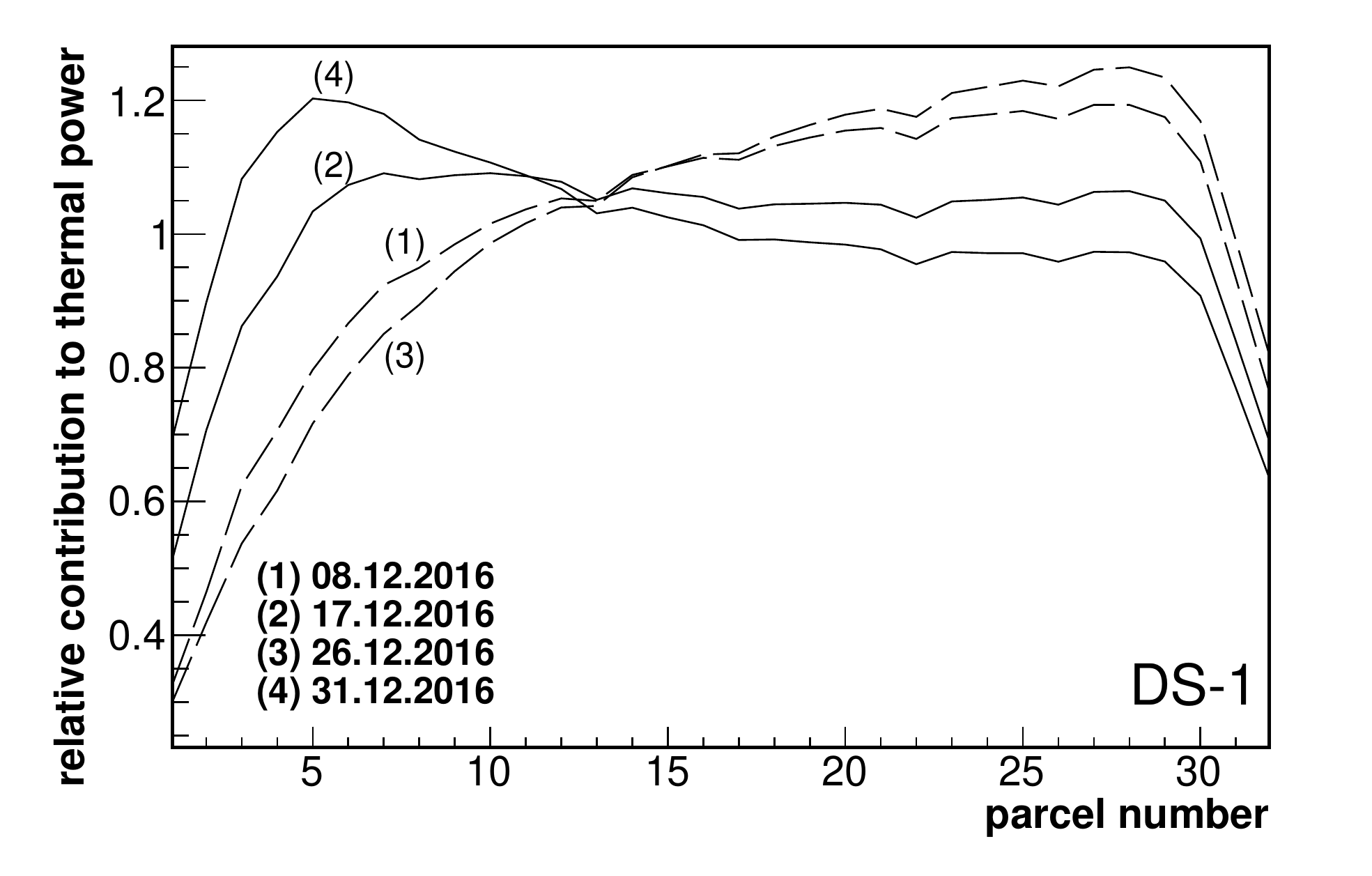}
    \caption{Relative distribution of thermal power along the z-axis during DS-1 from core simulation data for selected days over the one month period. The dashed lines (1) and (3) correspond to the days, where the bottom core instrumentation in Figure~\ref{fig:ncountsvsthermpower1} gives the highest contribution different to (2) and (4)}
    \label{fig:thermpoweralongzaxis1_simple}
    \end{figure}

Small discrepancies are observed between the total thermal power determination as compared to the ex-core instrumentation measuring neutrons at the top and bottom of the core. They occur because the maximum of the thermal power has moved up or down along the z-axis of the reactor. The effect is displayed in Figure~\ref{fig:thermpoweralongzaxis1_simple} where exemplary for a few days the relative contribution of the 32 parcels to the overall reactor power is displayed using the data from the reactor core simulation described in Section \ref{subsec:incoreinstrumentation}. %Dents in the lines are created in periodic distances due to material from the holding structure of the fuel assemblies, where neutrons can be absorbed. 
The dashed lines correspond to the days in Figure~\ref{fig:ncountsvsthermpower1} when the bottom instrumentation gives values above the total thermal power from heat. For those days the maximum of the power density is at the bottom of the reactor core.
The shift of the maximum is initiated, because at the beginning of a reactor cycle the thermal power is distributed nearly symmetrically with a maximum slightly below the middle of the core where the leakage of neutrons is small.
%The maximum is created because in the middle of the core the leakage of neutrons is small. 
At the end of the cycle the fuel in the middle of the reactor-core is mostly burned up leading to the maximum moving outwards. Additionally, when reducing the thermal power due to external constraints, the control rods are inserted into the core from the top causing a shift of the maximum to the bottom. Strong maxima on top of the core can also be produced at the end of a cycle because - if there is a negative temperature gradient within the core - more reactivity is freed in the upper half. 
%\textcolor{red}{("Streckbetrieb, keine englische Entsprechung?")}.

From the comparison to the thermal neutron fluence in A408, it was found that the best agreement could be achieved for the counters on top of the reactor core independent of the location of the maximum of the thermal power along the z axis. The neutrons seem to be more likely to escape the reactor core at the top, where the openings for the loop pipes are located. 
%\textcolor{red}{The neutrons are potentially transported outward by the flow of the cooling water with the speed of 16\,ms$^{-1}$}.

All in all, the thermal fluence in A408 was observed to be fully correlated to the reactor power and reactor core instrumentation meaning the reactor is by far the dominant source of thermal neutrons inside the room. Consequently, the reactor monitoring data can be used to predict the thermal neutron fluence inside room A408 at any given time. 

\subsubsection{Comparison of neutron fluences ON-OFF}
\label{subsec:fluences_ON-OFF}

The comparison of the neutron fluences during reactor ON and OFF time is important to disentangle the thermal power correlated neutron contribution from other steady-state neutron sources.

Due to the different normalisations, it is necessary to make some simplifying assumptions. In a rough approximation we consider the KBR reactor as a source of neutrons, and the neutron fluence $\mathit{\Phi}^{(1)}$ as being normalised to a unit of ``activity'' of this source. In Section \ref{subsec:meas-kbr} a linear correlation between the reactor power and the thermal neutron fluence within A408 was observed, justifying this approach.

For the estimate we used the time-resolved thermal power data $P_\text{KBR}$ already applied in the DS-3 measurement. The mean duty cycle of the KBR reactor during the overall DS-3 measurement (total measurement time of 34\,d) amounted to 88.9\%, i.e. the mean thermal power of the reactor was
\begin{equation}
\label{eq:ON_mean_power}
P_\text{KBR}^\text{DS-3 mean}=(0.89\pm0.02)\cdot P_\text{KBR}^\text{max} = (3.47\pm0.08)\,\text{GW}\,.
\end{equation}
Multiplying the values of $\mathit{\Phi}^{(1)}$ by the following factor of a mean thermal energy output
\begin{equation}
\label{eq:ON_mean_energy}
E_\text{KBR}^\text{DS-3 mean} = 3.47\,\text{GW} \cdot 24\,\text{h} = 83.3\,\text{GW\,h}\,,
\end{equation}
we obtain the mean neutron fluence $\mathit{\Phi}^{(1)}_\text{mean}$ which would hypothetically be present in A408 if the reactor would be running at the constant level of $P_\text{KBR}^\text{DS-3 mean}$ over the period of one day. Both quantities $\mathit{\Phi}^{(1)}_\text{mean}$ and  $\mathit{\Phi}^{(2)}$ have units of [$\text{cm}^{-2}\,\text{\,d}^{-1}$] and we define the ``$\text{ON}-\text{OFF}$'' difference,
\begin{equation}
\label{eq:ON-OFF_difference}
\mathit{\Phi}^\text{(res)}_\text{mean} (E_\text{n}) = \mathit{\Phi}^{(1)}_\text{mean} (E_\text{n}) - \mathit{\Phi}^{(2)} (E_\text{n})\,,
\end{equation}
as the residual neutron fluence $\mathit{\Phi}^\text{(res)}_\text{mean}$, caused solely by the fission neutrons from the reactor core.

The comparison of $\mathit{\Phi}^{(1)}_\text{mean}$ and $\mathit{\Phi}^{(2)}$ is summarized in Table~\ref{tab:integral_quantities_ON-OFF_comparison}. The uncertainties of $\mathit{\Phi}^\text{(res)}_\text{mean}$ in the individual energy regions were obtained by propagating the statistical uncertainties of $\mathit{\Phi}^{(1)}_\text{mean}$ and $\mathit{\Phi}^{(2)}$. In the thermal and intermediate regions, the absolute values of $\mathit{\Phi}^{(2)}$ amount to a tiny fraction of $\mathit{\Phi}^{(1)}_\text{mean}$ values, therefore $\mathit{\Phi}^\text{(res)}_\text{mean}$ retains the characteristics of a highly thermalized neutron field. For the fast neutron component, the residual neutron fluence rate is (6.0$\pm$4.2)\,cm$^{-2}$d$^{-1}$, which is compatible with zero within the uncertainties.% One should keep in mind that the large uncertainty quoted here (and mostly due to the uncertainty associated with the reactor ON component) is very likely an underestimate of the true uncertainty. We have only added the uncertainties of the ON and OFF components by quadrature without taking into account possible additional sources of uncertainties that could come from the normalization assumptions made in the analysis.

To estimate the residual neutron fluence for any given reactor power, the results in Table~\ref{tab:integral_quantities_ON-OFF_comparison} have to be divided by the mean thermal power from Eq. \ref{eq:ON_mean_power} and scaled with the respective thermal power. This was done for the maximum thermal power in Table~\ref{tab:MCneutronfluencerateatdiode}.

\begin{table*}[btp]
	\caption{Comparison of neutron fluence rates $\mathit{\Phi}^{(1)}_\text{mean}$ (reactor ON at constant mean level for time duration of 24\,h) and $\mathit{\Phi}^{(2)}$  (reactor OFF), and their difference $\mathit{\Phi}^\text{(res)}_\text{mean}$. All quantities are stated in units of [$\text{cm}^{-2}\,\text{\,d}^{-1}$]. The uncertainties are stated as standard uncertainties (1~standard deviation). The definition of the $E_\text{n}$ regions is identical to the one used in Table~\ref{tab:integral_quantities_ON}.}
	\label{tab:integral_quantities_ON-OFF_comparison}
			\begin{tabular}{lrrr}
			\hline\noalign{\smallskip}
			$E_\text{n}$ region	& \multicolumn{1}{c}{$\mathit{\Phi}^{(1)}_\text{mean}$} & \multicolumn{1}{c}{$\mathit{\Phi}^{(2)}$} & \multicolumn{1}{c}{$\mathit{\Phi}^\text{(res)}_\text{mean}$} \\
			\noalign{\smallskip}\hline\noalign{\smallskip}
			thermal				& $534.8 \pm 34.4$  & $4.47 \pm 0.67$  & $530.3 \pm 34.4$ \\
			intermediate		& $130.0 \pm 17.2$	& $4.19 \pm 1.15$  & $125.8 \pm 17.3$ \\
			fast				& $12.3  \pm 4.1$	& $6.35 \pm 0.96$  & $6.0 \pm 4.2$ \\ \hdashline
			total				& $677.2 \pm 27.0$	& $15.03\pm 0.99$  & $662.2 \pm 27.1$ \\
		\noalign{\smallskip}\hline
		\end{tabular}
	\end{table*}

\begin{figure}[h]
	\centering
	\includegraphics[width=1.05\columnwidth]{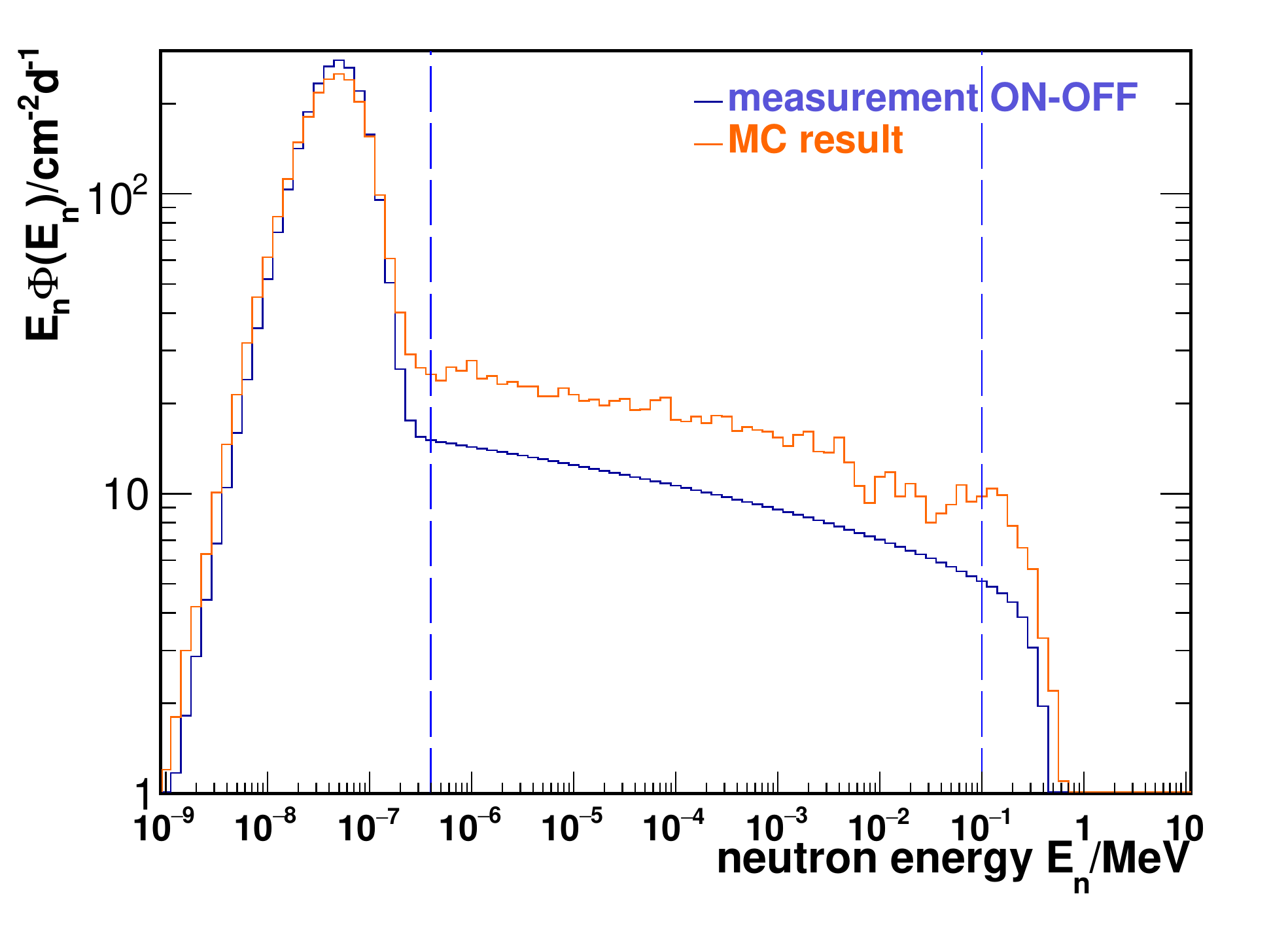}
	\caption{Neutron energy distributions of the residual fluence $\mathit{\Phi}^\text{(res)}_\text{mean} (E_\text{n})$, calculated via Eq.~\ref{eq:ON-OFF_difference} from the experimental BSS data analysis, and the calculated fluence $\mathit{\Phi}_\text{MC} (E_\text{n})$ based on MC simulations. The distribution $\mathit{\Phi}_\text{MC} (E_\text{n})$ was scaled to match the integral fluence of $\mathit{\Phi}^\text{(res)}_\text{mean}$ in the thermal neutron region. The vertical dotted lines indicate the limits of the individual $E_\text{n}$ regions, as defined in Table~\ref{tab:integral_quantities_ON}}
	\label{fig:ON-OFF_diff}
\end{figure}

The neutron energy distribution $\mathit{\Phi}^\text{(res)}_\text{mean} (E_\text{n})$ is shown in the lethargy representation in Figure~\ref{fig:ON-OFF_diff}. It is plotted together with the neutron energy distribution from the MC outcome $\mathit{\Phi}_\text{MC} (E_\text{n})$ which describes the neutrons arriving on the outside of room A408 from the KBR reactor core. The MC output is discussed in detail in the next section. The distribution $\mathit{\Phi}_\text{MC} (E_\text{n})$ was scaled in such a way that the fluences in the thermal region $E_\text{n}=[1.0\times10^{-9},\,4.0\times10^{-7}]$\,MeV match.

\subsection{Results from MC at KBR}
\label{subsec:resultreactorcoreMC}

\subsubsection{Propagation from reactor core to A408}
\begin{figure}[ht!]
    \includegraphics[width=0.51\textwidth]{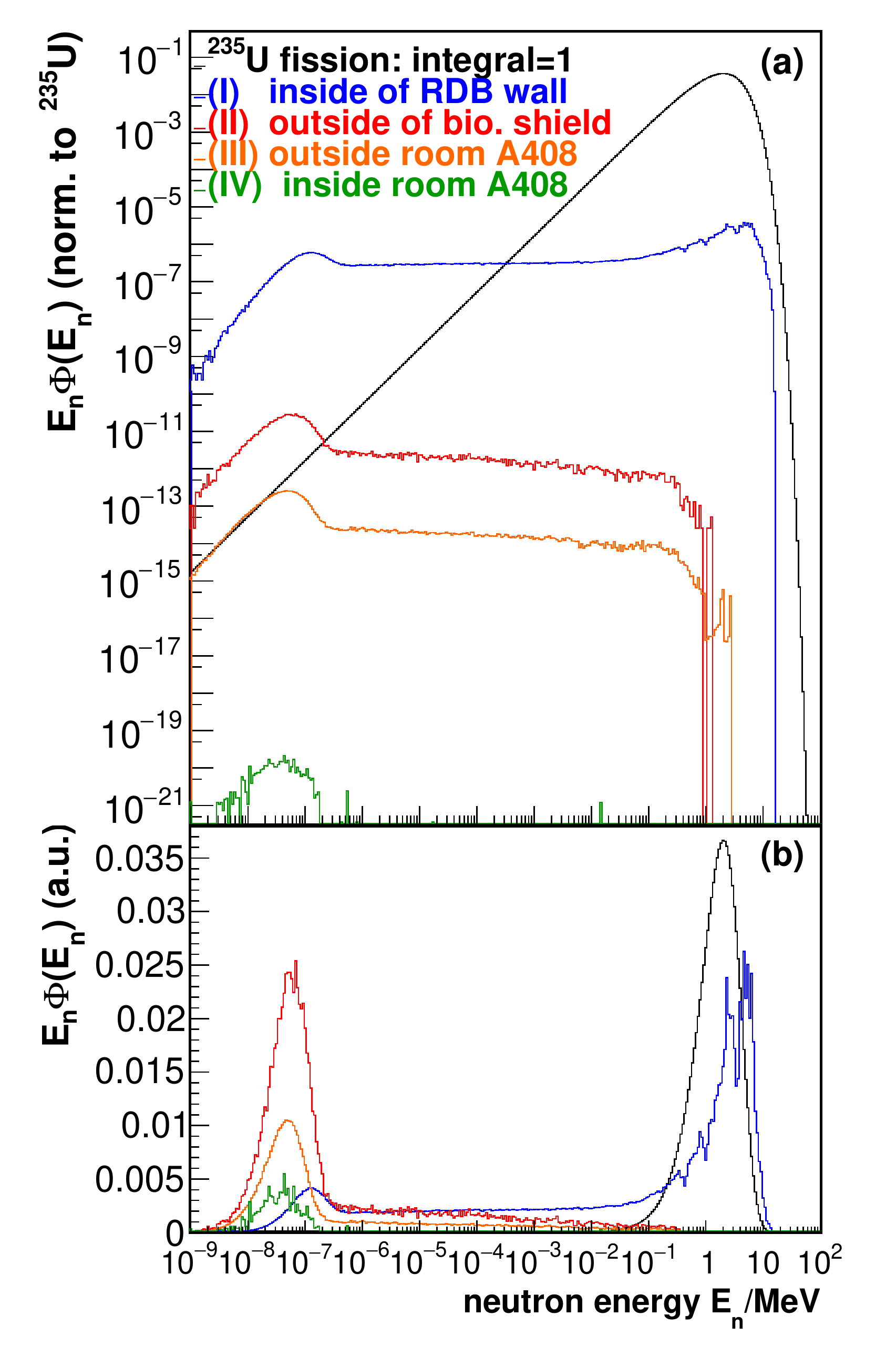}
    \caption{(a) MC neutron spectra of the propagation from the reactor core to room A408, the fission spectrum is normalized to 1 (b) same spectra as in (a) with a linear y-axis scaled arbitrarily to make spectral features visible}
    \label{fig:MCreactorcore_spectra_propagation}
    \end{figure}

In the MC simulation, the propagation of neutrons from the reactor core is split into four steps as indicated in Figure~\ref{fig:MCgeometry_display}. Before leaving the reactor core, most of the fission neutrons are moderated and absorbed inside the reactor fuel, fuel rod claddings and borated water (see Figure~\ref{fig:MCreactorcore_spectra_propagation}, blue spectrum). In the MC, absorption by fission fragments is not taken into account. 
Only remnants of fast neutrons can reach the inside of the RPV wall (marked as (I) in Figure~\ref{fig:MCgeometry_display}). The shape of the spectrum is dominated by the neutron absorption cross sections of light nuclei like oxygen (O) or B, which can be observed in Figure~\ref{fig:MCreactorcore_spectra_propagation}(b) with a linear y-axis. There are no longer neutrons above 17\,MeV and about 10\% of the neutrons are in the thermal energy regime. The thermal distribution is shifted towards higher energy with respect to the thermal distribution at room temperature as the mean temperature inside the RPV is at about 320\,$^{\circ}$C. The overall fluence of neutrons from the outer fuel assembly ring hitting the RPV wall is reduced by a factor of 1.4$\cdot$10$^{-4}$, while for the second outer ring it is 1$\cdot$10$^{-5}$, as these neutrons have to pass the additional layer of fuel assemblies where they can be absorbed or induce fission. The order of magnitude difference justifies the approach to only start neutrons from the outer volumes of the reactor core. 
Most neutrons arrive at the RPV wall on the level of the reactor core. Nearly no neutrons reach the area about 1\,m above or below the extension of the core. 

Next, the neutrons go through the steel of the RPV and several layers of concrete to the outside of the biological shield (marked as (II) in Figure~\ref{fig:MCgeometry_display}). This leads to a partly thermalization of the spectrum with about 70\% of the fluence inside the thermal peak. Also, the maximum neutron energy is lowered to $\sim$2.5\,MeV and the overall fluence is suppressed by an additional factor of 6.5$\cdot$10$^{-6}$ (see Figure~\ref{fig:MCreactorcore_spectra_propagation}, red spectrum). Many neutrons scatter back and forth between the different concrete layers of the biological shield, where they are thermalized and absorbed before they can leave the shield. 

Starting from the outside of the biological shield, only the neutrons arriving at the outside walls of A408 are of interest (location marked as (III) in Figure~\ref{fig:MCgeometry_display}). A reduction of 9$\cdot$10$^{-3}$ is expected from simple solid angle considerations, which is the suppression factor observed in the MC. The shape of the spectrum is slightly different due to the reflection of neutrons on the ceiling (see Figure~\ref{fig:MCreactorcore_spectra_propagation}, orange). About 72\% of the total fluence can be found inside the thermal peak already similar to the spectrum expected inside room A408.

In the last step, the neutrons are propagated through the wall of A408 interfacing the area around the biological shield. The neutrons leaving the wall on the other side are registered as well as those arriving inside an air plate and a sphere inside the room (marked as (IV) in Figure~\ref{fig:MCgeometry_display}). The last concrete wall leads to a complete thermalization of the spectrum (see Figure~\ref{fig:MCreactorcore_spectra_propagation}, green spectrum). 
The MC predicts a complete thermalized spectrum, while from the BSS measurement a thermal neutron fluence contribution of at least 80\% is expected, but also the neutron spectrum goes up to a few hundred keV even though with large uncertainties. Potentially, this discrepancy originates in the limited knowledge on the geometry and exact concrete composition of the wall.  
Thus, instead of the MC spectrum inside the room, the spectrum of neutrons hitting the outer wall was used to support the BSS analysis (see Section \ref{subsec:fluences_ON-OFF}) and compared to the measured spectrum in Figure~\ref{fig:ON-OFF_diff} achieving a good agreement. 

The neutron fluence is reduced by a factor of 10$^{-8}$ traveling through the outside wall of A408. Due to the already highly thermalized spectrum outside the room, a large fraction of neutrons is captured inside the wall, which is exploited in the measurement with the HPGe spectrometer CONRAD without shield (cf. Section \ref{subsec:conradoutsideshield}). 
Reducing the hydrogen content in the MC inside the concrete by a factor of two increases the number of observed neutrons leaving the wall by a factor of 4 with the spectrum still completely thermalized. This shows the relevance of knowing this number precisely.

All suppression factors are summarized in Table~\ref{tab:MCsuppressionfactorpropagation}. In the MC an overall reduction of 3.6$\cdot$10$^{-20}$ for neutrons entering through the wall of A408 with an area of 10\,m$^2$ is obtained, making it possible to access room A408 at any time even when the reactor is operational at full thermal power.

A significant amount of neutrons hitting the wall several times, scattering back inside the room, are observed. Comparing the total  number of neutrons entering through the adjoined wall to the space around the biological shield, about 10\% more neutrons leave the inclined middle piece of the wall than the straight wall pieces.

\begin{table*}[btp]
\caption{Supression of neutron fluence rate during propagation from reactor core to A408 in the MC corresponding to the respective surfaces.\label{tab:MCsuppressionfactorpropagation}}

\begin{tabular}{llll}
\hline\noalign{\smallskip}
location &area [m$^2$] &suppression factor to previous & maximum neutron energy\\\noalign{\smallskip}\hline\noalign{\smallskip}
reactor core &42.2 (cylinder) &1 & $>$20\,MeV\\
inner wall of RPV (I) & 62 (cylinder) & 1.6$\cdot$10$^{-4}$& 16.2\,MeV  \\
outside wall biological shield (II)& 355.2 (cylinder)& 6.5$\cdot$10$^{-6}$& 2.5\,MeV \\
outside wall A408 (III)& 18.4 (3 plates) &9$\cdot$10$^{-3}$& 1\,MeV \\
inside wall A408 (IV)&10.2 (3 plates)&4$\cdot$10$^{-8}$ &0.17\,eV\\ \noalign{\smallskip}\hline\noalign{\smallskip}
total & 1 & 3.6$\cdot$10$^{-20}$ &\\
\noalign{\smallskip}\hline
\end{tabular}
\end{table*}

\subsubsection{Normalization of MC}
\label{subsec:normmcoutcome}
To normalize the MC simulation, the number of neutrons per fission is required. 
The usable energy released per fission by the four main fission isotopes ($^{235}$U, $^{239}$Pu, $^{238}$U and $^{241}$Pu) weighted with the respective fission fractions (see Table~\ref{tab:MCsuppressionfactornormalization}) can be used to determine the number of fissions required to achieve a certain thermal power output. The fission fractions have been evaluated as mean value from a wide range of similar reactors in various states of the reactor cycle. The given uncertainty corresponds to the standard deviation of the mean value. Depending on the isotope, about 2.5 neutrons are released per fission (see Table~\ref{tab:MCsuppressionfactornormalization}). This leads to a total number of free neutrons for the maximum thermal reactor power of $(3.1\pm0.1)$\,$\cdot$\,10$^{20}$\,s$^{-1}$(3.9\,GW)$^{-1}$ which will be applied as normalization for the MC. For a reactor at the point of criticality, one neutron per fission will induce the next step in the fission chain, while the others are thermalized and absorbed in the materials of the reactor core with a small fraction leaving the reactor core. The MC follows the propagation of these neutrons going outwards from the core. %The MC starts with the released neutrons inside the fuel rods, where they will undergo the mentioned processes. 
%As the aim of the MC is to follow the propagation of the neutrons leaving the reactor core, all fission products are terminated immediately. 
%The normalization is considered to be valid for the whole reactor cycle. For the energy released per fission only changes within \textcolor{red}{2\%} are expected \cite{MCnorm:Kopeikin2004}.
The approximation in the MC that only neutrons from the outer ring of fuel assemblies and the second outer ring have a chance to escape the reactor core as described in Section \ref{sec:neutron-MC} has to be accounted for in the normalization. From the core simulation data with a high spatial resolution, as described in Section \ref{subsec:incoreinstrumentation}, it was found that consistently during the collection of DS-1 and DS-3 (16.3$\pm$0.1)\% of the thermal power is created by the outer ring of fuel assemblies and (30.4$\pm$0.1)\% by the second outer ring. The higher neutron flux for the second outer ring is partially compensated by their higher absorption probability, such that the contribution is a factor of 7 below the one of the outer ring.\newline
By scaling the MC in this way, for 40 years of operation with maximum thermal power (3.7$\pm$0.1)$\cdot$10$^{18}$\linebreak[0]\,cm$^{-2}$ neutrons are expected to hit the wall of the RPV. This number is consistent with the maximum level of 10$^{19}$\,cm$^{-2}$ for this range of time found in the safety guidelines for German PWRs \cite{siemens2003}.\newline
For the total thermal fluence rate inside room A408, (16$\pm$1)\,cm$^{-2}$d$^{-1}$(3.9\,GW)$^{-1}$ are expected in the MC. Comparing directly with the result for the thermal fluence rate for the reactor ON minus reactor OFF subtraction obtained from the BSS measurements (see Table~\ref{tab:MCneutronfluencerateatdiode}), the MC lies below the measurement by a factor of 38. This has to be contrasted with the suppression of the initial neutron flux from the reactor core by more than 20 orders of magnitude and the overall complexity of the simulation. Unknown factors in the geometry of the extended void space around the biological shield as well as of the exact composition of the steel enforced concrete further handicap a more exact reproduction of the measurement result.

As in the MC only thermal neutrons arrive inside A408, for the comparison to the measured spectrum instead the spectrum in front of A408 at the outside wall has been used (see Figure~\ref{fig:ON-OFF_diff}). In the MC, only 72\% of the neutrons are found inside the thermal peak as the last step of the thermalization is still missing. The agreement of the overall shape of the distributions is very good, keeping in mind again the complexity of the MC simulations. %In particular, the location of the peak at around 1\,MeV and the maximum energy agree very well. 
This confirms that the remaining neutrons at energies above 1\,MeV seen in the reactor ON spectrum in Figure~\ref{fig:unfolded_ON} are indeed not thermal power correlated and correctly removed by subtracting the reactor OFF spectrum. These neutrons are presumably created by muons passing through the concrete of the building. By generating neutrons in the MC inside the spent fuel assemblies in the cooling pond, it was confirmed that this is not the origin of the fast neutrons. In fact, from  the neutrons of the spent fuel assemblies, no contribution to the fluence inside room A408 is expected.

\begin{table*}[btp]
\caption{Reactor physics input for the calculation of the neutrons produced in the reactor core that is used for the normalization of the MC simulation.\label{tab:MCsuppressionfactornormalization}}
\begin{tabular}{llll}
\hline\noalign{\smallskip}
fission isotope& energy release per fission [MeV] \cite{MCnorm:Ma2013}& fission fraction \cite{An2016},\cite{Djurcic:2008ny}&number of neutrons per fission \cite{MCnorm:Kopeikin2004}  \\\noalign{\smallskip}\hline\noalign{\smallskip}
$^{235}$U &202.36$\pm$0.26 & 56.8$\pm$3.2& 2.432$\pm$0.004\\
$^{239}$Pu & 211.12$\pm$0.34&30.2$\pm$0.4 &2.875$\pm$0.006 \\
$^{238}$U &205.99$\pm$0.52 &7.6$\pm$2.4 & 2.937$\pm$0.007\\
$^{241}$Pu &214.26$\pm$0.33 &5.4$\pm$0.7 & 2.829$\pm$0.011\\
\noalign{\smallskip}\hline
\end{tabular}
\end{table*}

\section{CONRAD detector at KBR}
\label{subsec:conradoutsideshield}

\subsection{Measurement campaign with CONRAD detector}
\begin{figure}[h]
	\centering
	\includegraphics[width=0.65\columnwidth]{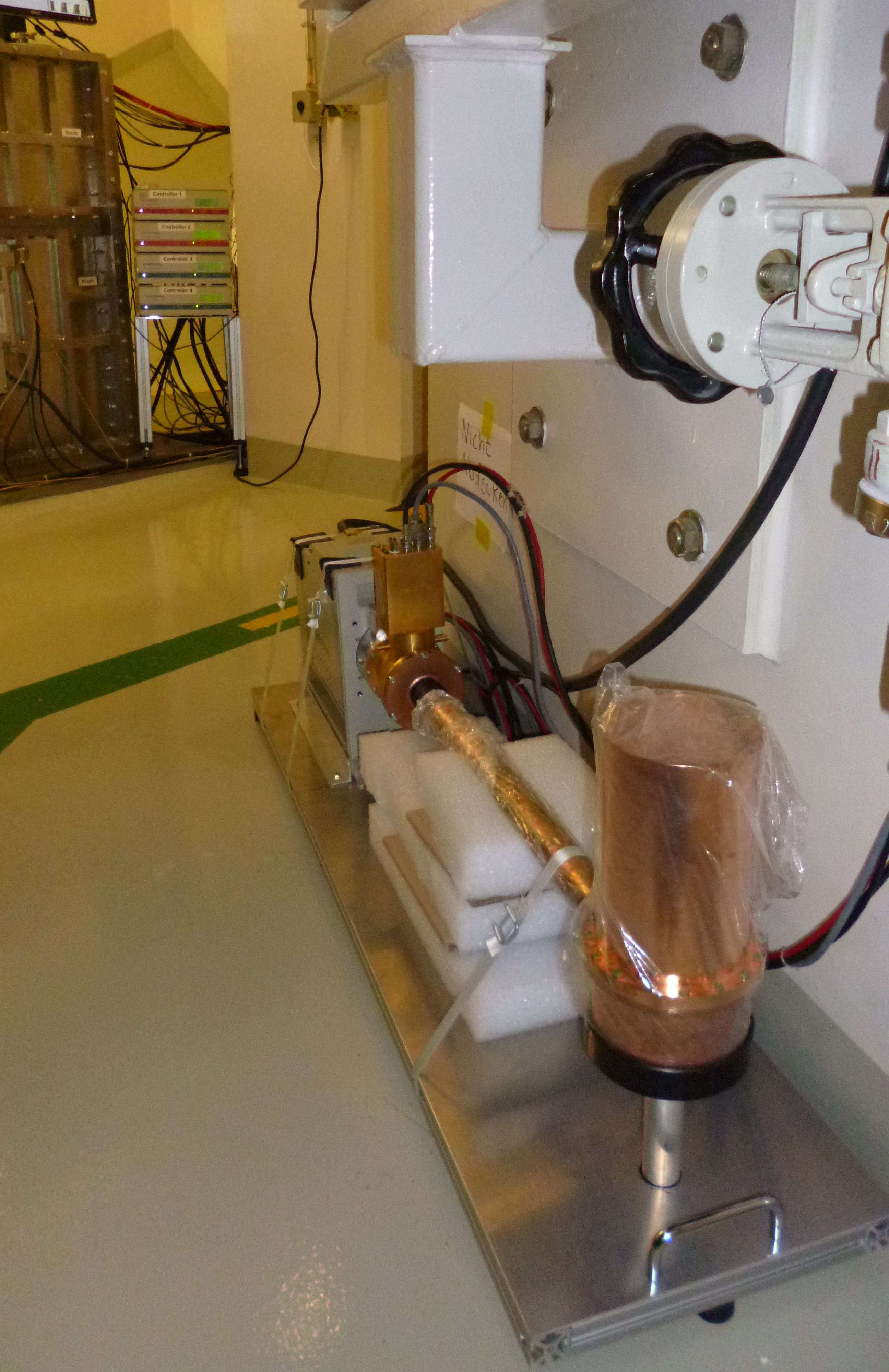}
	\caption{CONRAD detector at room A408 of KBR. The Cu cryostat with the HPGe diode has been placed near the closest wall to the reactor core}
	\label{fig:photo_conrad}
\end{figure}

The CONRAD HPGe spectrometer was deployed inside room A408 from 16.08.2018 to 12.10.2018 during reactor ON time. The detector was mounted onto an Aluminum (Al) plate and was placed next to the wall closest to the reactor core as can be seen in Figure~\ref{fig:photo_conrad} and Figure~\ref{fig:location_A408}, position 3. This results into a mean distance of 13.5\,m to the reactor core for this detector. 

Within $\sim$51\,d of live time, the energy spectrum in the range from 0.4 to 11.6\,MeV$_{ee}$ were recorded continuously except during a few hours of $^{228}$Th calibrations and pulser measurements carried out about every 10\,d. In this way, the stability of the energy scale and detection efficiency was checked as described in Section \ref{subsec:indirectndet}. Small fluctuations within a standard deviation of 
%up to 4 channels corresponding to 
about 1\,keV$_{ee}$ in the large energy range were observed. Thus, for each time interval between $^{228}$Th calibrations the measurement itself with the clearly present background lines was used for the energy calibration. The calibrated spectra were combined afterwards.

The acquired data set was split into daily and hourly time bins to compare to the thermal power available as described in Section \ref{subsec:absthermalpowe}.

\subsection{Measurement results with the CONRAD detector}
\subsubsection{HPGe energy spectrum acquired with reactor ON}
\begin{figure*}[ht!]
    \includegraphics[width=1.\textwidth]{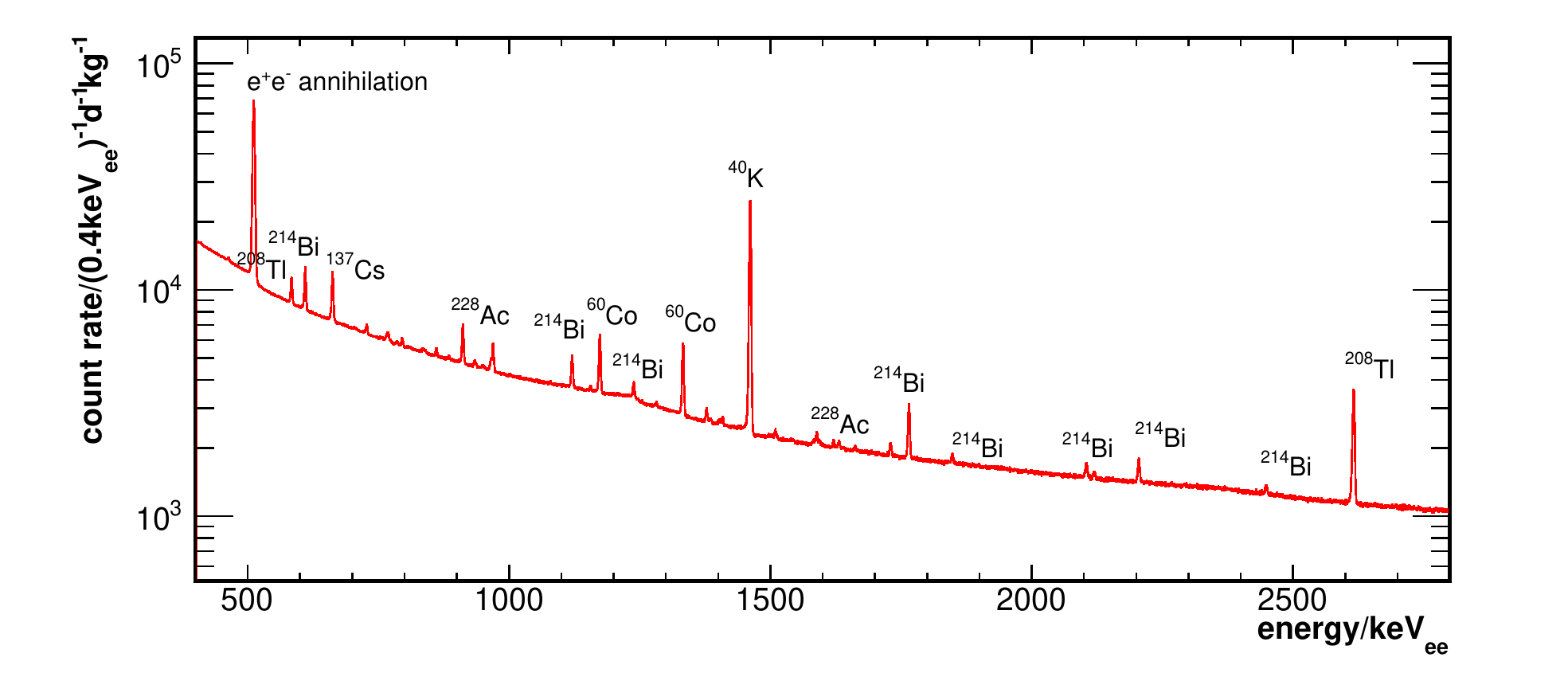}
    \caption{Spectrum of CONRAD detector inside A408 without any shield below 2700\,keV. It is dominated by natural radioactivity in the surrounding. For the strongest lines except the 511\,keV line the decaying isotope is given}
    \label{fig:conradbelow2700keV}
    \end{figure*}

In the measurement with the non-shielded HPGe spectrometer CONRAD inside A408, the contributions from natural radioactivity dominate the spectrum below \linebreak[4]2.7~MeV$_{ee}$. The spectrum is displayed in Figure~\ref {fig:conradbelow2700keV}. Contributions from the Th and U decay chains are visible as expected from the activity measurements of the concrete samples from the floor (cf. Section \ref{subsubnatrad}); $^{137}$Cs, $^{40}$K and $^{60}$Co were found as well. By doing a HPGe screening measurement of the Al board and the plastic holder below the detector, it could be confirmed that except for $^{40}$K, all measured contaminations originate from the interior of A408 and have not been brought inside with the detector setup.

\begin{figure*}[ht!]
    \includegraphics[width=1.\textwidth]{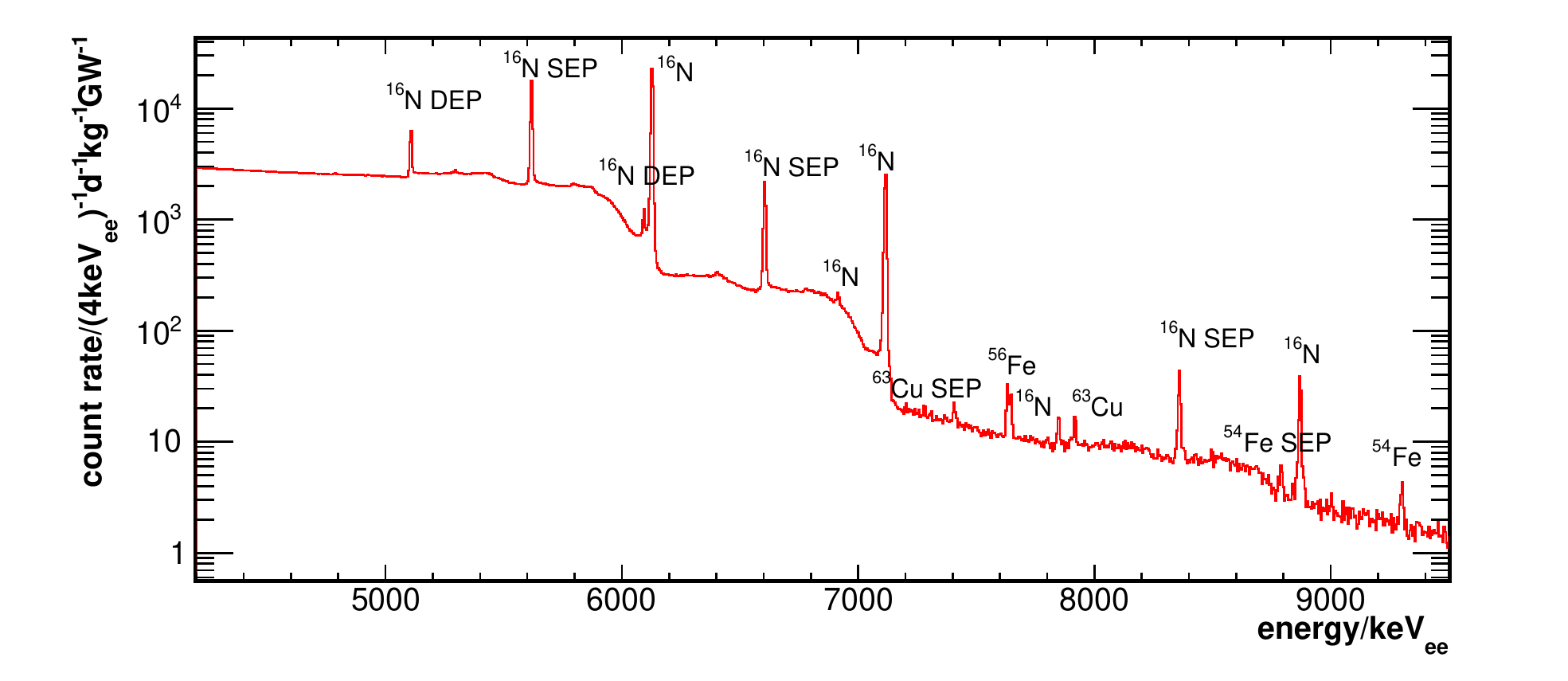}
    \caption{Spectrum of CONRAD detector inside A408 without any shield above 2700\,keV dominated by thermal power correlated contributions. The spectrum is normalized to the mean reactor power. For the strongest lines the decaying isotope is given}
    \label{fig:conradabove2700keV}
    \end{figure*}

Above 2.7\,MeV$_{ee}$, thermal power correlated contributions dominate. The strongest $\gamma$-lines in Figure~\ref{fig:conradabove2700keV} are created from the decay of the short-lived $^{16}$N with a half-life of 7.13\,s \cite{N16lineid}. The isotope is produced predominantly in (n,p) reactions on $^{16}$O in the water of the primary cooling cycle. For this activation process neutron energies higher than 10\,MeV are required \cite{N16crosSection}. The neutrons within the RPV have the highest contribution within this range (described in Section \ref{subsec:resultreactorcoreMC}). Thus, most of the longer living $^{16}$N ions will be produced there and transported to the outside with the flowing cooling water. The closest distance of the CONRAD detector to the loop pipes amounts to 3.8\,m. Not only the decay lines, but also single and double escape peaks (SEP and DEP) from this isotope were observed, where one or both of the $\gamma$-rays from e$^{+}$e$^{-}$ pair production within the HPGe diode leave the detector. 

While below the main lines of $^{16}$N, there is no chance to see other isotopes, at higher energies several $\gamma$-lines from neutron capture were identified \cite{capgamdatabase}.\linebreak[4] Predominately, $\gamma$-lines from neutron capture on Fe are seen created in the reinforced concrete wall. Moreover, lines from neutron capture on Cu can be found, which are created by the capture of thermal neutrons inside A408 in the Cu of the detector cryostat.  

All identified thermal power correlated $\gamma$-lines are listed in Table~\ref{tab:conradlines} together with the literature values from \cite{capgamdatabase} and \cite{N16lineid}. The count rates were determined by a counting method except for the double peak structure of $^{57}$Fe at 7631\,keV and 7646\,keV, which were fitted with two gaussian functions. In the same energy range at 7638\,keV, a line from $^{64}$Cu is expected, but overlaid by the Fe lines. To disentangle the two contributions, the count rate from the Cu $\gamma$-line has been calculated from the count rate of a clearly visible line from the same isotope at higher energies, which was corrected for by the relative branching ratio of the line and the detection efficiency due to the detector geometry derived from the MC. The result was subtracted from the count rate evaluated for the double peak structure.

For the CONUS experiment, 25\,cm of Pb are used to shield against natural radioactivity. The total absorption cross section of Pb is approximately constant from 2 to 10\,MeV \cite{xcomnist}, meaning all $\gamma$-rays inside the room are expected to be successfully suppressed by the shield. This is examined in detail in Section \ref{sec:gammainshield}.

%meaning that all $\gamma$-rays inside the room are successfully suppressed by the shield. Consequently, no increase in the background is observed comparing measurements at the MPIK laboratory to the ones at the power plant with the additional $\gamma$-ray background.
    
\begin{table*}[btp]
\caption{Identified thermal power correlated $\gamma$-lines in the spectrum of the non-shielded CONRAD detector inside A408 during reactor ON time. The energy and branching ratios (br) from literature are given as well. The count rates marked with * were measured as one-peak structure and the single count rates were extracted as described in the text. (1) and (2) mark the different ways the MC is scaled. For (1), the measured thermal neutron fluence inside A408 has been used ((597$\pm$39)\,cm$^{-2}$d$^{-1}$ for 3.9\,GW) and thus the uncertainties of the BSS measurement result dominate the uncertainties.
For (2), the MC has been scaled with the number of neutrons expected to arrive at the outside wall from the ab initio calculation starting at the reactor core ((477$\pm$20)$\cdot$10$^{6}$\,d$^{-1}$ for 3.9\,GW). Here, only statistic uncertainties and uncertainties on the literature values are included. \label{tab:conradlines}}

\begin{tabular}{llll}
\hline\noalign{\smallskip}
energy [keV$_{ee}$] &energy lit [keV] & count rate [cts/d/GW]& count rate MC [cts/d/GW]\\ \noalign{\smallskip}\hline\noalign{\smallskip}
\multicolumn{4}{l}{$^{53}$Fe(n,$\gamma$)$^{54}$Fe}\\ \noalign{\smallskip}\hline\noalign{\smallskip}
8790.9$\pm$0.6 & 8786.8$\pm$1.0 SEP& 9.6$\pm$0.6&  (2)9$\pm$1 \\
9301.0$\pm$0.6 & 9297.8$\pm$1.0 (br 100\%) & 11.3$\pm$0.5& (2)11$\pm$1 \\ \noalign{\smallskip}\hline\noalign{\smallskip}
\multicolumn{4}{l}{$^{56}$Fe(n,$\gamma$)$^{57}$Fe}\\ \noalign{\smallskip}\hline\noalign{\smallskip}
7280.0$\pm$0.7&7278.82$\pm$0.09 (br 20.69\%)& 11.5$\pm$1.1& (2)13$\pm$1\\
7632.8$\pm$0.1&7631.2$\pm$0.1 (br 100\%)& 137$\pm$4*& (2)111$\pm$1 \\
7646.6$\pm$0.1&7645.6$\pm$0.1 (br 86.21\%)& double peak& \\ \noalign{\smallskip}\hline\noalign{\smallskip}
\multicolumn{4}{l}{$^{63}$Cu(n,$\gamma$)$^{64}$Cu}\\ \noalign{\smallskip}\hline\noalign{\smallskip}
7406.2$\pm$0.3& 7405.26$\pm$0.08 SEP&27.2$\pm$1.3&(1)37$\pm$3\\
n.~d.&7638.00$\pm$0.09 (br 48.94\%)& 15.7$\pm$1.6*&(1)19$\pm$2\\
7916.9$\pm$0.2& 7916.26$\pm$0.08 (br 100\%)&29.7$\pm$1.0&(1)37$\pm$3\\  \noalign{\smallskip}\hline\noalign{\smallskip}
\multicolumn{4}{l}{$^{16}$O(n,p)$^{16}$N}\\ \noalign{\smallskip}\hline\noalign{\smallskip}
5106.96$\pm$0.01&5106.63$\pm$0.04 DEP&12612$\pm$146 &\\
5617.51$\pm$0.01&5617.63$\pm$0.04 SEP&49972$\pm$575 &\\
6093.28$\pm$0.04&6093.15$\pm$0.14 DEP&1605$\pm$21 &\\
6128.14$\pm$0.01&6128.63$\pm$0.04 (br 67.0\%)& 85086$\pm$979&\\
6604.36$\pm$0.01&6604.15$\pm$0.14 SEP&7163$\pm$83 &\\
6915.0$\pm$0.4&6915.5$\pm$0.6 (br 0.038\%)&155$\pm$4 &\\
7115.37$\pm$0.02&7115.15$\pm$0.14 (br 4.9\%)&10097$\pm$116 &\\
7848.4$\pm$0.3&7847.3$\pm$0.5 DEP& 31$\pm$1&\\
8359.7$\pm$0.1&8358.3$\pm$0.5 SEP&135$\pm$2 &\\
8870.8$\pm$0.1&8869.3$\pm$0.5 (br 0.076\%)&152$\pm$2 &\\
\noalign{\smallskip}\hline
\end{tabular}
\end{table*}

\subsubsection{Correlation to thermal power for the CONRAD detector measurements}
\begin{figure*}[ht!]
    \includegraphics[width=1.04\textwidth]{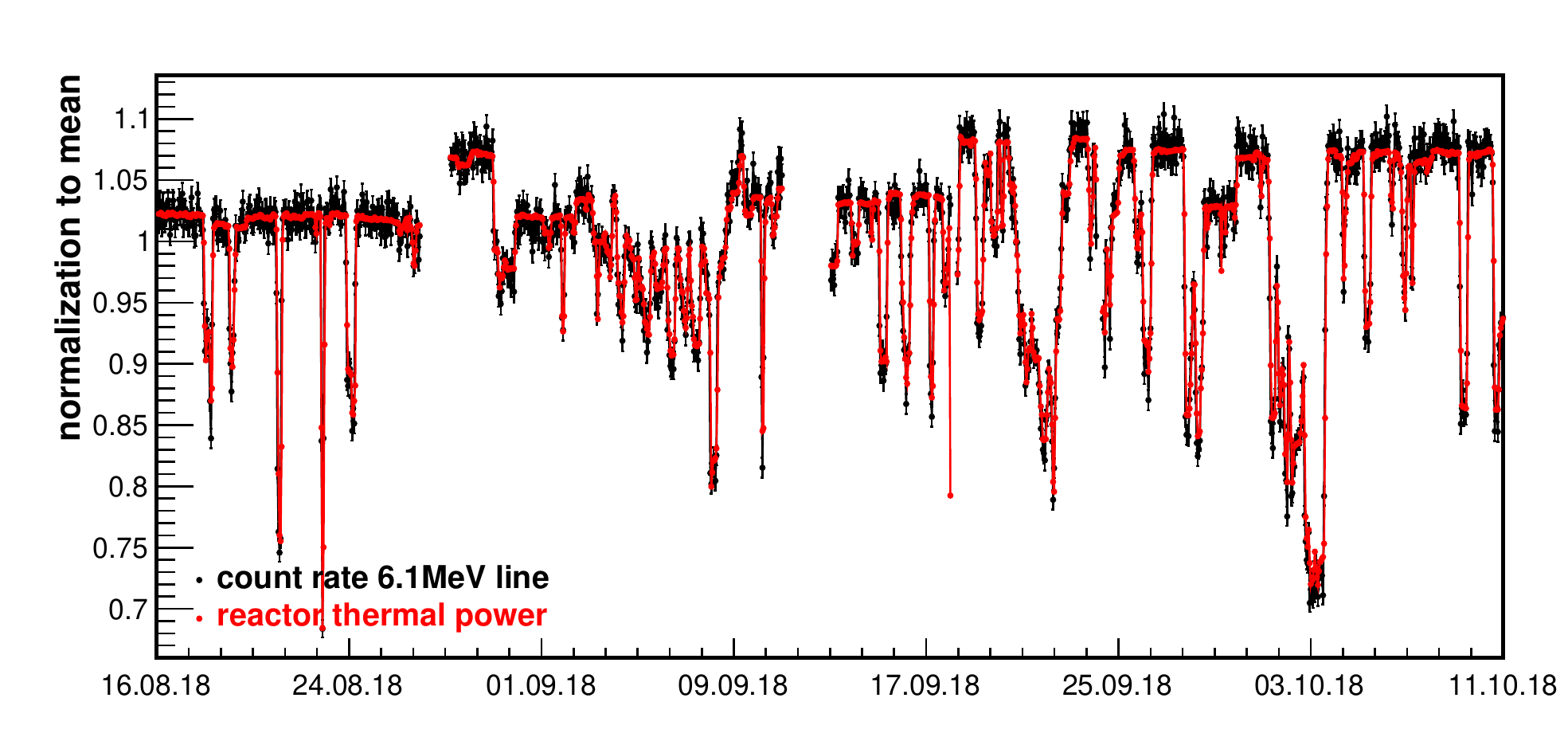}
    \caption{Correlation of thermal power to the count rate in the 6.1\,MeV $\gamma$-line of $^{16}$N. The curve is interrupted if no thermal power data is available and during the calibration of the detector}
    \label{fig:conrad_correlation_thermalpower}
    \end{figure*}

Due to the high statistics in the 6.1\,MeV line of $^{16}$N the thermal power reactor data (see Section \ref{subsec:absthermalpowe}) was compared to the peak count rate for an hourly binning.
The result in Figure~\ref{fig:conrad_correlation_thermalpower} shows a strong correlation, meaning that with the help of a HPGe spectrometer in the distance of 13.5\,m to the reactor core, the current thermal power can be precisely predicted for the specific case of A408, comparable to the thermal neutron fluence (cf. Section \ref{subsec:meas-kbr}). Similar correlations limited by statistics are observed for the $\gamma$-lines from neutron capture. 

\subsection{Comparison of the CONRAD detector measurements with MC}

The neutron capture inside the walls of A408 and the Cu cryostat can be reproduced in the MC. In the last step of the propagation of neutrons from the reactor core to A408, the detector is placed at the wall closest to the reactor core as in reality. For the direct approach simulating the neutrons passing the outside wall of A408, available statistics for the signals inside the HPGe diode are too low to get a reliable result.

Instead, two different sets of simulations were run. (1) First of all, thermal neutrons were started from the inside wall of A408 and normalized to the thermal power correlated thermal neutron fluence determined with the BSS (see Table~\ref{tab:MCneutronfluencerateatdiode}). For the Cu lines, an excellent agreement has been found, while for the Fe lines this MC configuration predicts a count rate of at least a factor of 5 lower. This means that most of the neutron capture reactions responsible for this signal already occur, while the neutrons travel through the walls before they enter room A408.

(2) Simulating neutrons passing through the outside wall of A408, the spectrum of the $\gamma$-rays leaving the respective wall to the inside of the room were registered. In a second step, mono-energetic $\gamma$-rays were started from this wall to determine the geometric detection efficiency of CONRAD. It lies at the order of 10$^{-5}$ to 10$^{-4}$ at the range of 7 to 10\,MeV$_{ee}$. The outcome of the simulation was scaled with the normalization from Section \ref{subsec:resultreactorcoreMC} and the reduction of the neutron fluence in step 1 to 3 from the propagation from the reactor core in Table~\ref{tab:MCsuppressionfactorpropagation}. Next, it has been added up with the results of (1) to take into account neutrons being captured within room A408 and not before entering it. All in all, for the neutron capture on $^{54}$Fe and $^{56}$Fe the same order of magnitude for the count rates have been found as in the measurement (see Table~\ref{tab:conradlines}). Especially, the agreement between MC and measured data is better than observed for the neutron fluence as stated in Section \ref{subsec:normmcoutcome}. This seems to indicate that especially the last step of the propagation through the wall predominately contributes to the lack of neutrons inside A408 expected in the MC.

\section{Expected impact of neutron-induced background signals on CONUS data}
\label{sec:impact-on-cevns}

\subsection{Reactor-induced neutrons at CONUS diodes}

Using the measured thermal power correlated neutron spectrum from Section \ref{subsec:fluences_ON-OFF}, neutrons are propagated through the CONUS shield. The neutrons arriving at the diodes as well as the induced energy depositions in the HPGe diode are registered. All results are normalized to the maximum reactor power of 3.9\,GW. The main uncertainties originate from the uncertainty on the initial spectrum. The results in this Section are given exemplary for C1. 

\begin{figure}[h!]
    \includegraphics[width=0.5\textwidth]{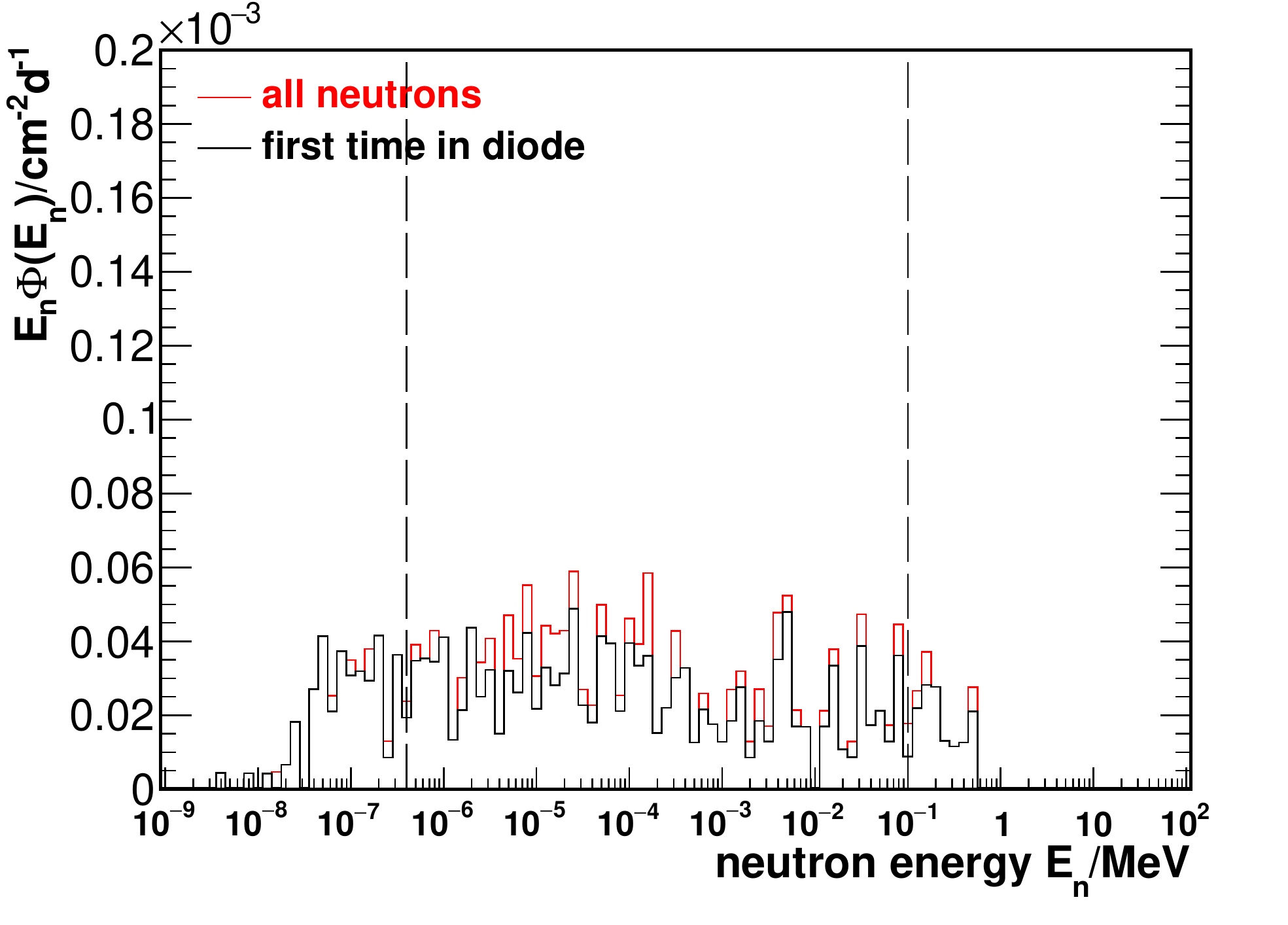}
    \caption{Spectrum of reactor-neutron induced neutrons at one CONUS HPGe diode predicted by MC}
    \label{fig:nspecatdiodeMC}
    \end{figure}

In Figure~\ref{fig:nspecatdiodeMC} the neutron fluence rate through the diode surface is displayed and Table~\ref{tab:MCneutronfluencerateatdiode} summarizes the rates in various energy ranges. Registering only neutrons entering the diode for the first time reduces the fluence by 16\% showing that the backscattering of neutrons around the diode cannot be neglected, especially in the intermediate and thermal energy region. The energy spectrum of these backscattered neutrons is plotted in Figure~\ref{fig:nspecatdiodeMC} as well.

\begin{figure}[h!]
    \includegraphics[width=0.45\textwidth]{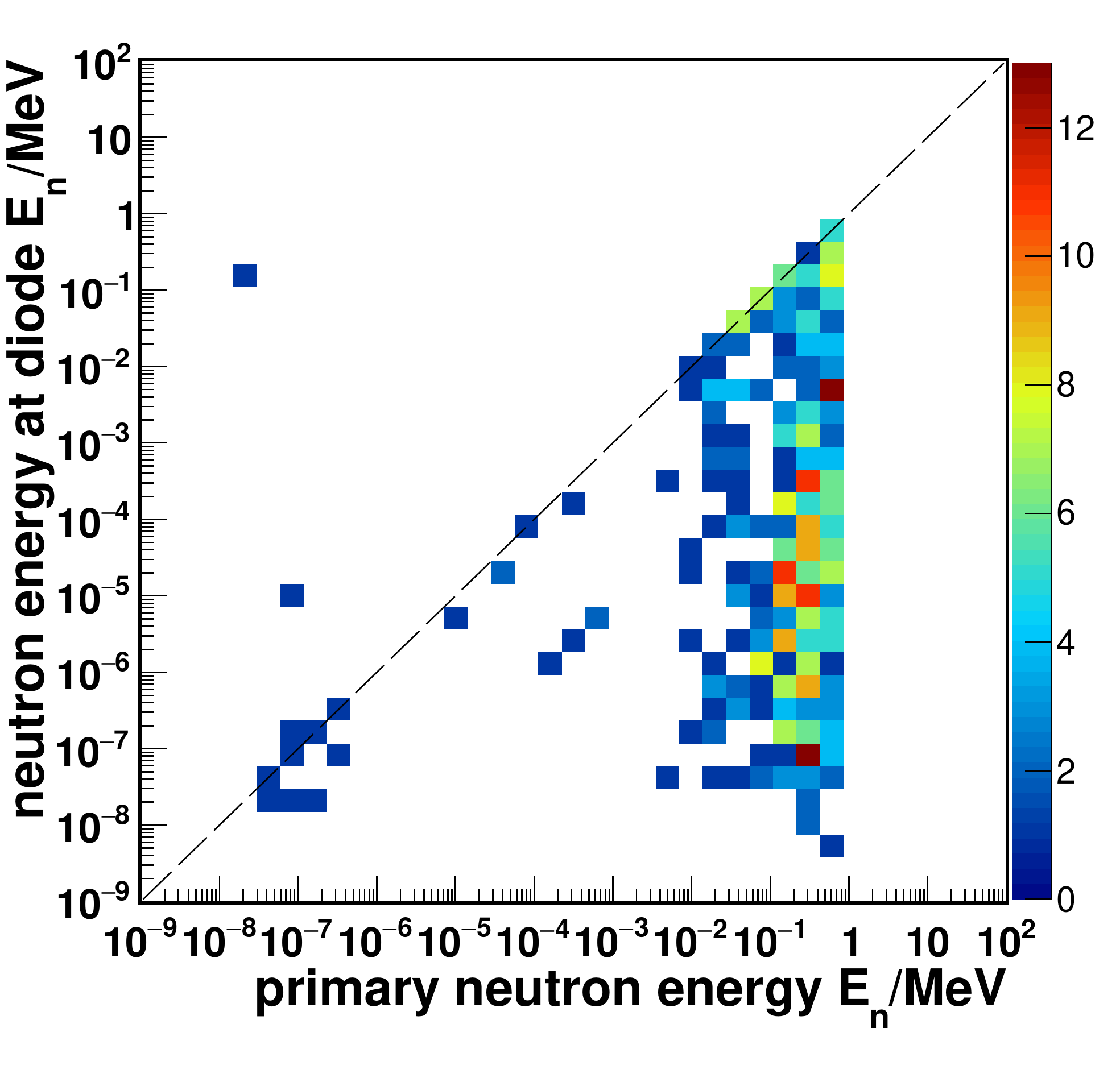}
    \caption{Neutrons arriving at one CONUS HPGe diode plotted against the energy of the primary neutrons. Dashed line at 45$^{\circ}$ for orientation}
    \label{fig:nspecatdiode_vs_nspecprim}
    \end{figure}

Less than 10\% of the neutrons at the diode have energies larger than 100\,keV, while the maximum neutron energy lies clearly below 1\,MeV. Those fast neutrons are the remnants of the fast neutrons started outside the shield, which is illustrated in Figure~\ref{fig:nspecatdiode_vs_nspecprim}. In this Figure, the projection on the y-axis is the spectrum of the neutrons arriving at the diode, while the projection on the x-axis corresponds to the part of input spectrum of the MC, where the neutrons did indeed reach the diode. Further, it shows that most of the intermediate or thermal neutrons at the diode are created in the thermalisation of these fast neutrons within the shield, while the contribution of neutrons with thermal or intermediate energies outside the shield is nearly negligible. The amount of inelastic neutron scattering processes inside the shield creating more than one neutron is also insignificant. 
There is a small fraction of neutrons at the diode at intermediate or fast energies that have been induced by thermal neutrons outside the shield. These are the entries in Figure~\ref{fig:nspecatdiode_vs_nspecprim} above the 45$^{\circ}$ diagonal. In the capture of thermal neutrons, highly energetic $\gamma$-rays with energies up to 10\,MeV can be released from the absorbing nuclei. Those can create fast neutrons by photonuclear reactions with heavy elements such as Pb. This has been confirmed by starting 8\,MeV $\gamma$-rays in the MC simulation at the innermost Pb layer and observing the neutrons arriving at the diode.

All in all, the CONUS shield proves to be highly effective to shield against the reactor neutron induced fluence inside A408 by reducing the overall fluence by a factor of 3$\cdot$10$^{-6}$. The thermal neutrons from outside the shield are completely absorbed.

    \begin{figure}[h!]
    \includegraphics[width=0.47\textwidth]{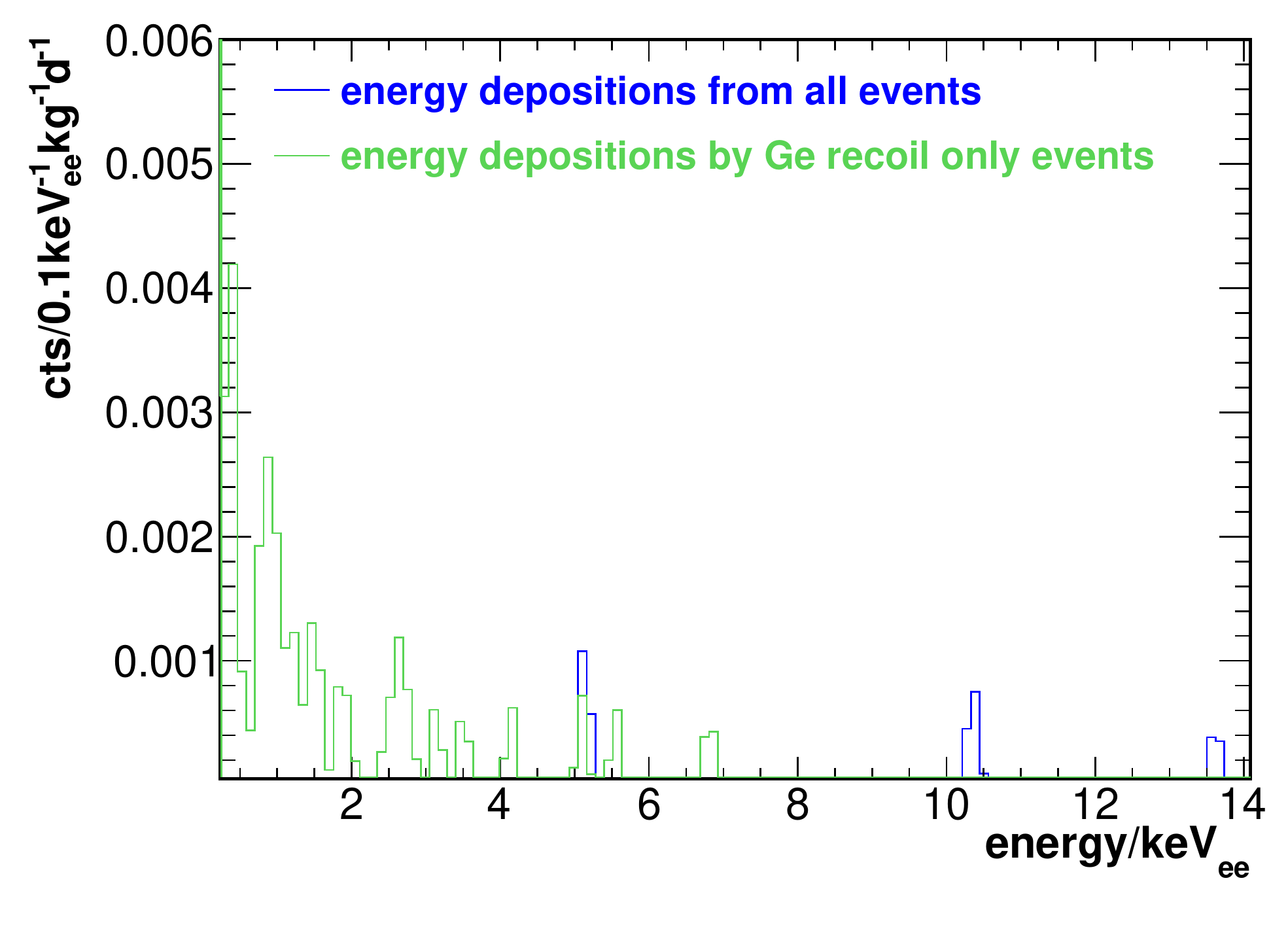}
    \caption{Expected MC spectrum within an HPGe diode from reactor neutrons at low energies. The outcome of the MC has been adapted to the detector response. Moreover, the spectrum of events consisting of only Ge recoils is displayed}
    \label{fig:Gespec_from_reactorn}
    \end{figure}

\begin{table*}[btp]
\caption{Measured reactor neutron induced fluence rate outside the shield compared to the neutron fluence rate at the surface of one CONUS HPGe diode inside the shield from MC for maximum thermal power of the reactor. The measured neutron spectrum inside A408 has been used as input for the simulation. The definition of the $E_\text{n}$ regions is identical to Table~\ref{tab:integral_quantities_ON}. \label{tab:MCneutronfluencerateatdiode}}
\begin{tabular}{llll}
\hline\noalign{\smallskip}
$E_\text{n}$ region &    n outside shield [cm$^{-2}$d$^{-1}$]&n arriving at diode [cm$^{-2}$d$^{-1}$]&first n [cm$^{-2}$d$^{-1}$] \\ \noalign{\smallskip}\hline\noalign{\smallskip}
thermal          & 597$\pm$39  & (0.39$\pm$0.04)$\cdot$10$^{-3}$  &   (0.36$\pm$0.04)$\cdot$10$^{-3}$     \\
intermediate &142$\pm$19 & (1.67$\pm$0.01)$\cdot$10$^{-3}$  &   (1.38$\pm$0.01)$\cdot$10$^{-3}$   \\
fast      & 7$\pm$5 & (0.17$\pm$0.03)$\cdot$10$^{-3}$ &     (0.14$\pm$0.03)$\cdot$10$^{-3}$  \\\noalign{\smallskip}\hdashline\noalign{\smallskip}
total                       & 745$\pm$30  &  (2.24$\pm$0.10)$\cdot$10$^{-3}$    &   (1.89$\pm$0.10)$\cdot$10$^{-3}$     \\
\noalign{\smallskip}\hline
\end{tabular}
\end{table*}

\subsection{Expected signal in p-type HPGe detectors}

The neutrons at the diode with energies from the spectrum of Figure~\ref{fig:nspecatdiodeMC} interact with Ge by neutron capture (mostly thermal neutrons), elastic scattering and inelastic scattering (mostly high energetic neutrons). In elastic and inelastic scattering processes recoils of Ge atoms can occur. 

\subsubsection{Detector response}
To determine the expected energy spectrum in the detector, the MC output has to be adapted to the true detector response.

The MC does not take into account that the energy deposition by a recoiling nucleus is not fully converted into ionization energy. This material-dependent loss is described by the quenching effect. It is included in the post-processing of the MC by correcting the energy deposition of recoils using the Lindhard theory with the fit parameters including the adiabatic correction from the measurement in \cite{Queching:Bscholz}. The energy loss parameter $k$ \cite{Queching:Bscholz} is set to 0.2 for a conservative approach in the estimation of a background contribution. %Especially at low energies, there is a limited number of measurements and thus an overall uncertainty of 20\% is assumed.

Moreover, using the parameters described in Section \ref{subsec:indirectndet}, in the post-processing of the MC the transition layer and dead layer are included as well as the efficiency loss towards the detection threshold. Finally, as there is no electronic response in the MC, the spectrum is folded with the energy resolution observed in the background measurements. 

\subsubsection{Reactor neutron-induced contribution in HPGe detector spectrum}
For CE$\nu$NS, the recoil of Ge atoms is observed. Thus, neutron-induced recoils are the most relevant \linebreak[4]background source. In total, (0.58$\pm$0.03)\,d$^{-1}$kg$^{-1}$ Ge recoils for the full reactor power occur by elastic and inelastic neutron scattering over the whole energy range down to 0\,keV$_{ee}$. For (37$\pm$3)\% of the recoils, no additional energy is deposited inside the diode and the signature in the detector corresponds to the one expected from CE$\nu$NS. This can happen if the neutron only interacts via elastic scattering and no additional particles were created by the neutron traveling through the shield. Alternatively, this is also possible if the $\gamma$-rays created in an inelastic scattering process leave the diode without further interaction. 

In Table~\ref{tab:MCninducedcountrateatdiode}, for the neutron-induced contribution the expected count rates are listed in the region of interest of CE$\nu$NS and at higher energies compared to the count rates observed at the experimental site during reactor OFF time. The expected spectrum in the low energy regime is displayed in Figure~\ref{fig:Gespec_from_reactorn} as well as the spectrum induced by recoil events without further energy depositions. The recoil events create an increasing background towards the detection threshold. Above the energy range displayed only a few events are scattered over the whole spectrum. Furthermore, in Figure~\ref{fig:Gespec_from_reactornplussignal} the expected CE$\nu$NS signal is shown. The signal \cite{Freedman:1973yd} has been adapted to the CONUS experiment in internal sensitivity studies and has been folded with the detector response as described for the neutron contribution with the same energy loss parameter for quenching.

The figures and the plot clearly show that the\linebreak[4] neutron-induced thermal power correlated background is negligible for maximum reactor power over the whole spectral range. The expected CE$\nu$NS signal exceeds the neutron background by at least one order of magnitude.

Comparing the integral ranges outside of the region of interest between reactor ON and OFF data for one month of reactor OFF data and six months of reactor ON data, the results are consistent with the findings here.   

\begin{table}[btp]
\caption{Expected count rate by reactor neutrons inside the CONUS shield for the maximum reactor thermal power of 3.9\,GW compared to the measured background during reactor OFF time. \label{tab:MCninducedcountrateatdiode}}
\begin{tabular}{lll}
\hline\noalign{\smallskip}
energy  &    MC & measurement \\
keV$_{ee}$&[kg$^{-1}$d$^{-1}$] & [kg$^{-1}$d$^{-1}$]\\ \noalign{\smallskip}\hline\noalign{\smallskip}
$[0.3,0.6]$& 0.006$\pm$0.002  & 12$\pm$1\\
$[0.6,11]$& 0.025$\pm$0.005   &148$\pm$2\\
$[11,400]$&  0.15$\pm$0.03  & 716$\pm$16\\
\noalign{\smallskip}\hline
\end{tabular}
\end{table}

\subsection{Comparison to muon-induced neutron background}

\begin{figure}[h]
    \includegraphics[width=0.49\textwidth]{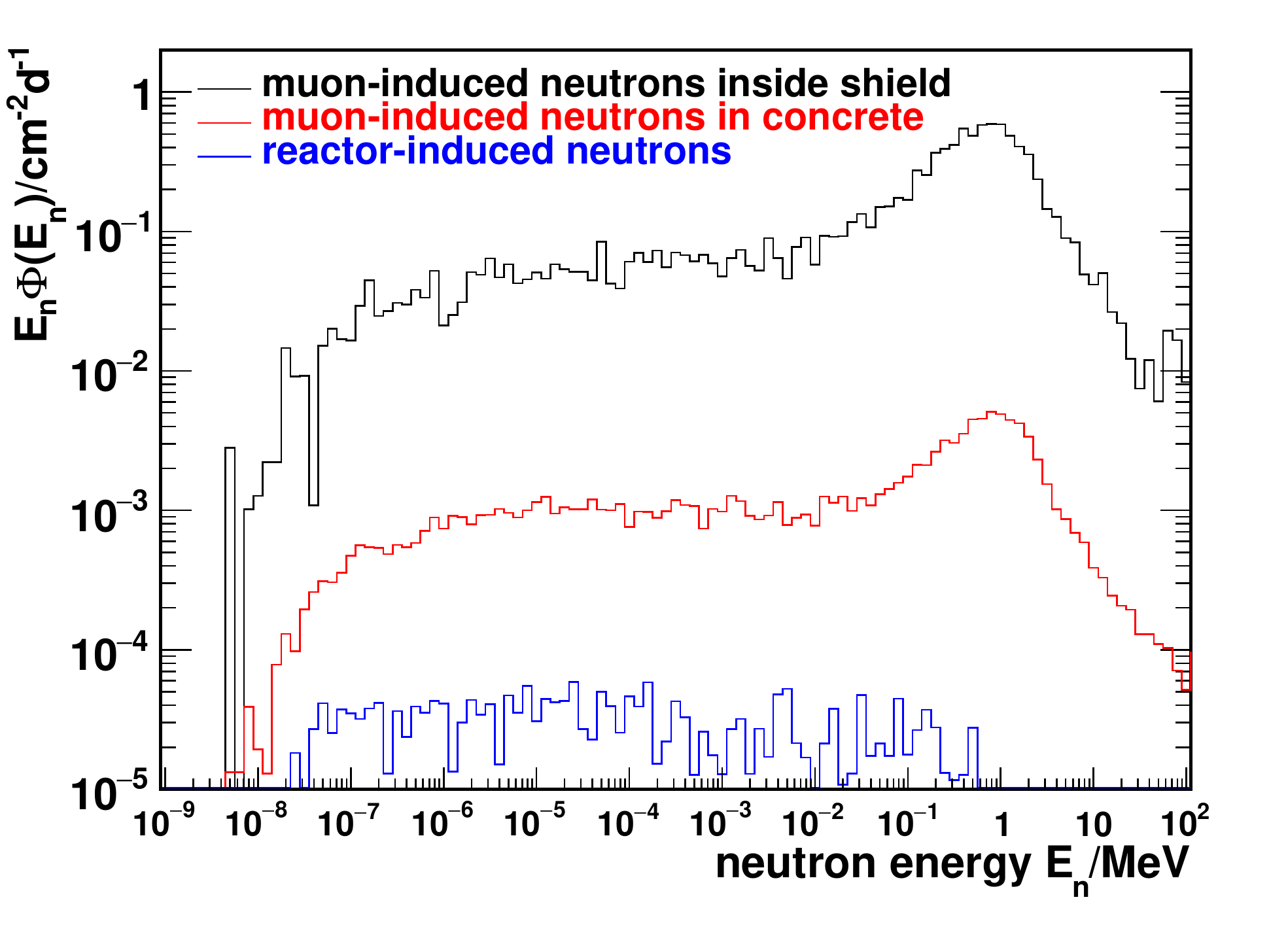}
    \caption{Comparison of MC neutron spectra at the HPGe diode from various sources}
    \label{fig:MC_comparisonnspec_various_sources}
    \end{figure}

The dominant source of neutrons within the CONUS shield are muon-induced neutrons. Contributions from the hadronic component of cosmic rays as well as neutrons from the ($\alpha$,n) reactions are considered negligible. Muon-induced neutrons are created in the concrete and steel of the reactor building as well as in the CONUS shield. The first contribution has been determined by the Bonner sphere measurement during reactor OFF time as described in Section \ref{sec:neutron-measurements} and is propagated through the shield in the same way as the reactor-induced neutrons. 
The latter one has been approximated by using the outcome from simulations in the underground laboratory at MPIK. The method has been described in \cite{hakenmueller:masterthesis} and has been applied to the CONUS shield in \cite{taup2017}. To correct for the thicker overburden, the results have been scaled with a factor of 1.62$^{-1}$ as described in Section \ref{sec:neutron-measurements}. Comparing the different neutron fluence rates at the diode, with (10.8$\pm$0.2)\,cm$^{-2}$d$^{-1}$ most of the neutrons at the diode have been created by muons inside the high-density materials of the shield, followed by (0.126$\pm$0.005)\,cm$^{-2}$d$^{-1}$ muon-induced neutrons inside the overburden and the tiny contribution of (0.0022$\pm$0.0001)\,cm$^{-2}$d$^{-1}$ from the reactor-induced neutron component, which again illustrates the successful suppression of neutrons from outside the shield. The spectral shape of the neutrons is depicted in Figure~\ref{fig:MC_comparisonnspec_various_sources}. Both, the neutron spectrum induced by muons inside the shield and in the concrete propagated through the shield, peak at fast neutron energies slightly below\linebreak[4]1~MeV. The higher neutron energy leads to a recoil spectrum that extends to higher energies. 

To reduce the muon-induced background at shallow depth, the CONUS shield contains an active muon veto with an efficiency of around 99\%, effectively reducing all prompt background components correlated to muons passing through the shield by about two orders of magnitude. This means that the prompt contribution of muon-induced neutrons inside the shield will be of the same order of magnitude as the one of the muon-induced neutrons in the concrete overburden.

%Additionally, metastable isotopes with a live time longer than the veto window of 320\,$\mu$s will be induced by neutron capture. They are barely reduced by the muon veto and thus are solemnly created by the muon-induced neutrons inside the shield. 

\begin{figure}[h]
    \includegraphics[width=0.49\textwidth]{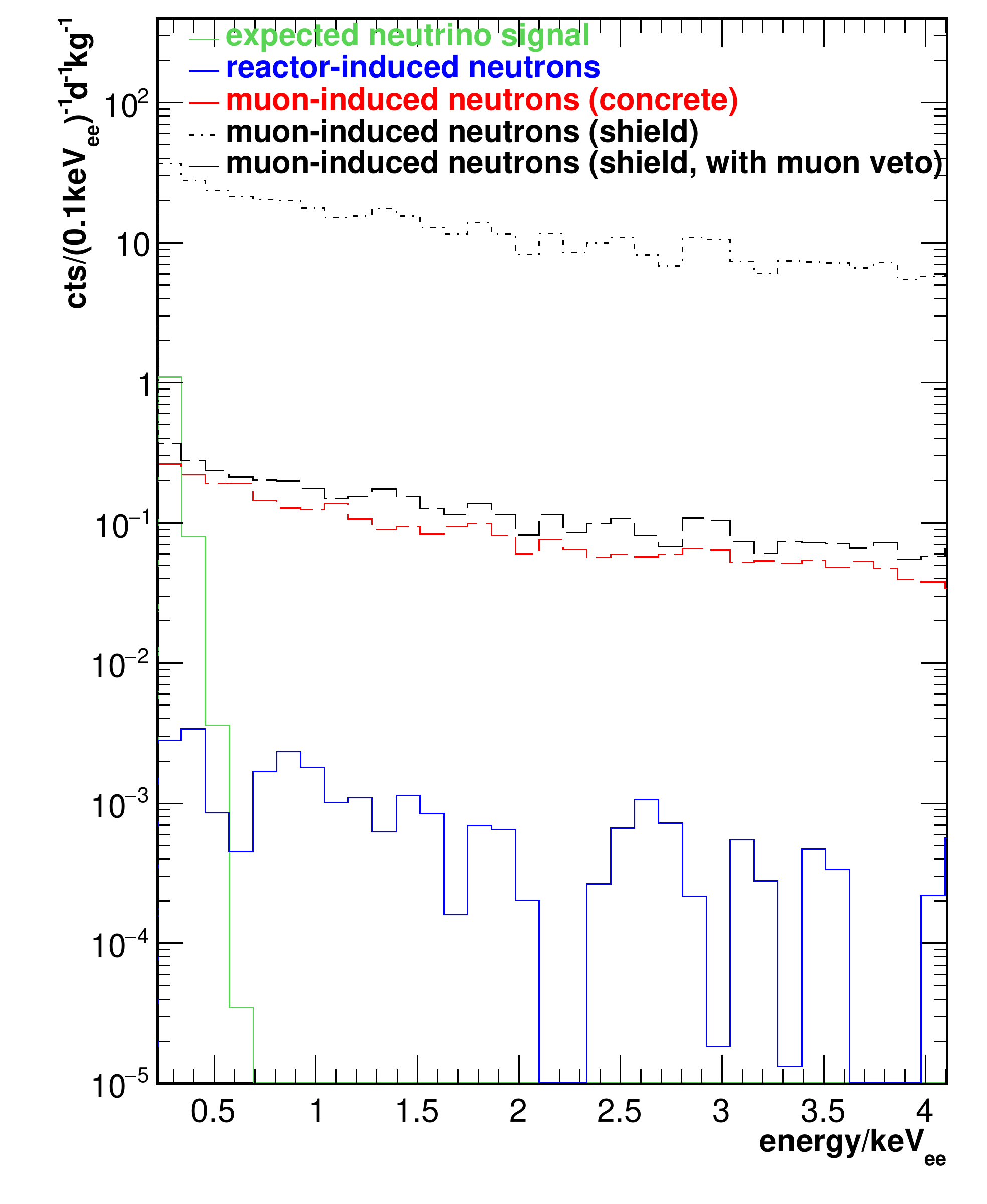}
    \caption{Comparison of the reactor-correlated neutron spectrum at the HPGe (from MC) and the expected CE$\nu$NS signal. The dashed line represents the steady-state muon-induced neutron contributions (from MC) from neutrons created in the concrete and inside the shield (reduced by the muon veto). The short dashed line corresponds to the muon-induced neutrons inside the shield without applied muon veto.}
    \label{fig:Gespec_from_reactornplussignal}
    \end{figure}

Figure~\ref{fig:Gespec_from_reactornplussignal} summarizes all the neutron-induced contributions and the CE$\nu$NS signal. The muon-induced neutron contributions created inside the concrete and inside the shield (displayed without and with applied muon veto) clearly dominate above the \linebreak[4]reactor-correlated contribution, but this is a steady-state background constant during reactor ON and OFF periods and thus can be distinguished from the CE$\nu$NS signal in this way. It will be included as input in the global fit for the background modeling of the CONUS experiment. 

\subsection{Reactor-induced $\gamma$-radiation inside CONUS shield}
\label{sec:gammainshield}
Mono-energetic $\gamma$-lines with the energies of two dominant $\gamma$-lines from $^{16}$N observed with the CONRAD detector above 2700\,keV were propagated through the CONUS shield. The wall adjoined to the space around the reactor core had been chosen as starting point. The simulation outcome was normalized by the number of $\gamma$-rays per day and 3.9\,GW. This amounts to (50.0$\pm$0.7)$\cdot$10${^3}$ for the 6129\,keV $\gamma$-line and to \linebreak[4](6.1$\pm$0.1)$\cdot$10${^3}$ for the 7115\,keV $\gamma$-line, such that the measured count rates inside the CONRAD detector can be reproduced (see Section \ref{tab:conradlines}). Adding up the contributions from the MC and including the detector response, a negligible integral count rate from 0 to 450\,keV of (11$\pm$2)$\cdot$10$^{-5}$\,kg$^{-1}$d$^{-1}$ is determined. This is seven orders of magnitude smaller than the measured background rate in Table~\ref{tab:MCninducedcountrateatdiode}. Thus, the 25\,cm of Pb inside CONUS completely shield this reactor-correlated contribution.

\section{Conclusions}\label{sec:conclusions}
The CONUS experiment is looking for CE$\nu$NS of reactor antineutrinos. It is located inside room A408 of the commercial pressurized water reactor KBR at a distance of only 17.1\,m from the reactor core where the expected antineutrino flux is very high %of the order of 10$^{13}$\,s$^{-1}$cm$^{-2}$. 
due to the close reactor core proximity. Because of neutrons emitted from the reactor core, but also due to the shallow depth of 24\,m w.e., the CONUS experiment might be exposed to a strong neutron-induced background. The most challenging contribution is the one correlated with the time-dependent thermal power: neutrons escaping the reactor core can mimic the CE$\nu$NS signal; radioactive decays of reactor neutron-induced $^{16}$N in the primary water loop cycle can generate events in the region of interest. To study their impact on the CONUS energy spectrum, a multiple approach measurement campaign was conducted. Especially the deployment of the Bonner sphere spectrometer NEMUS and the HPGe spectrometer CONRAD at A408 were crucial.

Bonner sphere measurements were carried out at different locations inside A408, including the exact position of the later installed CONUS detectors. The campaigns took place during reactor ON and OFF periods. The thermal power correlated neutron spectrum could be extracted from the difference of these two measurement periods. From their difference and combining the information of the different spheres that are part of the BSS, the thermal power correlated neutron spectrum could be extracted. A total fluence of (745$\pm$30)\,cm$^{-2}$d$^{-1}$ (for the maximum thermal power of 3.9\,GW) was determined. About 80\% of these neutrons are thermal, while the rest has a maximum energy below 1\,MeV. Due to the low statistics of the measurements and the large number of thermal neutrons, an accurate description of the spectrum, especially at higher neutron energies, requires the support of MC simulations. In this MC, the thermalization process was observed in detail in the propagation of the fission neutrons from the reactor core to room A408 in several steps. Regarding the spectral shape, it confirms the maximum neutron energy below 1\,MeV and reveals a slope in the lethargy representation of the spectrum. An overall reduction of the neutron fluence of 10$^{-20}$ is predicted. Even for this large suppression factor and the complex geometry with limited information available, this ab initio calculation reproduces the fluence measurement approximately within a factor of 38.

For the $\gamma$-ray spectrometric measurement, the non-shielded CONRAD detector was
placed inside A408 close to the CONUS setup and operated during the reactor ON time.
At high energies above 2.6\,MeV$_{ee}$, the spectrum is dominated by $\gamma$-radiation emitted by decays of isotopes produced via (n,p) reactions on $^{16}$O in the water of the primary loop, but there are also $\gamma$-rays from neutron capture on Fe isotopes inside the concrete of the reactor building, as well as on Cu isotopes inside the detector cryostat. At energies below 2.6 MeV$_{ee}$\,there are contributions from natural radioactivity. Their concentrations were deduced via screening measurements using concrete samples taken from A408. Moreover, the concentrations in K, U, Th are comparable to the values obtained in standard concrete.
The neutron capture $\gamma$-lines on the Cu cryostat of the detector could be reproduced accurately by simulating thermal neutrons within A408 and normalizing the MC with the BSS results. This confirms the result and validates the MC. While the amount and geometry of the Cu is very well known, there are only estimations of the iron content inside the walls. The double peak structure of the decay of $^{57}$Fe could be reproduced within 20\%, for the other iron $\gamma$-line peaks even a better agreement with the measurement is achieved. 

All the measurements carried out inside A408 can be linked to the evolution of the thermal reactor power over time. There are several experimental methods applied by KBR to monitor the thermal power and the neutron flux around the reactor core. Using this information, the thermal neutron count rate from the BSS measurements as well as the count rate of the dominant $^{16}$N $\gamma$-line are found to be highly correlated to the reactor progression over time. Especially the high statistical precision of the $\gamma$-line at 1\% within one hour of data taking makes it possible to reproduce the thermal power accurately on fine-grained time intervals. Thus, the KBR thermal power monitoring systems can be used to predict quantitatively the expected neutron fluence at the CONUS specific location in A408 at any time.

Finally, the impact of the measurement results on the CONUS experiment has been evaluated. The measurement results for the neutron fluence and the $\gamma$-line background within A408 have been employed as input of MC simulations. This part of the MC simulation is very well validated due to long-term efforts by the MPIK. Neutrons and $\gamma$-rays were propagated through the well-known geometry of the CONUS shield towards the HPGe diodes, which have been implemented in detail in the MC simulation. The dead layer and transition layer, electronic detection efficiency and quenching have been taken into account to be able to reproduce the detector response comparable to measurements. 

The $\gamma$-line background is successfully suppressed by 25\,cm of Pb with a nearly constant attenuation cross section for $\gamma$-lines of 2 to 10\,MeV within the CONUS shield. This leads to a negligible contribution of the order of 10$^{-4}$\,kg$^{-1}$d$^{-1}$ over the whole spectral range. 
The neutrons are especially moderated and captured by layers of borated PE. It has been found that the reactor-correlated neutron fluence at the diode is two orders of magnitude smaller than the non-reactor correlated neutron background from muon-induced neutrons in the shield and the concrete of the surroundings. 
The expected spectrum fully folded with the detector response is at least one order of magnitude smaller than the expected CE$\nu$NS signal assuming a realistic ionization quenching factor of about 0.2 in Ge at 77\,K. %The neutron-induced recoils in Ge contribute to an extend of \textcolor{red}{less than 1\%} to the background in the region of interest assuming the same quenching factor as before.

The unique approach with multiple experimental efforts combined with MC simulations of the reactor neutron-related background has allowed to confine the impact of this troublesome background on the CONUS spectra. It will be of sub-dominant order for the experiment. The approach further pinned down the difficulties of such measurements, but also the great importance for all experiments situated close to the reactor core looking for fundamental neutrino interactions within and beyond the standard model of particle physics.

%It is a factor of $\approx$1.5 below the one encountered on Earth's surface \cite{Wiegel2002b}.About 80\% of the neutron fluence is thermal and the maximal energy lies below 1\,MeV with a slope in between in lethargy representation of the spectrum.

\begin{acknowledgements}
We are grateful to Jochen Schreiner for the HPGe spectroscopy measurements of the concrete samples and Andreas Klischies for providing information on the concrete structure of the reactor building. We would like to thank Frank K\"ock for the support concerning the computer infrastructure of the MPIK. Furthermore, we appreciate the help of Michael Reissfelder for technical support. J.~H. is supported by the IMPRS-PTFS, T.~R. is supported by the IMPRS-PTFS and the research training group GRK 1940 of the Heidelberg University.
\end{acknowledgements}

\end{document}